\documentclass[a4paper,11pt]{article}

\usepackage{jheppub}

\usepackage{bm}
\usepackage{booktabs}
\usepackage{mathtools}
\usepackage{cancel}
\usepackage{dcolumn}
\usepackage{slashed}
\usepackage{empheq}

\usepackage{multirow}

\usepackage{tikz}

\newcommand{\pill}[1]{%
  \tikz[baseline=(pill.base)]{
    \node[
      draw=black!75,
      rounded corners=2.5pt,
      inner xsep=4.2pt,
      inner ysep=1.5pt,
      line width=0.7pt,
      font=\ttfamily\bfseries
    ] (pill) {#1};
  }%
}

\usetikzlibrary{quantikz}

\hypersetup{
  colorlinks=true,
  citecolor=blue,
  linkcolor=blue,
  urlcolor=blue
}

\newcommand{\VB}{V_{\mathcal B}}

\newcommand{\dd}{\mathrm{d}}

\newcommand{\I}{i}

\newcommand{\bea}{\begin{eqnarray}}
\newcommand{\eea}{\end{eqnarray}}

\arxivnumber{2608.13537}

\title{\boldmath
Kinematic fingerprints of a nucleon-triggered
$V_{\mathcal B}$ in rare
$\eta^{(\prime)}\to\pi^0(\eta)\gamma\gamma$
decays on nucleon targets}

\author[a]{Yaroslav Balytskyi}

\affiliation[a]{
Department of Physics and Astronomy,
Wayne State University,\\
Detroit, Michigan 48201, USA}

\emailAdd{ybalytsk@uccs.edu}

\abstract{The nucleon-triggered vector boson $V_{\mathcal B}$ is a hypothetical particle motivated by a statistically significant discrepancy between the KLOE measurement, $\mathrm{BR}^{\eta\rightarrow\pi^{0}\gamma\gamma}_{\text{KLOE}}  = \left(0.98\pm0.11_{\text{stat}}\pm0.14_{\text{syst}}\right)\times10^{-4}$, and the current world average, $\left(2.55\pm0.22\right)\times10^{-4}$, dominated by MAMI photoproduction data. This hypothetical new interaction is activated by external nucleons, causing deviations from the Standard Model predictions, while leaving leptonic measurements, such as KLOE and BESIII, unaffected. We embed this mechanism in representative MAMI- and JEF-like photoproduction settings, deriving distinctive and directly testable kinematic and cut-dependent signatures to quantify small differences between the two regimes. One such signature is a small predicted population of events beyond the nominal on-shell boundaries of the $\gamma\gamma$ and $\pi^0\gamma$ spectra, containing 4--23 events in a sample of 1200, depending on the selection window. These events correlate with the recoil proton kinematics, and the polar angle relative to the on-shell hypothesis can reach $\lesssim -5^\circ$ at MAMI energies, but only $\approx -0.2^\circ$ at JEF. For identical selection windows, the JEF-like effective branching fraction exceeds the MAMI-like prediction by less than $1\%$, while widening the window increases both predictions by up to $\sim4\%$. Both energy regimes produce a similar upward shift of the mean reconstructed mass, $\langle M_{\pi^0\gamma\gamma}\rangle-m_\eta\simeq +2$--$7~\mathrm{MeV}$. Finally, we identify representative $V_{\mathcal B}$-induced topologies that could contribute to recently reported discrepancies in the angular distributions of $\gamma d\to\pi^0\eta d$ and $\gamma d\to\pi^0\pi^0 d$, leaving their quantitative investigation to future work.

}


\begin{document}

\maketitle
\flushbottom

\section{Introduction}

The decays of neutral pseudoscalar mesons $\eta$ and $\eta^{\prime}$ are sensitive probes of low-energy quantum chromodynamics (QCD) and potential physics beyond the Standard Model (BSM)~\cite{Gan:2020aco}. Notably, the recent measurement by the KLOE Collaboration~\cite{Babusci:2026} of the $\eta\rightarrow\pi^0\gamma\gamma$ decay, $\mathrm{BR}^{\eta \rightarrow \pi^0 \gamma\gamma}_{\mathrm{KLOE}\text{, 2026}} = (0.98 \pm 0.11_{\mathrm{stat}} \pm 0.14_{\mathrm{syst}}) \times 10^{-4}$, where $\eta$ was produced via \(e^+e^- \rightarrow \phi \rightarrow \eta \gamma \rightarrow (\pi^0 \gamma\gamma) + \gamma\), is consistent with their previous result~\cite{DiMicco:2006}: $\mathrm{BR}^{\eta \rightarrow \pi^0 \gamma\gamma}_{\mathrm{KLOE}\text{, 2006}} = (0.84 \pm 0.27 \pm 0.14) \times 10^{-4}$, but is $\approx 5.5\sigma$ different from the current world average~\cite{PDG:2026}, $\mathrm{BR}_{\text{PDG, 2026}}^{\eta \rightarrow \pi^0 \gamma\gamma} = (2.55 \pm 0.22) \times 10^{-4}$. As noted by the KLOE Collaboration~\cite{Babusci:2026}:``This result agrees with a preliminary KLOE measurement, but is a factor of two smaller than the current world average.''

In turn, the current world average for this decay is dominated by the measurement performed by the A2 Collaboration at MAMI~\cite{Nefkens:2014} via photoproduction, $\gamma + p \rightarrow \eta + p \rightarrow \left(\pi^0\gamma\gamma\right) + p$,
reporting a partial decay width: $\Gamma(\eta \rightarrow \pi^0 \gamma\gamma)_{\text{MAMI, 2014}} = (0.33 \pm 0.03)\,\mathrm{eV}$. With the most recent value, $\Gamma_\eta=1.31(5)\,\textrm{keV}$~\cite{PDG:2026}, this corresponds to the branching ratio: $\mathrm{BR}^{\eta\to\pi^0\gamma\gamma}_{\text{MAMI, 2014}}
=(2.52\pm0.25)\times10^{-4}$.

It is important to note that this result has been consistent with the previous measurements performed by Crystal Ball@AGS through the charge-exchange reaction, $
\pi^- + p \rightarrow \eta + n \rightarrow \left(\pi^0\gamma\gamma\right) + n$, 
which were reported in 2005~\cite{Prakhov:2005}, $
\mathrm{BR}^{\eta\rightarrow\pi^0\gamma\gamma}_{\text{Crystal Ball@AGS, 2005}} = (3.5 \pm 0.7 \pm 0.6) \times 10^{-4}$, 
and in 2008~\cite{Prakhov:2008}, $\mathrm{BR}^{\eta\rightarrow\pi^0\gamma\gamma}_{\text{Crystal Ball@AGS, 2008}} = (2.21 \pm 0.24 \pm 0.47) \times 10^{-4}$, 
while an independent reanalysis of the Crystal Ball data produced a consistent value of $\mathrm{BR}^{\eta\rightarrow\pi^0\gamma\gamma}_{\text{Crystal Ball@AGS, reanalysis}}  = (2.7 \pm 0.9 \pm 0.5) \times 10^{-4}$~\cite{Knecht:2004}.

\begin{figure}[!t]
    \centering
    \includegraphics[width=\textwidth]{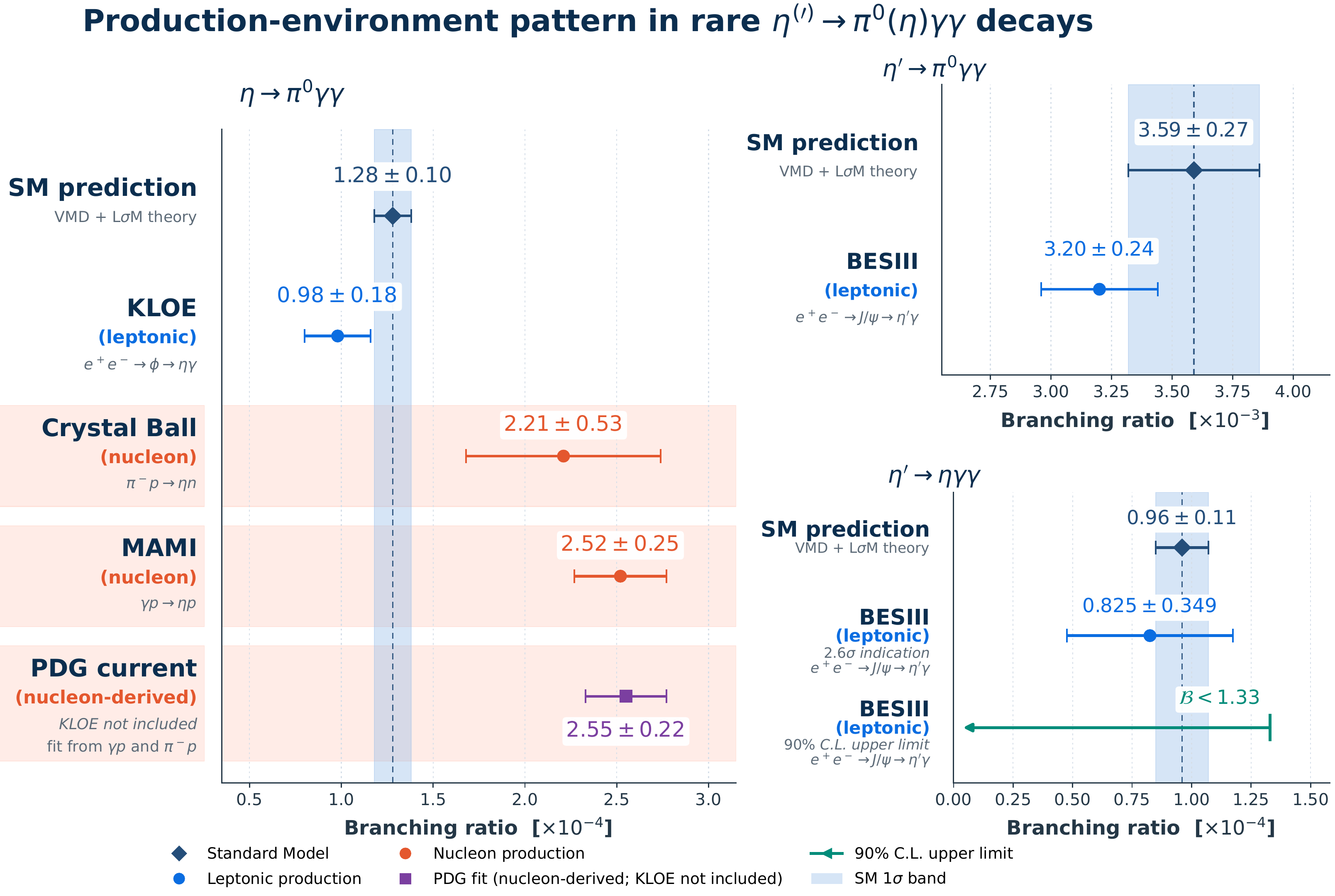}
\caption{The production-dependent pattern that serves as a motivation for the $V_\mathcal{B}$ scenario~\cite{Balytskyi:2026}. }
\label{EnvironmentPatternFig}
\end{figure}

As noted in~\cite{Balytskyi:2026}, except for the older GAMS-2000 experiment \cite{Alde:1984,Landsberg:1985} that used the same charge-exchange reaction and reported a higher branching ratio, $\mathrm{BR}^{\eta\rightarrow\pi^0\gamma\gamma}_{\text{GAMS-2000, 1984}} = (7.1 \pm 1.4) \times 10^{-4}$, the branching-ratio measurements for the $\eta\rightarrow\pi^0\gamma\gamma$ decay, obtained either by photoproduction or charge-exchange reaction, are \underline{self-consistent} and converge to roughly: $\mathrm{BR}^{\eta\rightarrow\pi^0\gamma\gamma}_{\text{Nucleon target}} \approx \left(2.2 - 2.7\right)\times 10^{-4}$, resulting in the current world average~\cite{PDG:2026}.

On the other hand, the decays $\eta^\prime \to \pi^0\gamma\gamma$ and $\eta^\prime \to \eta\gamma\gamma$ have so far been studied only in the leptonic production environment, $e^+e^- \to J/\psi \to \gamma\eta^\prime$, by BESIII~\cite{ablikim2017observation,ablikim2019search}. BESIII measured \(\mathrm{BR}_{\textrm{BESIII}}^{\eta^\prime \to \pi^0 \gamma\gamma} = \left(3.20 \pm 0.07 \pm 0.23\right) \times 10^{-3}\), and reported the corresponding diphoton invariant-mass distribution
~\cite{ablikim2017observation}. This result exceeds the previous GAMS-2000 upper bound, $\mathrm{BR}^{\eta^\prime\to\pi^0\gamma\gamma}<8\times10^{-4}$ at $90\%$ CL~\cite{alde1987neutral}. In the $\eta^\prime\to\eta\gamma\gamma$ channel, BESIII obtained a fitted branching fraction of $\mathrm{BR}_{\mathrm{BESIII}}^{\eta^\prime\to\eta\gamma\gamma} = (0.825\pm0.349)\times10^{-4}$, with a signal significance of $2.6\sigma$~\cite{ablikim2019search}. Since this was not a statistically significant observation, BESIII specified an upper bound of \(\mathrm{BR}_{\textrm{BESIII}}^{\eta^\prime \to \eta \gamma\gamma} < 1.33 \times 10^{-4}\) at \(90\%\) CL~\cite{ablikim2019search}.

The pattern in the $\eta^{\left(\prime\right)}\rightarrow\pi^0\left(\eta\right)\gamma\gamma$ data, identified in~\cite{Balytskyi:2026}, is summarized in Fig.~\ref{EnvironmentPatternFig}. In leptonic production environments, the measured branching fractions agree well with the Standard Model (SM) predictions, represented here by Vector Meson Dominance (VMD) combined with the Linear Sigma Model (L$\sigma$M)~\cite{escribano2020theoretical}. By contrast, measurements performed in nucleon-induced production environments exhibit statistically significant deviations from both the corresponding leptonic measurements and the Standard Model predictions.

The $V_\mathcal{B}$ mechanism, proposed in~\cite{Balytskyi:2026} to accommodate the data pattern in Fig.~\ref{EnvironmentPatternFig}, explicitly distinguishes two distinct experimental classes:
\begin{align}
\text{leptonic production:}\quad
&e^+e^-\to\phi\to\eta\gamma,
\qquad
 e^+e^-\to J/\psi\to\gamma\eta^{\left(\prime\right)},
\label{eq:leptonic-class}
\\
\text{nucleon-target production:}\quad
&\gamma p\to p\,\eta^{(\prime)},
\qquad
\pi^-p\to n\,\eta^{(\prime)},
\label{eq:nucleon-class}
\end{align}
and contributes only when the experiment is performed on nucleon targets.

The existing BESIII data set contains approximately
$1.1\times10^{7}$ $\eta$ mesons produced through
$J/\psi\to\gamma\eta$~\cite{ablikim2023improved}, potentially enabling the future dedicated measurement of $\eta\to\pi^{0}\gamma\gamma$. Since both BESIII and KLOE produce $\eta$ mesons in $e^+e^-$-initiated reactions without external nucleons, the $V_{\mathcal B}$ contribution is absent in both experimental environments. Thus, the model predicts that the BESIII result should be consistent with the KLOE measurement~\cite{DiMicco:2006,Babusci:2026}, $\mathrm{BR}^{\text{predicted}}_{\mathrm{BESIII}}(\eta\to\pi^{0}\gamma\gamma)\simeq\mathrm{BR}_{\mathrm{KLOE}}(\eta\to\pi^{0}\gamma\gamma)$.

On the other hand, the current experiment at JLab Eta Factory
(JEF)~\cite{JEFproposal,Gan:JEFII2026}, for which
$\eta\to\pi^0\gamma\gamma$ is the flagship channel, has unique
opportunities for testing the $V_{\mathcal B}$ mechanism. At both JEF and MAMI, $\eta$-mesons are produced by photoproduction on protons~\cite{Nefkens:2014}, so the $V_{\mathcal B}$ mechanism must be active in both experimental environments.

However, since the mechanism involves an interaction between the decay
products and an external nucleon, the magnitude of the observed
deviation from the corresponding $e^+e^-$-initiated measurement
depends on the production kinematics and event-selection criteria. JEF and MAMI should therefore exhibit the same qualitative enhancement relative to the leptonic-production baseline, although the branching fractions extracted by the two experiments need not be identical, since the experiments operate at significantly different representative incident-photon energies,
$E_{\gamma}^{\mathrm{MAMI}}\sim1.4\,\mathrm{GeV}$ and
$E_{\gamma}^{\mathrm{JEF}}\sim11\,\mathrm{GeV}$.

In~\cite{Balytskyi:2026}, the $V_{\mathcal B}$ contribution was
computed in the on-shell approximation,
\begin{equation}
\label{OffShellCondition}
\begin{aligned}
M_{\pi^0\gamma\gamma}&=m_\eta
&&\text{for }\eta\to\pi^0\gamma\gamma,\\
M_{\pi^0\gamma\gamma}&=m_{\eta'}
&&\text{for }\eta'\to\pi^0\gamma\gamma,\\
M_{\eta\gamma\gamma}&=m_{\eta'}
&&\text{for }\eta'\to\eta\gamma\gamma
\end{aligned}
\end{equation}

The present work relaxes this assumption and quantifies the
dependence of the predicted observables on the production kinematics
and selection cuts. The most distinctive signatures, discussed below, are events populating regions beyond the nominal on-shell endpoints of the $\gamma\gamma$ and $\pi^0(\eta)\gamma$ invariant-mass spectra, with the resulting tail populations correlated with the recoil-nucleon kinematics. Additional signatures include cut-dependent shifts in the reconstructed meson mass, $M_{\eta^{(\prime)}}^{\mathrm{rec}}$, and in the extracted branching fraction.

If the $V_{\mathcal B}$ interpretation of the KLOE--MAMI discrepancy is correct, the same effective operator also permits the $V_{\mathcal B}$-mediated two-nucleon exchange amplitudes in coherent $\gamma d\to\pi^0\eta d$ and $\gamma d\to\pi^0\pi^0d$ photoproduction. These channels are particularly interesting since both exhibit unexpectedly flat deuteron angular distributions that are not reproduced by Standard Model calculations, as discussed in detail below.

For $\gamma d\to\pi^{0}\eta d$, marked discrepancies have been observed between the experimental results~\cite{Ishikawa:2022,Figueiredo:2026} and theoretical predictions~\cite{MartinezTorres:2023}. As emphasized in~\cite{Ishikawa:2024review}:``No theoretical calculations reproduce rather flat angular distributions of deuteron emission.''. Moreover~\cite{Ishikawa:2026review}, ``The rather uniform distributions may suggest existence of two-baryon correlated states (dibaryons) in the intermediate states.'' A closely related pattern appears in the coherent $\gamma d\to\pi^{0}\pi^{0}d$ photoproduction~\cite{Ishikawa:2017Pi0Pi0,Ishikawa:2019Pi0Pi0,Jude:2022Pi0Pi0,Fix:2005DoublePion,Egorov:2015Pi0Pi0}, for which~\cite{Ishikawa:2019Pi0Pi0} notes: ``The measured angular distribution of deuteron emission is rather flat, which cannot be reproduced by the kinematics of quasi-free $\pi^0\pi^0$ production with deuteron coalescence.''

Further in the text, we identify representative $V_{\mathcal B}$-mediated two-nucleon diagrams that could contribute to both reactions by generating an additional interaction between the constituent nucleons and redistributing the large momentum transfer. We defer a quantitative calculation of these contributions and their interference with the conventional amplitudes to future work.

The remainder of the paper is organized as follows. Section~\ref{VB_rationale} reviews the $V_{\mathcal B}$ model and its underlying rationale, and Section~\ref{VB_Kinematic_Fingerprints} develops the analytical framework for relaxing the on-shell condition in Eq.~\eqref{OffShellCondition} to treat the resulting off-shell kinematics. Section~\ref{KinematicRubicon} introduces the kinematic Rubicon as an additional discriminator of the $V_{\mathcal B}$ scenario and derives the corresponding on-shell kinematic boundaries. Section~\ref{Numerical_Results} presents the corresponding numerical predictions. Guided by these results, Section~\ref{VB_search_strategy} identifies the most promising experimental search strategies and the sharpest kinematic signatures for testing the $V_{\mathcal B}$ hypothesis. In Section~\ref{VB_Angular_Distributions}, we present representative $V_{\mathcal B}$ diagrams contributing to coherent $\gamma d\to\pi^{0}\eta d$ and $\gamma d\to\pi^{0}\pi^{0}d$ photoproduction, while deferring their detailed quantitative evaluation to future work. We summarize our conclusions and outline directions for further study in Section~\ref{Conclusions}. Appendices~\ref{Appendix:SM_benchmark} and~\ref{Appendix:PhaseSpaceFactorization} collect the auxiliary results.

\section{$V_\mathcal{B}$ model and its rationale}
\label{VB_rationale}

The $V_{\mathcal B}$ mechanism, proposed in~\cite{Balytskyi:2026},
introduces a leptophobic vector boson with mass
$m_{V_{\mathcal B}}\simeq1.5$--$5.0~\mathrm{GeV}$, and the
low-energy interaction: 
\begin{equation}
\mathcal{L}^{V_{\mathcal B}}_{\rm eff}
=
\frac{g_{\rm eff}}{4}
(\bar N N)\,
P\,V_{\!\mathcal B}^{\mu\nu}
\widetilde F_{\mu\nu},
\label{eq:nucleon-operator}
\end{equation}
where $P\in\{\pi^0,\eta,\eta'\}$,
$\bar N N\equiv\bar p p+\bar n n$, and
$\widetilde F^{\mu\nu}
=\tfrac12\varepsilon^{\mu\nu\alpha\beta}F_{\alpha\beta}$.

A representative tree-level mediator realization of
Eq.~\eqref{eq:nucleon-operator} contains a real CP-even scalar
$S$ coupled to nucleons, a real CP-odd dark pion $\pi_D$, and
a trilinear interaction connecting the visible and dark
pseudoscalar sectors:
\begin{align}
\mathcal L^{V_{\mathcal B}}_{\rm mediator}\supset\;
&\underbrace{g_S\,S\,\left(\bar N N\right)}_{\text{nucleon portal}}
+\underbrace{c_{SP}\,S\,P\,\pi_D}_{\text{trilinear bridge}}
+\underbrace{
\frac{C_{V_{\mathcal B}\gamma}}{f_{\pi_D}}\,
\pi_D\,V_{\!\mathcal B}^{\mu\nu}\widetilde F_{\mu\nu}
}_{\text{mixed dark anomaly}} 
\label{UV_complete}
\end{align}

\begin{figure}[!t]
\centering
\includegraphics[width=\textwidth]{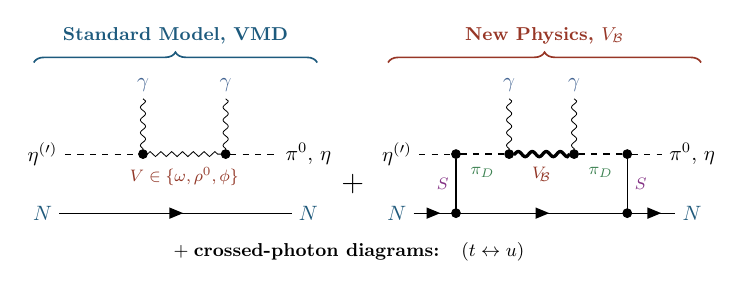}
\caption{The dominant Standard Model contribution mediated by $V\in\{\omega,\rho^0,\phi\}$, in which the nucleon remains a spectator, and the nonfactorizable $V_{\mathcal B}$ contribution, in which energy--momentum can be transferred between the final-state particles and the external nucleon.}
\label{fig:SM_VB_cross}
\end{figure}

Imposing the renormalized vacuum condition
$\langle S\rangle_{\rm vac}=0$ prevents the trilinear
interaction from generating a nucleon-independent
$P\pi_D$ mixing term. Therefore, tree-level exchange of $S$ and $\pi_D$ generates the nucleon-dependent interaction responsible for production on nucleon targets at MAMI, while the corresponding nucleon-independent interaction, relevant for KLOE and BESIII, is absent at this order. Integrating out $S$ and $\pi_D$ at tree level yields the nucleon-triggered contact operator in Eq.~\eqref{eq:nucleon-operator}:
\begin{equation}
\mathcal L^{V_{\mathcal B}}_{\rm eff}\supset
\frac{g_S\,c_{SP}\,C_{V_{\mathcal B}\gamma}}
     {m_S^2m_{\pi_D}^2f_{\pi_D}}\,
(\bar N N)\,P\,
V_{\!\mathcal B}^{\mu\nu}\widetilde F_{\mu\nu},
\qquad
g_{\rm eff}
=
\frac{4g_S\,c_{SP}\,C_{V_{\mathcal B}\gamma}}
     {m_S^2m_{\pi_D}^2f_{\pi_D}} 
\label{eq:mediator-matching}
\end{equation}

The rationale for the mediator realization in Eq.~\eqref{UV_complete} can be summarized as follows. As emphasized in~\cite{Balytskyi:2026}, it occupies a relatively weakly constrained region of parameter space. Existing fifth-force, nuclear, astrophysical, and collider searches do not directly exclude a new nucleophilic scalar with $m_S\sim1~\mathrm{GeV}$~\cite{joseph2026,epelbaum2002,kamiya2015,xu2013,dev2020,hardy2025,epelbaum2009,winkler2019,boiarska2019,batell2019,kling2023}. As a result, $\mathcal O(1)$ nucleon-level Yukawa coupling $g_S$ remains phenomenologically viable and can generate an effective interaction strong enough to explain the sharp $2.6$-fold ($5.5\sigma$) KLOE--MAMI discrepancy. The trilinear bridge in Eq.~\eqref{UV_complete} is similar to the conventional $f_0\left(500\right)\pi\pi$ coupling, except that the scalaron couples the Standard Model pseudoscalar to a dark-sector pseudoscalar. Being nonderivative, this interaction does not introduce additional momentum suppression near the threshold, the kinematic regime probed at MAMI~\cite{Nefkens:2014}. Finally, the mixed dark anomaly term in Eq.~\eqref{UV_complete} can arise as a gauged Wess--Zumino--Witten-type interaction in a confining dark sector of the hidden-valley (HV) type~\cite{strassler2007,kuwahara2023,asadi2026}. Such a realization does not require a direct coupling of $V_{\mathcal B}$ to Standard Model quark or baryon currents and therefore avoids the bounds on the corresponding couplings.

The Standard Model and $V_{\mathcal B}$ amplitudes are added coherently, with the corresponding topologies illustrated in Fig.~\ref{fig:SM_VB_cross}. The complete Standard Model amplitude,
summarized in Appendix~\ref{Appendix:SM_benchmark}, is dominated by vector-meson
exchange (VMD), $V\in\{\omega,\rho^0,\phi\}$, and only this leading contribution is displayed in the figure. In this factorizable contribution, the nucleon remains a spectator, and the parent $\eta^{(\prime)}$ meson is on shell, so the reconstructed invariant mass of the corresponding $\pi^0(\eta)\gamma\gamma$ final state satisfies the on-shell condition in Eq.~\eqref{OffShellCondition}. By contrast, the nonfactorizable $V_{\mathcal B}$ contribution allows energy--momentum transfer between the external nucleon and the final-state particles, and we quantify the corresponding effects further in the text.

Since \(V_{\mathcal B}\) is a \rm{GeV}-scale particle, the kinematic region of interest satisfies \(M_{V_{\mathcal B}}^2\gg |t|,|u|\), and therefore its propagator can be expanded, for \(x\in\{t,u\}\), as: 

\begin{equation}
\frac{1}{
M_{V_{\mathcal B}}^2-x-\I M_{V_{\mathcal B}}\Gamma_{\VB}
}
=
\frac{1}{M_{V_{\mathcal B}}^2}
\left[
1+\mathcal{O}\!\left(
\frac{x}{M_{V_{\mathcal B}}^2},
\frac{\Gamma_{\VB}}{M_{V_{\mathcal B}}}
\right)
\right],
\label{eq:contact-expansion}
\end{equation}
and the \(V_{\mathcal B}\) exchange can be treated as a contact-like interaction.

Since \(V_{\mathcal B}\) is assumed to be isoscalar~\cite{Balytskyi:2026}, its contribution arising from the two insertions of the interaction in
Eq.~\eqref{eq:nucleon-operator}, represented diagrammatically in
Fig.~\ref{fig:SM_VB_cross}, can be parametrized in the isospin limit by a single effective
coefficient: 
\begin{equation}
C_{\mathrm{eff}}
\equiv
\frac{g_{\mathrm{eff}}^2 C_N}
     {4\pi M_{V_{\mathcal B}}^2},
\qquad
\bigl\langle p(n)\bigl|(\bar N N)^2\bigr|p(n)\bigr\rangle
=
C_N(2m_N)^6,
\label{eq:Ceff-definition}
\end{equation}
where \(C_N\) represents the nonperturbative nucleon matrix element, including target-dependent nuclear corrections.

The resulting contact-like $V_\mathcal{B}$ contribution is added coherently to the conventional VMD amplitude. After combining the \(t\)- and \(u\)-channel contributions, it takes the form: 
\begin{equation}
\label{MVBAmplitude}
\begin{split}
\mathcal{M}_{V_{\mathcal B}}
={}&
\underbrace{
\frac{g_{\mathrm{eff}}^2 C_N}
     {4\pi M_{V_{\mathcal B}}^2}
}_{C_{\mathrm{eff}}}
\frac{
4(2m_N)^6
g_{\omega\eta^{(\prime)}\gamma}
g_{\omega\pi^0(\eta)\gamma}
}{
\alpha_{\mathrm{em}}
}\times
\left[
\left(
P\!\cdot\!(q_1+q_2)-2m_{\eta^{(\prime)}}^2
\right)\{a\}
-2\{b\}
\right],
\end{split}
\end{equation}
where the Lorentz structures \(\{a\}\) and \(\{b\}\) are defined in the same way as for the VMD contribution, summarized in Appendix~\ref{Appendix:SM_benchmark}.

In the next Section~\ref{VB_Kinematic_Fingerprints}, we examine the distinctive kinematic consequences of the \(V_{\mathcal B}\) mechanism and quantify them in Section~\ref{Numerical_Results}.

\section{Kinematic fingerprints of \(V_{\mathcal B}\)} \label{VB_Kinematic_Fingerprints}

To isolate the dependence on the $\pi^0\left(\eta\right)\gamma\gamma$ invariant mass, we adopt the following symmetric window as an idealized proxy:
\begin{equation} 
\left|M_X-m_{\eta^{(\prime)}}\right| \leq \Delta_{\rm cut} \quad\Longleftrightarrow\quad M_X^2\in \left[ \left(m_{\eta^{(\prime)}}-\Delta_{\rm cut}\right)^2, \left(m_{\eta^{(\prime)}}+\Delta_{\rm cut}\right)^2 \right], 
\label{eq:continuum-mass-window} 
\end{equation}
where $\Delta_{\rm cut}$ denotes the half-width of the $\eta^{(\prime)}$ selection window. 

We use Eq.~\eqref{eq:continuum-mass-window} to quantify the sensitivity of our results to the assumed mass window. This proxy is not intended to reproduce an experiment-specific reconstruction. A detector-level analysis would additionally account for detector resolution and acceptance, kinematic-fit criteria, background rejection, and other experiment-specific selections. In Section~\ref{Numerical_Results}, we present results for three representative half-widths, $\Delta_{\rm cut}\in\{50,75,100\}~\mathrm{MeV}$.

The total amplitude associated with Fig.~\ref{fig:SM_VB_cross} consists of the Standard Model contribution proceeding through the $\eta^{(\prime)}$ pole, and the $V_{\mathcal B}$ contribution that can populate configurations with $M\neq m_{\eta^{(\prime)}}$, including the kinematic region outside the on-shell condition in Eq.~\eqref{OffShellCondition}. We parametrize the total amplitude as:

\begin{equation}
\mathcal{M}^{\text{Nucleon}}(M,\Omega)
=
\underbrace{
\frac{\mathcal{N}_{\rm SM}(M,\Omega)}
{M^2-m_{\eta^{(\prime)}}^2
-\I m_{\eta^{(\prime)}}\Gamma_{\eta^{(\prime)}}}
}_{\mathcal{M}_{\rm SM}(M,\Omega)}
+
\mathcal{M}_{V_{\mathcal B}}(M,\Omega),
\label{eq:total-amplitude}
\end{equation}
where $\Omega$ collectively denotes the remaining phase-space variables. The squared amplitude is therefore:
\begin{align}
\left|\mathcal{M}^{\rm Nucleon}(M,\Omega)\right|^2
={}&
\left|\mathcal{M}_{\rm SM}(M,\Omega)\right|^2
+
2\operatorname{Re}\!
\left[
\mathcal{M}_{\rm SM}^{\dagger}(M,\Omega)
\mathcal{M}_{V_{\mathcal B}}(M,\Omega)
\right]
\nonumber\\
&+
\left|\mathcal{M}_{V_{\mathcal B}}(M,\Omega)\right|^2 
\label{eq:total-amplitude-squared}
\end{align}
To examine the invariant-mass support of the interference term, we define: 
\begin{equation}
\Delta \equiv M^2-m_{\eta^{(\prime)}}^2,
\qquad
\gamma_{\eta^{(\prime)}}\equiv
m_{\eta^{(\prime)}}\Gamma_{\eta^{(\prime)}},
\qquad
\mathcal{C}(M,\Omega)\equiv
\mathcal{N}_{\rm SM}^{\dagger}(M,\Omega)
\mathcal{M}_{V_{\mathcal B}}(M,\Omega)
\end{equation}
The interference contribution can then be written as: 
\begin{equation}
2\operatorname{Re}\!
\left[
\mathcal{M}_{\rm SM}^{\dagger}(M,\Omega)
\mathcal{M}_{V_{\mathcal B}}(M,\Omega)
\right]
=
2\operatorname{Re}\!
\left[
\frac{\mathcal{C}(M,\Omega)}
{\Delta+\I\gamma_{\eta^{(\prime)}}}
\right]
\label{eq:interference-finite-width}
\end{equation}
Using the Sokhotski--Plemelj formula: 
\begin{equation}
\lim_{\gamma\to0^+}
\frac{1}{\Delta+\I\gamma}
=
\operatorname{PV}\frac{1}{\Delta}
-\I\pi\delta(\Delta),
\label{eq:Sokhotski}
\end{equation}
in the narrow-width limit, we obtain: 
\begin{align}
2\operatorname{Re}\!
\left[
\mathcal{M}_{\rm SM}^{\dagger}(M,\Omega)
\mathcal{M}_{V_{\mathcal B}}(M,\Omega)
\right]
\underset{\Gamma_{\eta^{(\prime)}}\to0}{\longrightarrow}{}&
\underbrace{
2\operatorname{PV}
\left[
\frac{\operatorname{Re}\left[\mathcal{C}(M,\Omega)\right]}
{M^2-m_{\eta^{(\prime)}}^2}
\right]
}_{\text{Neglected in this work}}
\nonumber\\
&+
2\pi\,
\operatorname{Im}\left[\mathcal{C}
\bigl(m_{\eta^{(\prime)}},\Omega\bigr)\right]
\delta\!\left(M^2-m_{\eta^{(\prime)}}^2\right)
\label{eq:interference-NWA}
\end{align}
The principal-value term is odd in $\Delta=M^2-m_{\eta^{(\prime)}}^2$ about $\Delta=0$, and when integrated over a symmetric invariant-mass window in $\Delta$, it cancels at leading order, assuming that $\operatorname{Re}\mathcal{C}(M,\Omega)$ varies slowly across the resonance.

The mass-selection interval in Eq.~\eqref{eq:continuum-mass-window}, however, is symmetric in $M$, not in $M^2$. For the phenomenologically most relevant channel, $\eta\rightarrow\pi^0\gamma\gamma$, the resulting residual integral of the principal-value kernel in Eq.~\eqref{eq:interference-NWA} is:
\begin{align}
    \int^{\left(m_{\eta}+\Delta_{\rm cut}\right)^2}_{\left(m_{\eta}-\Delta_{\rm cut}\right)^2}\rm{d}M^2 \operatorname{PV}
\left[
\frac{1}
{M^2-m_{\eta}^2}
\right] = -\ln\left[1 - \frac{2\Delta_{\rm cut}}{\Delta_{\rm cut} + 2m_\eta }\right] = \frac{\Delta_{\rm cut}}{m_\eta} + \rm{O}\left[\left(\frac{\Delta_{\rm cut}}{m_\eta}\right)^3\right]
\end{align}
The $\pi\delta(M^2-m_\eta^2)$ kernel in Eq.~\eqref{eq:interference-NWA} instead integrates to $\pi$. We define the geometric window-asymmetry factor: 
\begin{align}
 \epsilon_{\rm{PV}-\delta }^{\rm window} \equiv \frac{I_{\rm PV}}{\pi} = -\frac{\ln\left[1 - \frac{2\Delta_{\rm cut}}{\Delta_{\rm cut} + 2m_\eta }\right]}{\pi}   = 
\begin{cases}
    0.029, \ \text{for} \ \Delta_{\rm cut} = 50\ \rm{MeV} \\
    0.044, \ \text{for} \ \Delta_{\rm cut} = 75\ \rm{MeV} \\
    0.058, \ \text{for} \ \Delta_{\rm cut} = 100\ \rm{MeV}
\end{cases}
\end{align}
These two terms multiply $\operatorname{Re}\mathcal{C}(M,\Omega)$ and $\operatorname{Im}\mathcal{C}(M,\Omega)$, respectively. Schematically, the corresponding ratio is the following:  
\begin{align}
\left|
\frac{I_{\rm interference}^{\rm PV}}
{I_{\rm interference}^{\delta}}
\right|
\sim
\epsilon_{\rm{PV}-\delta }^{\rm window}
\left|
\frac{\operatorname{Re}\mathcal{C}(M,\Omega)}
{\operatorname{Im}\mathcal{C}(M,\Omega)}
\right|
\end{align}
Therefore, assuming that this ratio is not parametrically enhanced and that $\mathcal{C}(M,\Omega)$ varies slowly within the mass-selection interval, we expect the residual principal-value contribution to be suppressed at the few-percent level. 

Within the accuracy of this work, we neglect this residual asymmetric correction, and treat the interference as effectively being on shell and localized at the $\eta^{(\prime)}$ pole by the $\delta$-function term in Eq.~\eqref{eq:interference-NWA}. In other words, the interference is localized on the common phase space of the pole and continuous terms, i.e. on shell.

The purely resonant Standard Model term is also localized at the pole $\eta^{\left(\prime\right)}$, since:

\begin{equation}
\frac{1}{\Delta^2+\gamma_{\eta^{(\prime)}}^2} \ 
\underset{\Gamma_{\eta^{(\prime)}}\to0}{\longrightarrow}
\
\frac{\pi}{m_{\eta^{(\prime)}}\Gamma_{\eta^{(\prime)}}}
\delta\!\left(M^2-m_{\eta^{(\prime)}}^2\right)
\label{eq:Breit-Wigner-NWA}
\end{equation}
Therefore, up to the suppressed principal value term in
Eq.~\eqref{eq:interference-NWA}, which we neglect, both the Standard Model contribution and its interference with the new $V_{\mathcal{B}}$ amplitude are confined to the $\eta^{(\prime)}$ resonance.

The only unsuppressed contribution extending into the off-shell region is therefore
$\lvert\mathcal{M}_{V_{\mathcal B}}(M,\Omega)\rvert^2$.
This continuum contribution is the root cause of the distinct kinematic fingerprints of the $V_{\mathcal B}$ mechanism, analyzed in detail below. We therefore decompose the full nucleon-activated $V_{\mathcal B}$ contribution, shown in Fig.~\ref{fig:SM_VB_cross}, into its pole and continuum components:
\begin{equation}
\left|\mathcal{M}^{\rm Nucleon}(M,\Omega)\right|^2
\simeq
\begin{gathered}[t]
\underbrace{
\left.
\Biggl[
\left|\mathcal{M}_{\rm SM}(M,\Omega)
+\mathcal{M}_{V_{\mathcal B}}(M,\Omega)\right|^2
-
\left|\mathcal{M}_{V_{\mathcal B}}(M,\Omega)\right|^2
\Biggr]
\right|_{M=m_{\eta^{(\prime)}}}
}_{\mathclap{\substack{
\displaystyle
\mathcal W_{\rm pole}(\Omega):
\ \text{\pill{universal}\enspace pole component}\\
\text{production-, cut-, and beam-energy-independent;
identical for JEF and MAMI}
}}}
\\[1.5ex]
\kern-3em
+
\underbrace{
\left|\mathcal{M}_{V_{\mathcal B}}(M,\Omega)\right|^2
}_{\mathclap{\substack{
\displaystyle
\mathcal W_{\rm cont}(M,\Omega):
\ \text{\pill{production-dependent}\enspace continuum}\\
\text{encodes cut- and beam-energy dependence;
off-shell fingerprints differ between JEF and MAMI}
}}}
\kern2em
\end{gathered}
\label{eq:pole-continuum-decomposition},
\end{equation}
where the amplitudes evaluated at \(M=m_{\eta^{(\prime)}}\) are understood with the resonant propagator already integrated out in the narrow-width approximation.

To describe the kinematics of the continuum component
$\mathcal W_{\rm cont}(M,\Omega)$ of the nucleon-activated
$V_{\mathcal B}$ mechanism, we isolate the reconstructed three-body
subsystem $X\equiv\pi^0(\eta)\gamma\gamma$ within the full
$2\rightarrow4$ reaction
$\gamma+p\rightarrow p+\pi^0(\eta)\gamma\gamma$:
\begin{equation}
\gamma(q)+p(p)
\longrightarrow
p(p^\prime)+X(Q),
\qquad
X(Q)\equiv
\Bigl[
\bigl(\pi^0(k)\ \text{or}\ \eta(k)\bigr)
+\gamma(q_1)+\gamma(q_2)
\Bigr]_{\rm reconstructed}
\label{eq:continuum-process}
\end{equation}

In the proton-rest frame, $p=(m_p,\vec{\boldsymbol{0}})$ and $q^2=0$. Denote the total initial four-momentum by $P$, and the corresponding momentum relations have the form: 
\begin{equation}
\begin{aligned}
P&\equiv q+p=p^\prime+Q,
&\qquad P^2&\equiv s=m_p^2+2m_pE_\gamma,\\
Q&\equiv k+q_1+q_2,
&\qquad Q^2&\equiv M_X^2,
\end{aligned}
\label{eq:continuum-momenta}
\end{equation}
where $m_p$ is the proton mass and $E_\gamma$ is the incident photon
energy in this frame.

Note that the operator in Eq.~\eqref{eq:nucleon-operator} does not modify the $\eta^{(\prime)}$ production at tree level~\cite{Balytskyi:2026}, and therefore, $\eta^{(\prime)}$ enters the nucleon-assisted topology in Fig.~\ref{fig:SM_VB_cross} as an on-shell state. Its subsequent interaction with the nucleon permits four-momentum exchange between the reconstructed $\pi^0(\eta)\gamma\gamma$ subsystem and the recoil proton, so that its reconstructed momentum need not satisfy $Q^2=m_{\eta^{(\prime)}}^2$. Consequently, $M_X=\sqrt{Q^2}$ is a continuously varying invariant mass and coincides with $m_{\eta^{(\prime)}}$ only in the on-shell approximation.

In the narrow-width limit and for an ideal detector, the pole component is localized at $M_X=m_{\eta^{(\prime)}}$. Any nonzero selection window containing the pole therefore retains the complete pole contribution, making $\mathcal W_{\rm pole}$ independent of $\Delta_{\rm cut}$.

By contrast, the fraction of $\mathcal W_{\rm cont}$ retained by the
selection is inherently continuously variable with $\Delta_{\rm cut}$. Continuum events inside the window pass the same invariant-mass selection and enter the measured sample as $\eta^{(\prime)}$ candidates. Although these events cannot be separated from pole events using the invariant-mass cut alone, their presence can be identified statistically through the kinematic fingerprints described below.

Using the phase-space factorization derived in
Appendix~\ref{Appendix:PhaseSpaceFactorization}, the four-body phase space can be factorized as:  
\begin{equation}
\mathrm{d}\Phi_4
\left(P;p^\prime,p_{\pi^0\left(\eta\right)},q_1,q_2\right)
=
\mathrm{d}\Phi_2\left(P;p^\prime,Q\right)
\frac{\mathrm{d}M_X^2}{2\pi}
\mathrm{d}\Phi_3\left(Q;p_{\pi^0\left(\eta\right)},q_1,q_2\right)
\label{eq:four-body-phase-space-factorization}
\end{equation}

At fixed $M_X$, we define the $V_{\mathcal B}$-driven partial width
associated with the reconstructed three-body subsystem $X$ as:
\begin{align}
\Gamma_{V_{\mathcal B}}(M_X)
&\equiv
\frac{1}{2!}\frac{1}{2M_X}
\int
\mathrm d\Phi_3(Q;k,q_1,q_2)
\nonumber\\[-0.2em]
&\qquad\times
\left|
\mathcal M_{V_{\mathcal B}}
\left(
M_X,
m_{\pi^0(\eta)\gamma_1}^2,
m_{\pi^0(\eta)\gamma_2}^2,
m_{\gamma_1\gamma_2}^2
\right)
\right|^2
\nonumber\\[0.5em]
&=
\frac{1}{2!}\frac{1}{256\pi^3M_X^3}
\int_{\mathrm{Dalitz}(M_X)}
\mathrm d m_{\pi^0(\eta)\gamma_1}^2\,
\mathrm d m_{\pi^0(\eta)\gamma_2}^2
\nonumber\\[-0.2em]
&\qquad\times
\left|
\mathcal M_{V_{\mathcal B}}
\left(
M_X,
m_{\pi^0(\eta)\gamma_1}^2,
m_{\pi^0(\eta)\gamma_2}^2,
m_{\gamma_1\gamma_2}^2
\right)
\right|^2,
\label{eq:VB-running-width}
\end{align}
where the integration is performed over the physical Dalitz region at fixed $M_X$, bounded by:
\begin{equation}
m_{\pi^0(\eta)\gamma_1}^2
+m_{\pi^0(\eta)\gamma_2}^2
+m_{\gamma_1\gamma_2}^2
=
M_X^2+m_{\pi^0(\eta)}^2,
\label{eq:continuum-dalitz-relation}
\end{equation}
and the factor $1/2!$ accounts for the two identical photons in the final state. 

The production-level two-body phase-space in
Eq.~\eqref{eq:four-body-phase-space-factorization} is:

\begin{equation}
\mathrm d\Phi_2(P;p^\prime,Q)
=
\frac{\lvert\vec p_{\rm CM}\rvert}
{16\pi^2\sqrt{s}}\,
\mathrm d\Omega
\label{eq:production-two-body-phase-space}
\end{equation}
Integrating over the full solid angle,
$\int\mathrm d\Omega=4\pi$, and simplifying gives:
\begin{align}
\Omega_2(E_\gamma,m_p,M_X)
&\equiv
\int\mathrm d\Phi_2(P;p^\prime,Q)
\nonumber\\
&=
\frac{1}{8\pi s}
\sqrt{
\left[s-\left(m_p-M_X\right)^2\right]
\left[s-\left(m_p+M_X\right)^2\right]
}
\label{eq:integrated-production-phase-space}
\end{align}
This results in the integrated width driven by the $\mathcal W_{\rm cont}(M,\Omega)$ term in Eq.~\eqref{eq:pole-continuum-decomposition}:
\begin{equation}
\Gamma^{\rm eff}_{V_{\mathcal B}}
\left(E_\gamma,\Delta_{\rm cut}\right)
=
\frac{
\displaystyle
\int_{M_-^2}^{M_+^2}
\frac{\mathrm d M_X^2}{2\pi}\,
\Omega_2\!\left(E_\gamma,m_p,M_X\right)\,
\Gamma_{V_{\mathcal B}}\!\left(M_X\right)
}{
\displaystyle
\int_{M_-^2}^{M_+^2}
\frac{\mathrm d M_X^2}{2\pi}\,
\Omega_2\!\left(E_\gamma,m_p,M_X\right)
},
\qquad
M_\pm\equiv m_{\eta^{(\prime)}}\pm\Delta_{\rm cut}
\label{eq:effective-continuum-width}
\end{equation}
The total observed width within the selected reconstructed-mass
window is:
\begin{align}
\Gamma_{\rm pole}
&\equiv
\frac{1}{2!}\,
\frac{1}{256\pi^3m_{\eta^{(\prime)}}^3}
\int_{\mathrm{Dalitz}\left(m_{\eta^{(\prime)}}\right)}
\mathrm d m_{\pi^0(\eta)\gamma_1}^2\,
\mathrm d m_{\pi^0(\eta)\gamma_2}^2
\nonumber\\[-0.2em]
&\quad\times
\mathcal W_{\rm pole}\!\left(
m_{\eta^{(\prime)}},
m_{\pi^0(\eta)\gamma_1}^2,
m_{\pi^0(\eta)\gamma_2}^2,
m_{\gamma_1\gamma_2}^2
\right),
\label{eq:pole-width}
\\[0.5em]
\Gamma^{\rm total}_{\rm observed}
\left(E_\gamma,\Delta_{\rm cut}\right)
&=
\Gamma_{\rm pole}
+
\Gamma^{\rm eff}_{V_{\mathcal B}}
\left(E_\gamma,\Delta_{\rm cut}\right)
\label{eq:total-observed-width}
\end{align}
where the first term, \(\Gamma_{\rm pole}\), contains the pure Standard Model contribution combined with its interference with the $V_{\mathcal B}$ amplitude, being effectively on shell in Eq.~\eqref{eq:pole-continuum-decomposition}. 

Note that in the limit $\Delta_{\rm cut}\rightarrow 0$,  the prescription defined by Eq.~\eqref{eq:effective-continuum-width} and Eq.~\eqref{eq:total-observed-width} reduces to an on-shell approximation in Eq.~\eqref{OffShellCondition}. Therefore, we proceed in two steps. First, we fit the coefficient in Eq.~\eqref{eq:Ceff-definition} to the published MAMI differential decay width~\cite{Nefkens:2014} in an on-shell approximation. This is consistent with the MAMI extraction of the differential decay width for $\eta\rightarrow\pi^0\gamma\gamma$, which used a signal template effectively corresponding to an on-shell $\eta$~\cite{Nefkens:2014}: ``The centroid and width of the Gaussian were fixed to the values obtained from the previous fit to the MC simulation for $\gamma p\rightarrow\eta p\rightarrow\pi^0\gamma\gamma p$.'' Second, we use the resulting fitted coefficient to estimate the kinematic and finite-window effects that arise when the on-shell condition in Eq.~\eqref{OffShellCondition} is relaxed.

In the ongoing experimental analyses at JEF~\cite{JEFproposal,Gan:JEFII2026}, it would be valuable to perform two parallel extractions: one using the conventional on-shell-$\eta$ signal template, and the other one without imposing an on-shell-$\eta$ assumption, keeping
$M_X=m_{\pi^0\gamma\gamma}$ as an unconstrained observable. Otherwise, the continuum contribution, parameterized as $\Gamma^{\rm eff}_{V_{\mathcal B}}
\left(E_\gamma,\Delta_{\rm cut}\right)$ in Eq.~\eqref{eq:total-observed-width}, may be diluted or partially absorbed by the fitted background.

In addition, to characterize the selected event sample by its average reconstructed mass, we define the continuum-weighted mean mass as:
\begin{equation}
\begin{gathered}
\left\langle
M_X\,\Gamma_{V_{\mathcal B}}\!\left(M_X\right)
\right\rangle_{V_{\mathcal B}}
\left(E_\gamma,\Delta_{\rm cut}\right)
\equiv
\frac{
\displaystyle
\int_{M_-^2}^{M_+^2}
\frac{\mathrm d M_X^2}{2\pi}\,
\Omega_2\!\left(E_\gamma,m_p,M_X\right)\,
M_X\,\Gamma_{V_{\mathcal B}}\!\left(M_X\right)
}{
\displaystyle
\int_{M_-^2}^{M_+^2}
\frac{\mathrm d M_X^2}{2\pi}\,
\Omega_2\!\left(E_\gamma,m_p,M_X\right)
},
\\[0.8ex]
M_\pm\equiv
m_{\eta^{(\prime)}}\pm\Delta_{\rm cut}
\end{gathered}
\label{eq:continuum-mean-mass}
\end{equation}
The average reconstructed mass of the sample, taking into account both the pole and continuum parts, is thus:
\begin{align}
\left\langle M_X\right\rangle_{\rm reconstructed}
\left(E_\gamma,\Delta_{\rm cut}\right)
&=
\frac{
m_{\eta^{(\prime)}}\,\Gamma_{\rm pole}
+
\left\langle M_X\,\Gamma_{V_{\mathcal B}}\!\left(M_X\right)\right\rangle_{V_{\mathcal B}}
\left(E_\gamma,\Delta_{\rm cut}\right)
}{
\Gamma^{\rm total}
\left(E_\gamma,\Delta_{\rm cut}\right)
}
\label{eq:average-reconstructed-mass}
\end{align}
Finally, we evaluate the corresponding combined pole-plus-continuum
differential decay widths,
\begin{equation}
\frac{\dd\Gamma^{\rm total}_{\rm observed}
\left(E_\gamma,\Delta_{\rm cut}\right)}{\dd m^2_{\gamma\gamma}}
\qquad\text{and}\qquad
\frac{\dd\Gamma^{\rm total}_{\rm observed}
\left(E_\gamma,\Delta_{\rm cut}\right)}{\dd m^2_{\pi^0\left(\eta\right)\gamma}},
\end{equation}
Their corresponding complete kinematic supports are: 
\begin{equation}
0\leq m^2_{\gamma\gamma}\leq
\left(M_+-m_{\pi^0\left(\eta\right)}\right)^2,
\qquad
m_{\pi^0\left(\eta\right)}^2\leq m^2_{\pi^0\left(\eta\right)\gamma}\leq M_+^2,
\qquad
M_+\equiv m_{\eta^{\left(\prime\right)}}+\Delta_{\rm cut}
\label{eq:continuous-spectrum-support}
\end{equation}
For a strictly on-shell decay \(\eta^{\left(\prime\right)}\to \pi^0\left(\eta\right)\gamma\gamma\), the corresponding supports reduce to: 
\begin{equation}
s_{\gamma\gamma}\leq\left(m_{\eta^{\left(\prime\right)}}-m_{\pi^0\left(\eta\right)}\right)^2,
\qquad
s_{\pi^0\left(\eta\right)\gamma}\leq m_{\eta^{\left(\prime\right)}}^2
\label{eq:on-shell-spectrum-support}
\end{equation}
The parts of the continuum support lying outside these on-shell
kinematic bounds:
\begin{equation}
\left(m_{\eta^{\left(\prime\right)}}-m_{\pi^0\left(\eta\right)}\right)^2<m^2_{\gamma\gamma}
\leq\left(M_+-m_{\pi^0\left(\eta\right)}\right)^2,
\qquad
m_{\eta^{\left(\prime\right)}}^2<m^2_{\pi^0\left(\eta\right)\gamma}\leq M_+^2,
\label{eq:forbidden-tail-regions}
\end{equation}
define kinematic tails that are forbidden for an isolated
on-shell $\eta^{\left(\prime\right)}$ decay but accessible to the nucleon-assisted continuum contribution. Their observation, or exclusion at the predicted rate, provides a direct test of the $V_{\mathcal B}$ mechanism.

\section{Crossing kinematic Rubicon}
\label{KinematicRubicon}

The regions in Eq.~\eqref{eq:forbidden-tail-regions} for which any of $m_{\gamma_1\gamma_2}^2$, $m_{\pi^0\left(\eta\right)\gamma_1}^2$, or
$m_{\pi^0\left(\eta\right)\gamma_2}^2$ crosses its corresponding kinematic Rubicon are of particular interest. The true on-shell pole contribution in Eq.~\eqref{eq:pole-continuum-decomposition}, $\mathcal{W}_{\text{pole}}$, has exactly zero kinematic support in these regions.

As discussed further in Section~\ref{VB_search_strategy}, a genuine Rubicon crossing generated by the nucleon-assisted topology in Fig.~\ref{fig:SM_VB_cross} is correlated with the four-momentum transferred to the recoil nucleon. This correlation provides an additional discriminator between genuine continuum events and on-shell pole events migrated beyond their nominal kinematic endpoints due to possible detector imperfections, potentially making the Rubicon regions particularly clean experimental probes of the continuum contribution.

Can $\gamma_1\gamma_2$ cross the  kinematic Rubicon in Eq.~\eqref{eq:forbidden-tail-regions} simultaneously with $\pi^0\left(\eta\right)\gamma_1$? Crossing both Rubicons \textit{simultaneously} means:  
\begin{equation}
    m^2_{\pi^0\left(\eta\right)\gamma_1} \ge m^2_{\eta^{\left(\prime\right)}}, \  m^2_{\gamma_1\gamma_2} \ge \left(m_{\eta^{\left(\prime\right)}} - m_{\pi^0\left(\eta\right)}\right)^2
\end{equation}
For the off-shell part of the $\mathcal{W}_{\text{cont}}$ amplitude, the kinematic constraint is: 
\begin{equation}
m_{\pi^0\left(\eta\right)\gamma_1}^2
+ m_{\pi^0\left(\eta\right)\gamma_2}^2
+ m_{\gamma_1\gamma_2}^2
= M_X^2 + m_{\pi^0\left(\eta\right)}^2 \le \left(m_{\eta^{\left(\prime\right)}} +\Delta_{\text{cut}} \right)^2+m_{\pi^0\left(\eta\right)}^2,
\end{equation}
Consider the $\pi^0\left(\eta\right)$ rest frame, with the $\pi^0\left(\eta\right)$ momentum being: 
\begin{equation}
p_{\pi^0\left(\eta\right)}=(m_{\pi^0\left(\eta\right)},\vec{\mathbf{0}})
\end{equation}
and the $\gamma_1$ and $\gamma_2$ having the energies $E_1$ and $E_2$, and the relative angle $\theta$ between them. In this reference frame: 
\begin{equation}
    m^2_{\pi^0\left(\eta\right)\gamma_{1, 2}} = m^2_{\pi^0\left(\eta\right)} + 2m_{\pi^0\left(\eta\right)}E_{1,2}
\end{equation}
And thus, the condition for $m^2_{\pi^0\left(\eta\right)\gamma_{1}} \ge m^2_{\eta^{\left(\prime\right)}}$ is: 
\begin{equation}
    E_1 \ge \frac{m^2_{\eta^{\left(\prime\right)}}-m^2_{\pi^0\left(\eta\right)}}{2m_{\pi^0\left(\eta\right)}},
    \label{E1conditionRubicon1}
\end{equation}
The $\gamma_1\gamma_2$ invariant mass is:
\begin{equation}
    m^2_{\gamma_1\gamma_2}
    =
    2E_1E_2\left(1-\cos\left(\theta\right)\right)
\end{equation}
For fixed $E_1$, the smallest value of $E_2$ capable of crossing the
$\gamma_1\gamma_2$ Rubicon is obtained when the photons are antiparallel,
$\cos\left(\theta\right)=-1$, and thus: 
\begin{equation}
    E_2
    \ge
    \frac{
    \left(
    m_{\eta^{\left(\prime\right)}}
    -
    m_{\pi^0\left(\eta\right)}
    \right)^2
    }{4E_1}.
    \label{eq:E2-simultaneous-Rubicon}
\end{equation}
The minimum value of $M_X$ compatible with the simultaneous crossing both $\pi^0\left(\eta\right)\gamma_1$ and $\gamma_1\gamma_2$ Rubicons is determined by minimizing:
\begin{equation}
\begin{split}
M_X^2\left(E_1\right)
={}&
m_{\pi^0\left(\eta\right)}^2
+
2m_{\pi^0\left(\eta\right)}
\left(E_1+E_2\right)
+
2E_1E_2
\left(1-\cos\left(\theta\right)\right) = 
\\
={}&
m_{\pi^0\left(\eta\right)}^2
+
\left(
m_{\eta^{\left(\prime\right)}}
-
m_{\pi^0\left(\eta\right)}
\right)^2
+
2m_{\pi^0\left(\eta\right)}E_1
+
\frac{
m_{\pi^0\left(\eta\right)}
\left(
m_{\eta^{\left(\prime\right)}}
-
m_{\pi^0\left(\eta\right)}
\right)^2
}{
2E_1
}
\end{split}
\label{eq:MX-simultaneous-Rubicon}
\end{equation}
subject to the constraint in Eq.~\eqref{E1conditionRubicon1}. Its derivative
with respect to $E_1$ is:
\begin{equation}
    \frac{\mathrm{d}M_X^2}{\mathrm{d}E_1}
    =
    2m_{\pi^0\left(\eta\right)}
    -
    \frac{
    m_{\pi^0\left(\eta\right)}
    \left(
    m_{\eta^{\left(\prime\right)}}
    -
    m_{\pi^0\left(\eta\right)}
    \right)^2
    }{2E_1^2} \ge \frac{2m_{\eta^{\left(\prime\right)}}m_{\pi^0\left(\eta\right)}\left(m_{\eta^{\left(\prime\right)}} + 2m_{\pi^0\left(\eta\right)}\right)}{\left(m_{\eta^{\left(\prime\right)}} + m_{\pi^0\left(\eta\right)}\right)^2} > 0,
\end{equation}
and the minimum value of $M_X$ is obtained by saturating
Eq.~\eqref{E1conditionRubicon1}:
\begin{equation}
    E_1
    =
    \frac{
    m_{\eta^{\left(\prime\right)}}^2
    -
    m_{\pi^0\left(\eta\right)}^2
    }{
    2m_{\pi^0\left(\eta\right)}
    }, \ 
    E_2
    =
    \frac{
    m_{\pi^0\left(\eta\right)}
    \left(
    m_{\eta^{\left(\prime\right)}}
    -
    m_{\pi^0\left(\eta\right)}
    \right)
    }{
    2\left(
    m_{\eta^{\left(\prime\right)}}
    +
    m_{\pi^0\left(\eta\right)}
    \right)
    },
\end{equation}
which corresponds to the minimum reconstructed invariant mass:
\begin{equation}
    M_{X,\min}^2
    =
    \frac{
    2m_{\eta^{\left(\prime\right)}}^3
    }{
    m_{\pi^0\left(\eta\right)}
    +
    m_{\eta^{\left(\prime\right)}}
    }
\end{equation}
Therefore, the condition when the Rubicon is crossed by $\pi^0\gamma_1$ and $\gamma_1\gamma_2$ simultaneously is: 
\begin{equation}
    \Delta_{\text{cut}} \ge \sqrt{ \frac{2m^3_{\eta^{\left(\prime\right)}}}{m_{\pi^0\left(\eta\right)} + m_{\eta^{\left(\prime\right)}}}} - m_{\eta^{\left(\prime\right)}} = \begin{cases}
        146 \ \rm{MeV}, for \ \eta\rightarrow\pi^0\gamma\gamma \\
        310 \ \rm{MeV}, for \ \eta^\prime\rightarrow\pi^0\gamma\gamma \\
        123 \ \rm{MeV}, for \ \eta^\prime\rightarrow\eta\gamma\gamma 
    \end{cases}
\end{equation}

Now, consider the condition when $\pi^0\left(\eta\right)\gamma_1$ crosses the Rubicon simultaneously with $\pi^0\left(\eta\right)\gamma_2$. To achieve this, the photons need to be parallel, to make $m^2_{\gamma_1\gamma_2} = 0$, and in the $\pi^0\left(\eta\right)$ reference frame: 
\begin{equation}
    E_1 = E_2 \ge \frac{m^2_{\eta^{\left(\prime\right)}} - m^2_{\pi^0\left(\eta\right)}}{2m_{\pi^0\left(\eta\right)}},
\end{equation}
corresponding to the minimal reconstructed mass: 
\begin{equation}
\begin{split}
M_{X,\min}^2
={}&
m_{\pi^0\left(\eta\right)}^2
+
2m_{\pi^0\left(\eta\right)}
\left(E_1+E_2\right) = 
2m_{\eta^{\left(\prime\right)}}^2
-
m_{\pi^0\left(\eta\right)}^2
\end{split}
\end{equation}
which leads to the selection cut: 
\begin{equation}
    \Delta_{\text{cut}} \ge \sqrt{2m^2_{\eta^{\left(\prime\right)}} - m_{\pi^0\left(\eta\right)}^2} - m_{\eta^{\left(\prime\right)}}  = \begin{cases}
        215 \ \rm{MeV}, for \ \eta\rightarrow\pi^0\gamma\gamma \\
        390 \ \rm{MeV}, for \ \eta^\prime\rightarrow\pi^0\gamma\gamma \\
        281 \ \rm{MeV}, for \ \eta^\prime\rightarrow\eta\gamma\gamma 
    \end{cases}
\end{equation}
Therefore, for our chosen $\Delta_{\text{cut}} \in \{50, 75, 100\}\ \rm{MeV}$, the kinematically forbidden regions of $\gamma_1\gamma_2$, $\pi^0\left(\eta\right)\gamma_1$, and $\pi^0\left(\eta\right)\gamma_2$ in Eq.~\eqref{eq:forbidden-tail-regions} are \textit{mutually exclusive}. In other words, if the observed event $\pi^0\left(\eta\right)\gamma\gamma$ from $\mathcal{W}_{\text{cont}}$ is crossing the kinematic Rubicon, it happens exclusively in one of three channels, and the three forbidden-tail fractions can be added without double counting.
The fraction of events crossing the $\gamma\gamma$ kinematic Rubicon is:
\begin{equation}
F_{\gamma\gamma}^{\rm Rubicon}
\left(E_\gamma,\Delta_{\rm cut}\right)
=
\frac{
\displaystyle
\int_{\left(
m_{\eta^{\left(\prime\right)}}
-m_{\pi^0\left(\eta\right)}
\right)^2}^{
\left(
m_{\eta^{\left(\prime\right)}}
+\Delta_{\rm cut}
-m_{\pi^0\left(\eta\right)}
\right)^2}
\frac{
\dd\Gamma_{\rm observed}^{\rm total}
\left(E_\gamma,\Delta_{\rm cut}\right)
}{
\dd m_{\gamma\gamma}^2
}
\dd m_{\gamma\gamma}^2
}{
\displaystyle
\int_{0}^{
\left(
m_{\eta^{\left(\prime\right)}}
+\Delta_{\rm cut}
-m_{\pi^0\left(\eta\right)}
\right)^2}
\frac{
\dd\Gamma_{\rm observed}^{\rm total}
\left(E_\gamma,\Delta_{\rm cut}\right)
}{
\dd m_{\gamma\gamma}^2
}
\dd m_{\gamma\gamma}^2
}
\label{eq:gamma-gamma-rubicon-fraction}
\end{equation}
Similarly, the fraction crossing one of the
$\pi^0\left(\eta\right)\gamma$ Rubicons is:
\begin{equation}
F_{\pi^0\left(\eta\right)\gamma}^{\rm Rubicon}
\left(E_\gamma,\Delta_{\rm cut}\right)
=
\frac{
\displaystyle
\int_{m_{\eta^{\left(\prime\right)}}^2}^{
\left(
m_{\eta^{\left(\prime\right)}}
+\Delta_{\rm cut}
\right)^2}
\frac{
\dd\Gamma_{\rm observed}^{\rm total}
\left(E_\gamma,\Delta_{\rm cut}\right)
}{
\dd m_{\pi^0\left(\eta\right)\gamma}^2
}
\dd m_{\pi^0\left(\eta\right)\gamma}^2
}{
\displaystyle
\int_{m_{\pi^0\left(\eta\right)}^2}^{
\left(
m_{\eta^{\left(\prime\right)}}
+\Delta_{\rm cut}
\right)^2}
\frac{
\dd\Gamma_{\rm observed}^{\rm total}
\left(E_\gamma,\Delta_{\rm cut}\right)
}{
\dd m_{\pi^0\left(\eta\right)\gamma}^2
}
\dd m_{\pi^0\left(\eta\right)\gamma}^2
}
\label{eq:meson-gamma-rubicon-fraction}
\end{equation}
Finally, the number of events crossing one of the kinematic Rubicons is: 
\begin{equation}
F_{\text{off-shell}}^{\rm Rubicon}
\left(E_\gamma,\Delta_{\rm cut}\right) = F_{\gamma\gamma}^{\rm Rubicon}
\left(E_\gamma,\Delta_{\rm cut}\right) + 2F_{\pi^0\left(\eta\right)\gamma}^{\rm Rubicon}
\left(E_\gamma,\Delta_{\rm cut}\right),
\end{equation}
where the factor of $2$ takes into account the exchange symmetry
$\gamma_1\leftrightarrow\gamma_2$. The two integration regions defined by $m_{\pi^0\left(\eta\right)\gamma_1}^2
\geq m_{\eta^{\left(\prime\right)}}^2$ and $m_{\pi^0\left(\eta\right)\gamma_2}^2
\geq m_{\eta^{\left(\prime\right)}}^2$, which  give equal contributions. Since these regions are mutually exclusive for our chosen values of $\Delta_{\rm cut}$, they can be added without double-counting. The identical-photon symmetry factor $1/2!$ is
already included in the corresponding phase-space integrals.

In what follows, we focus on the two decays,
$\eta\rightarrow\pi^0\gamma\gamma$ and
$\eta^\prime\rightarrow\pi^0\gamma\gamma$, since, as shown in~\cite{Balytskyi:2026}, the $V_{\mathcal B}$ contribution changes the $\eta^\prime\rightarrow\eta\gamma\gamma$ decay width by $\sim 1\%$ in production on a nucleon, and can therefore be neglected.

In the next Section~\ref{Numerical_Results}, we will show numerically that, for both decays, the fraction of events crossing the $\gamma_1\gamma_2$ Rubicon is much larger than the fraction crossing the $\pi^0\gamma_1$ Rubicon. This is because for the latter crossing, the second photon, $\gamma_2$, must become soft, which we demonstrate analytically below. Consider the $M_X$ rest frame, in which:
\begin{equation}
    p_X
    =
    \left(M_X,\vec{\mathbf 0}\right)
\end{equation}
Since
$p_{\pi^0}+p_{\gamma_1}=p_X-p_{\gamma_2}$, we obtain:
\begin{equation}
\begin{split}
    m_{\pi^0\gamma_1}^2
    =
    \left(p_X-p_{\gamma_2}\right)^2
    =
    M_X^2-2M_XE_{\gamma_2} \ge m^2_{\eta^{\left(\prime\right)}},
\end{split}
\end{equation}
and therefore, crossing the $\pi^0\gamma_1$ Rubicon requires:
\begin{equation}
    E_{\gamma_2}
    \leq
    \frac{
    M_X^2-m_{\eta^{\left(\prime\right)}}^2
    }{
    2M_X
    }
    =
    \frac{
    \left(
    M_X-m_{\eta^{\left(\prime\right)}}
    \right)
    \left(
    M_X+m_{\eta^{\left(\prime\right)}}
    \right)
    }{
    2M_X
    }
    \simeq \delta M_X \equiv 
    M_X-m_{\eta^{\left(\prime\right)}}.
\end{equation}

\begin{figure*}[t]
    \centering
    \includegraphics[width=\textwidth]
        {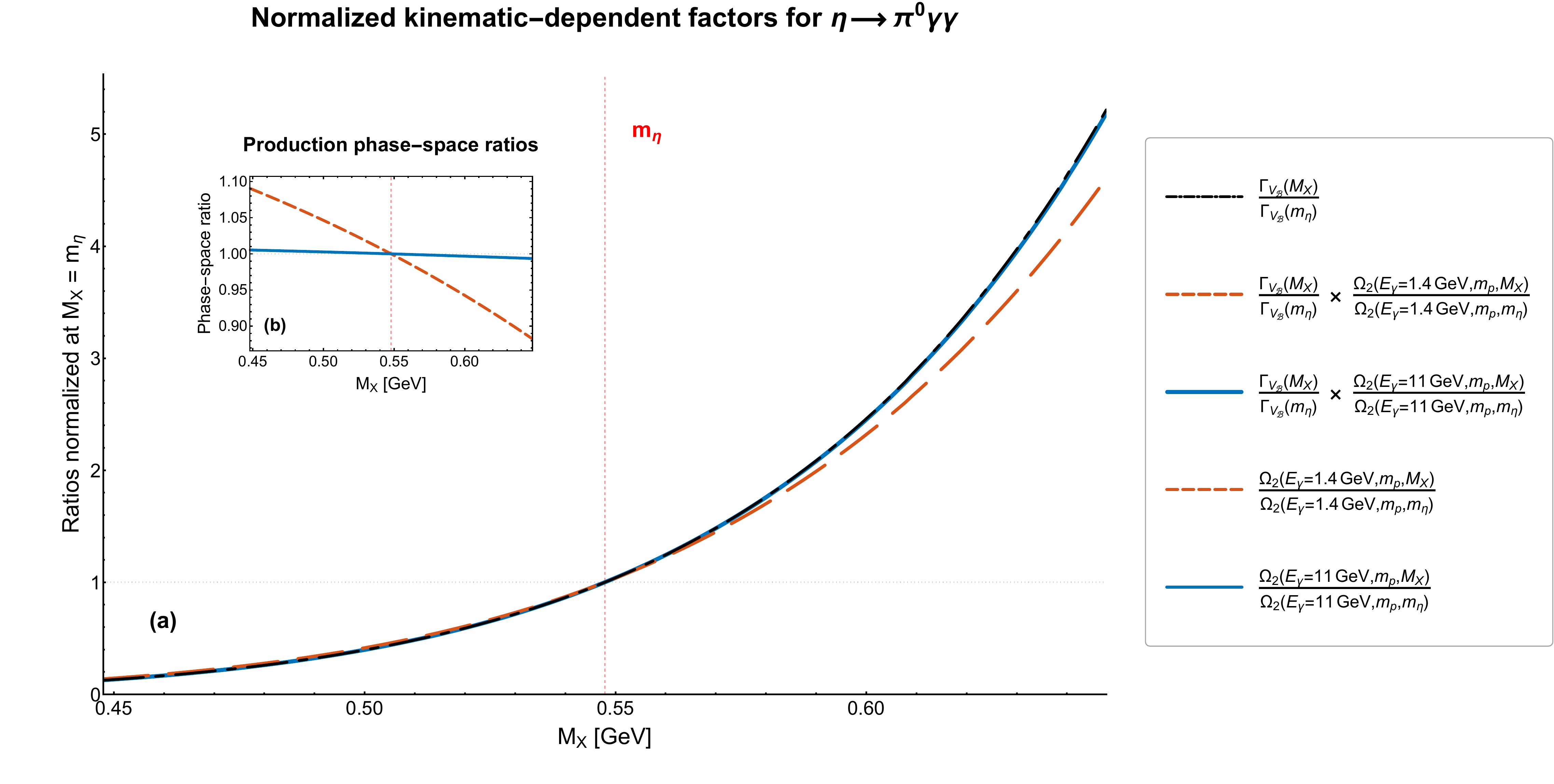}
    \caption{$M_X$-dependent kinematic factors for $\eta\to\pi^0\gamma\gamma$ normalized at $M_X=m_\eta$. Main panel \textbf{(a)}: the normalized  continuous width $\Gamma_{V_{\mathcal B}}(M_X)/\Gamma_{V_{\mathcal B}}(m_\eta)$, Eqn.~\eqref{eq:effective-continuum-width}, and its product with the normalized production phase-space factor for MAMI-like, $E_\gamma=1.4~\mathrm{GeV}$, and JEF-like, $E_\gamma=11~\mathrm{GeV}$, kinematics. Inset \textbf{(b)}: the corresponding normalized production phase-space factors. Although the production phase space decreases with $M_X$, the increase in $\Gamma_{V_{\mathcal B}}(M_X)$ dominates, leading to an increase in the total kinematic weight and contributing to a positive shift in the reconstructed mass.}
    \label{fig:eta_pi0_kinematic_factors}
\end{figure*}

\begin{figure*}[t]
    \centering
    \includegraphics[width=\textwidth]
        {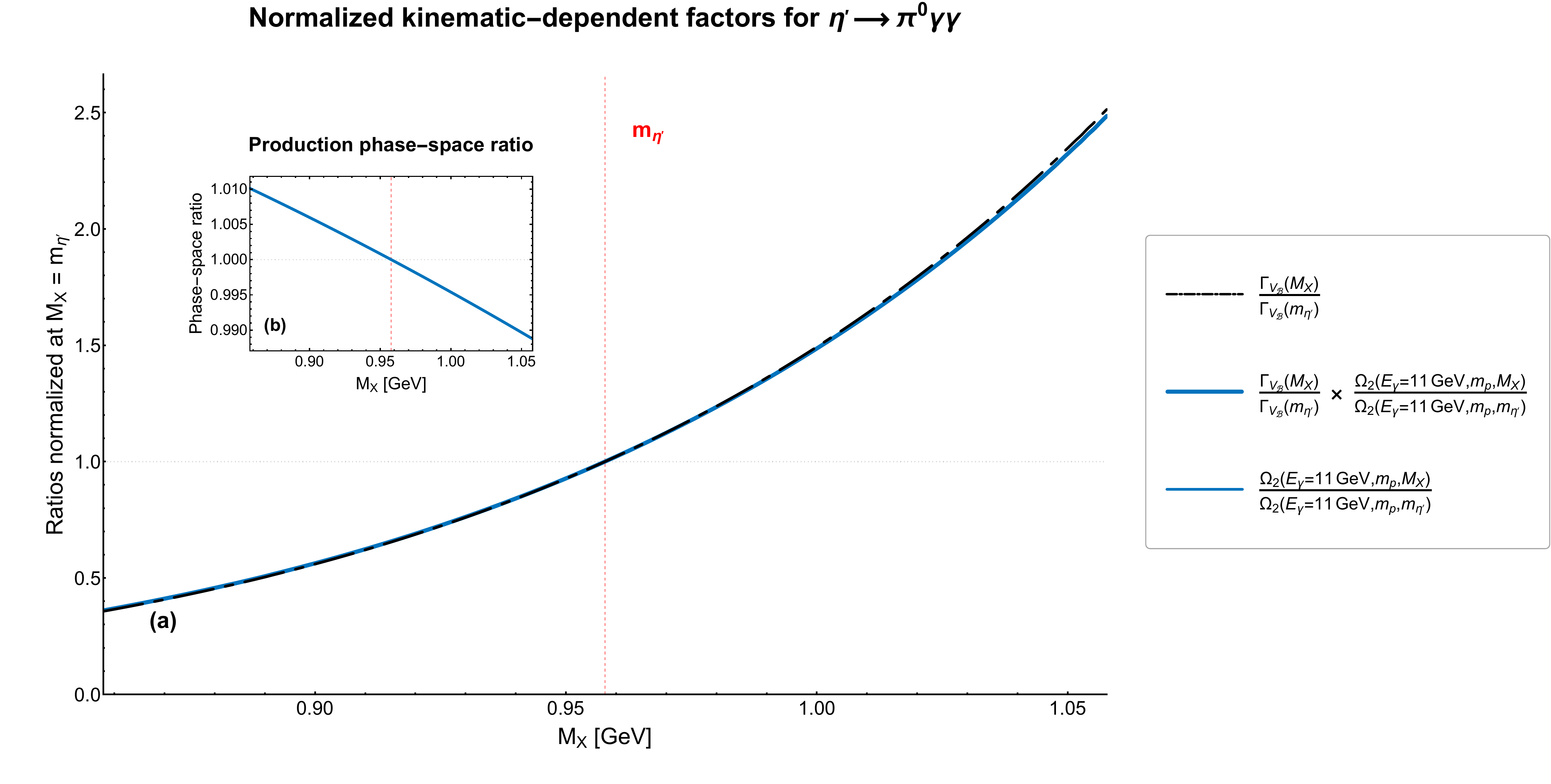}
    \caption{$M_X$-dependent kinematic factors for $\eta^\prime\to\pi^0\gamma\gamma$ normalized at $M_X=m_{\eta^\prime}$. Main panel \textbf{(a)}: the normalized  continuous width $\Gamma_{V_{\mathcal B}}(M_X)/\Gamma_{V_{\mathcal B}}(m_{\eta^\prime})$, Eqn.~\eqref{eq:effective-continuum-width}, and its product with the normalized production phase-space factor for JEF-like, $E_\gamma=11~\mathrm{GeV}$, kinematics. Inset \textbf{(b)}: the corresponding normalized production phase-space factors. Although the production phase space decreases with $M_X$, the increase in $\Gamma_{V_{\mathcal B}}(M_X)$ dominates, leading to an increase in the total kinematic weight and contributing to a positive shift in the reconstructed mass.
    }
    \label{fig:eta_prime_pi0_kinematic_factors}
\end{figure*}

As a result, near the on-shell point, the maximum allowed energy of the second photon is proportional to the off-shell displacement, 
$\delta M_X$, which leads to phase space suppression. Using the phase-space factorization derived in Appendix~\ref{Appendix:PhaseSpaceFactorization}, we obtain:
\begin{equation}
\begin{split}
\mathrm{d}\Phi_3
\left(
p_X;
p_{\pi^0},p_{\gamma_1},p_{\gamma_2}
\right)
={}&
\frac{
\mathrm{d}m_{\pi^0\gamma_1}^2
}{
2\pi
}
\,
\mathrm{d}\Phi_2
\left(
p_X;
p_{\pi^0\gamma_1},p_{\gamma_2}
\right)
\times
\mathrm{d}\Phi_2
\left(
p_{\pi^0\gamma_1};
p_{\pi^0},p_{\gamma_1}
\right)
\\
={}\hspace{4.5cm}&\hspace{-4.5cm}
\frac{1}{2048\pi^5}
\frac{
\lambda^{1/2}
\left(
M_X^2,m_{\pi^0\gamma_1}^2,0
\right)
}{
M_X^2
}
\times
\frac{
\lambda^{1/2}
\left(
m_{\pi^0\gamma_1}^2,m_{\pi^0}^2,0
\right)
}{
m_{\pi^0\gamma_1}^2
}
\,
\mathrm{d}m_{\pi^0\gamma_1}^2
\mathrm{d}\Omega
\mathrm{d}\Omega^*
\\
={}\hspace{4.5cm}&\hspace{-4.5cm}
\frac{1}{2048\pi^5}
\underbrace{\left(
M_X^2-m_{\pi^0\gamma_1}^2
\right)}_{=2M_XE_{\gamma_2}}
\frac{
m_{\pi^0\gamma_1}^2-m_{\pi^0}^2
}{
M_X^2m_{\pi^0\gamma_1}^2
}
\times
\mathrm{d}m_{\pi^0\gamma_1}^2
\mathrm{d}\Omega
\mathrm{d}\Omega^*
\propto
\mathcal{O}\left(\delta M_X\right)
\end{split},
\label{eq:pi0-gamma-body-phase-space-factorization}
\end{equation}
where \(\lambda(x,y,z)=x^2+y^2+z^2-2xy-2xz-2yz\)
is the Källén function, and the $\lvert V_\mathcal{B}\lvert^2$ matrix element is polynomial in particle masses. Therefore, while crossing the $\pi^0\gamma_1$ Rubicon requires the second photon to be soft in the $M_X$ rest frame, there is no such restriction and phase-space suppression for crossing the $\gamma_1\gamma_2$ Rubicon. We will confirm this conclusion numerically in the next Section~\ref{Numerical_Results}.

Finally, as shown numerically in Section~\ref{Numerical_Results}, the model predicts an \underline{upward} shift in the mean of the invariant-mass distribution, $\langle M_{\pi^0\gamma\gamma}\rangle-m_{\eta^{(\prime)}}>0$. Over the mass windows considered, the growth of $\Gamma_{V_{\mathcal B}}(M_X)$ with $M_X$ more than compensates for the decrease in the production phase-space factor, and thus biases the selected distribution toward $M_X>m_{\eta^{(\prime)}}$, as shown in Figs.~\ref{fig:eta_pi0_kinematic_factors} and~\ref{fig:eta_prime_pi0_kinematic_factors}.

\section{Numerical results}
\label{Numerical_Results}

\begin{figure*}[t]
    \centering
    \includegraphics[width=0.49\textwidth]
        {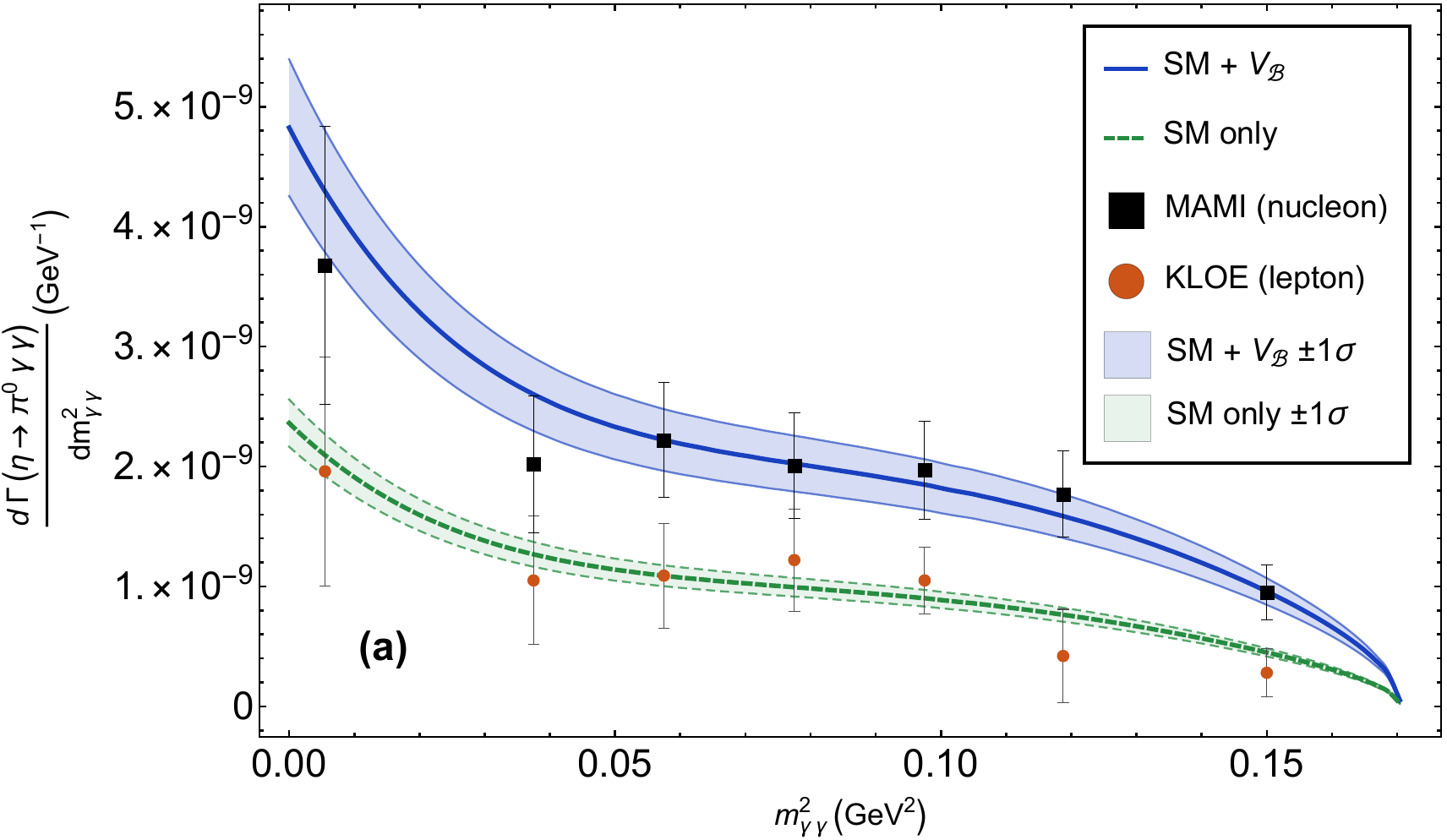}
    \hfill
    \includegraphics[width=0.49\textwidth]
        {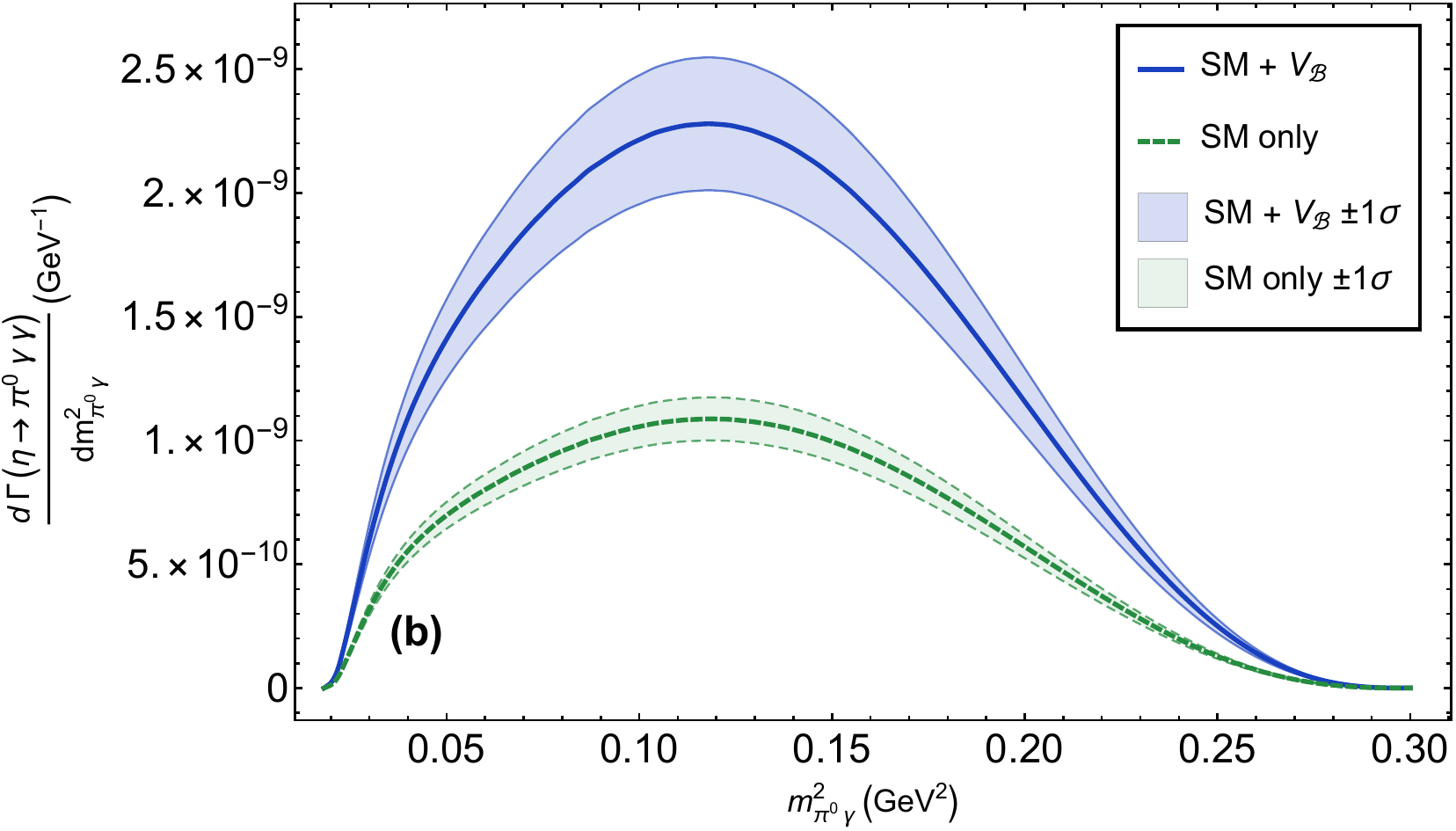}
\caption{Invariant-mass distributions for
$\eta\rightarrow\pi^0\gamma\gamma$ in the on-shell approximation $M_X = M_{\pi^0\gamma\gamma}=m_\eta$. Left \textbf{(a)}: the $\gamma\gamma$ invariant-mass spectrum. Right \textbf{(b)}: the $\pi^{0}\gamma$ invariant-mass spectrum.}
    \label{fig:eta_pi0_spectra}
\end{figure*}

\begin{figure*}[t]
    \centering
    \includegraphics[width=0.49\textwidth]
        {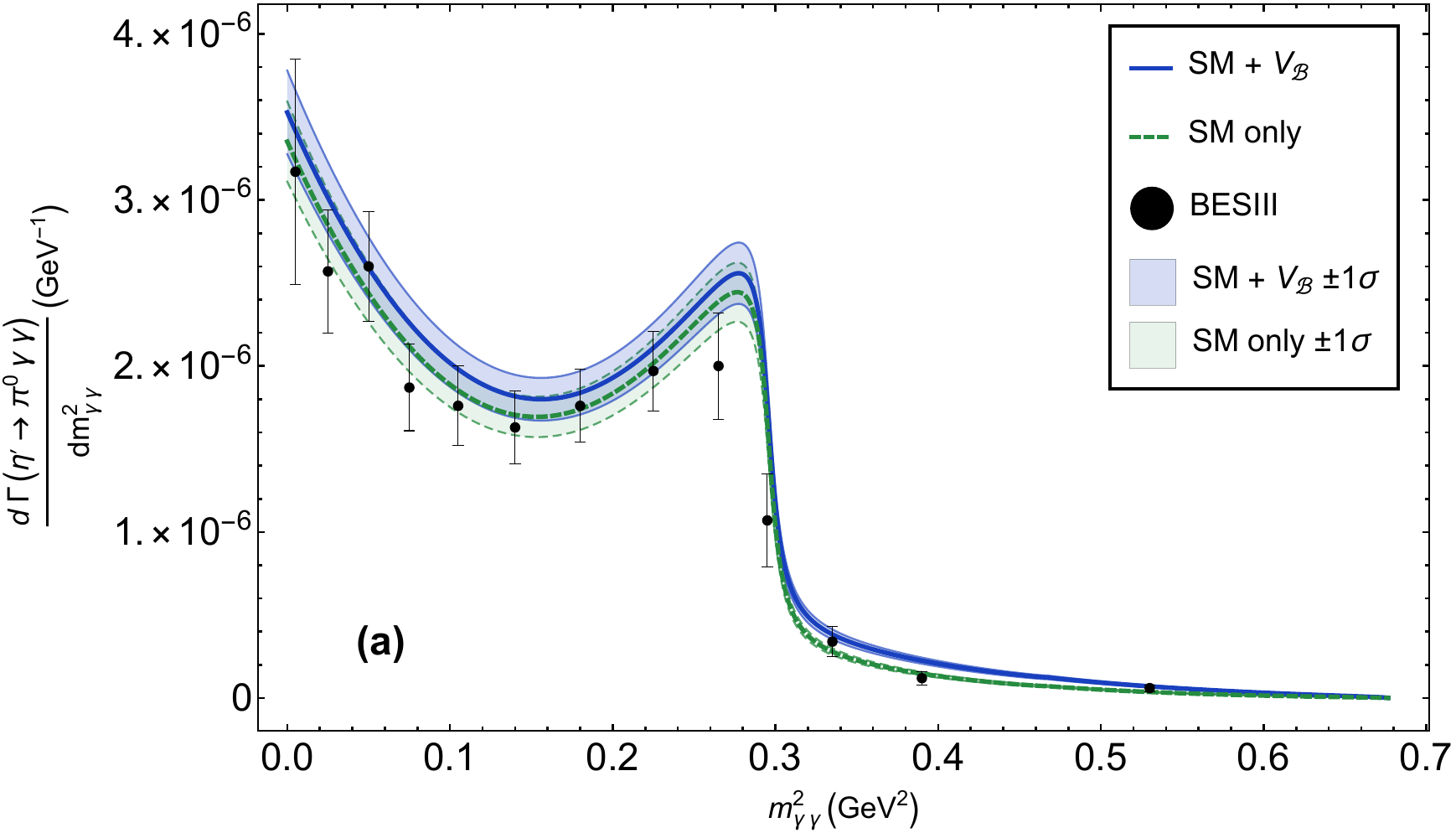}
    \hfill
    \includegraphics[width=0.49\textwidth]
        {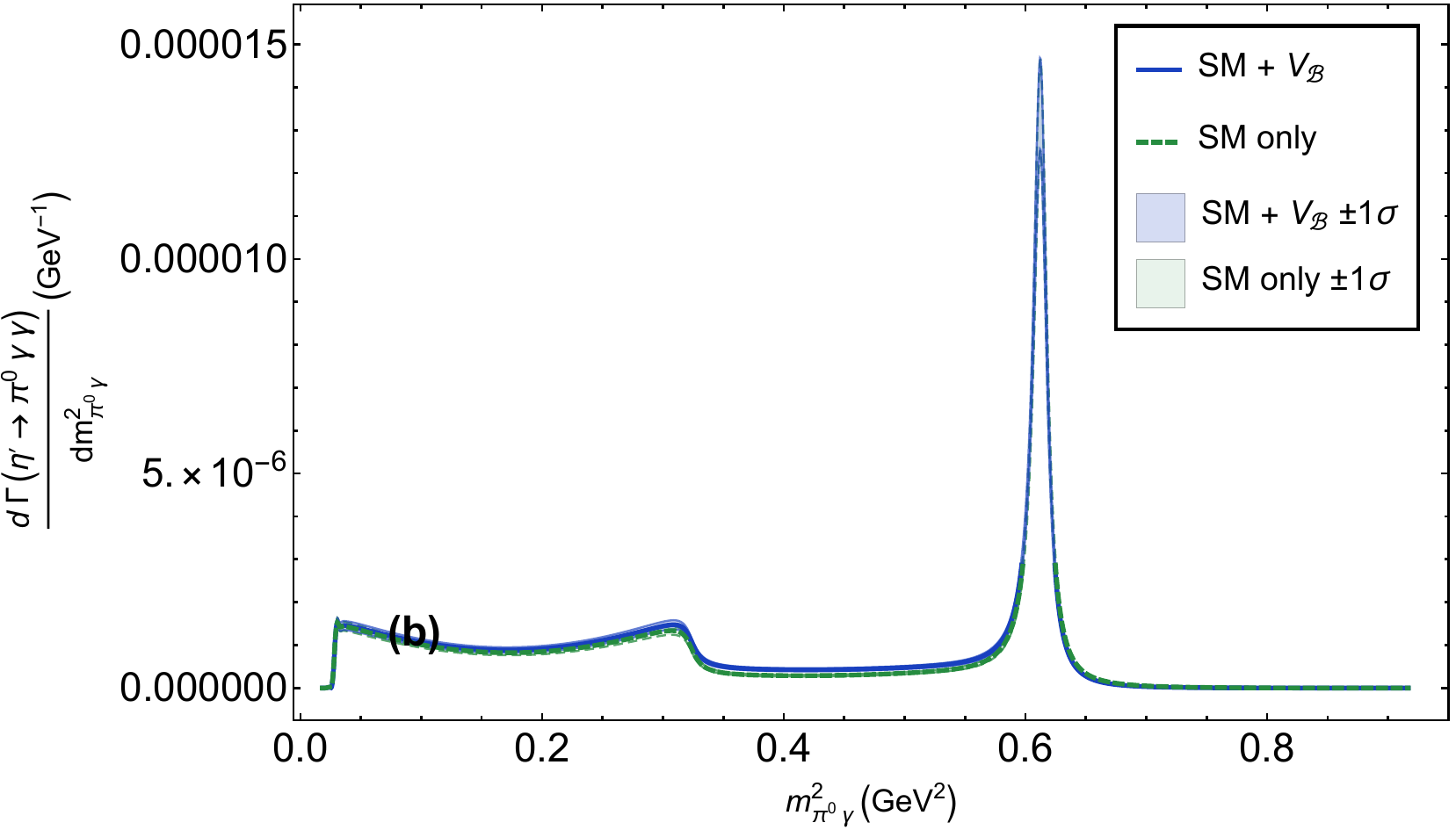}

\caption{Invariant-mass distributions for
$\eta^\prime\rightarrow\pi^0\gamma\gamma$ in the on-shell approximation $M_X = M_{\pi^0\gamma\gamma}=m_{\eta^\prime}$. Left \textbf{(a)}: the $\gamma\gamma$ invariant-mass spectrum. Right \textbf{(b)}: the $\pi^{0}\gamma$ invariant-mass spectrum.}
    \label{fig:eta_prime_pi0_spectra}
\end{figure*}

\begin{figure*}[t]
    \centering
    \includegraphics[width=0.49\textwidth]
        {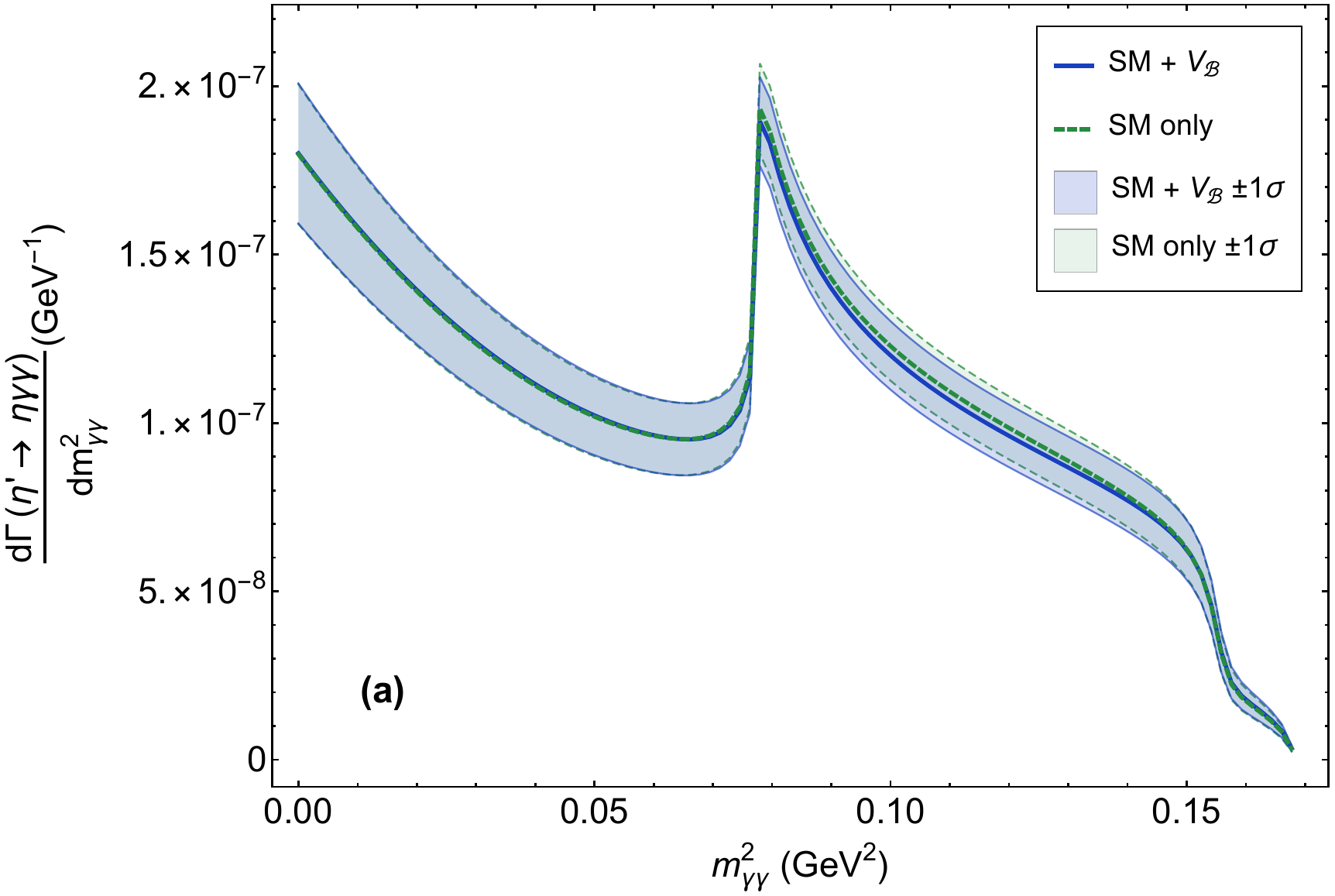}
    \hfill
    \includegraphics[width=0.49\textwidth]
        {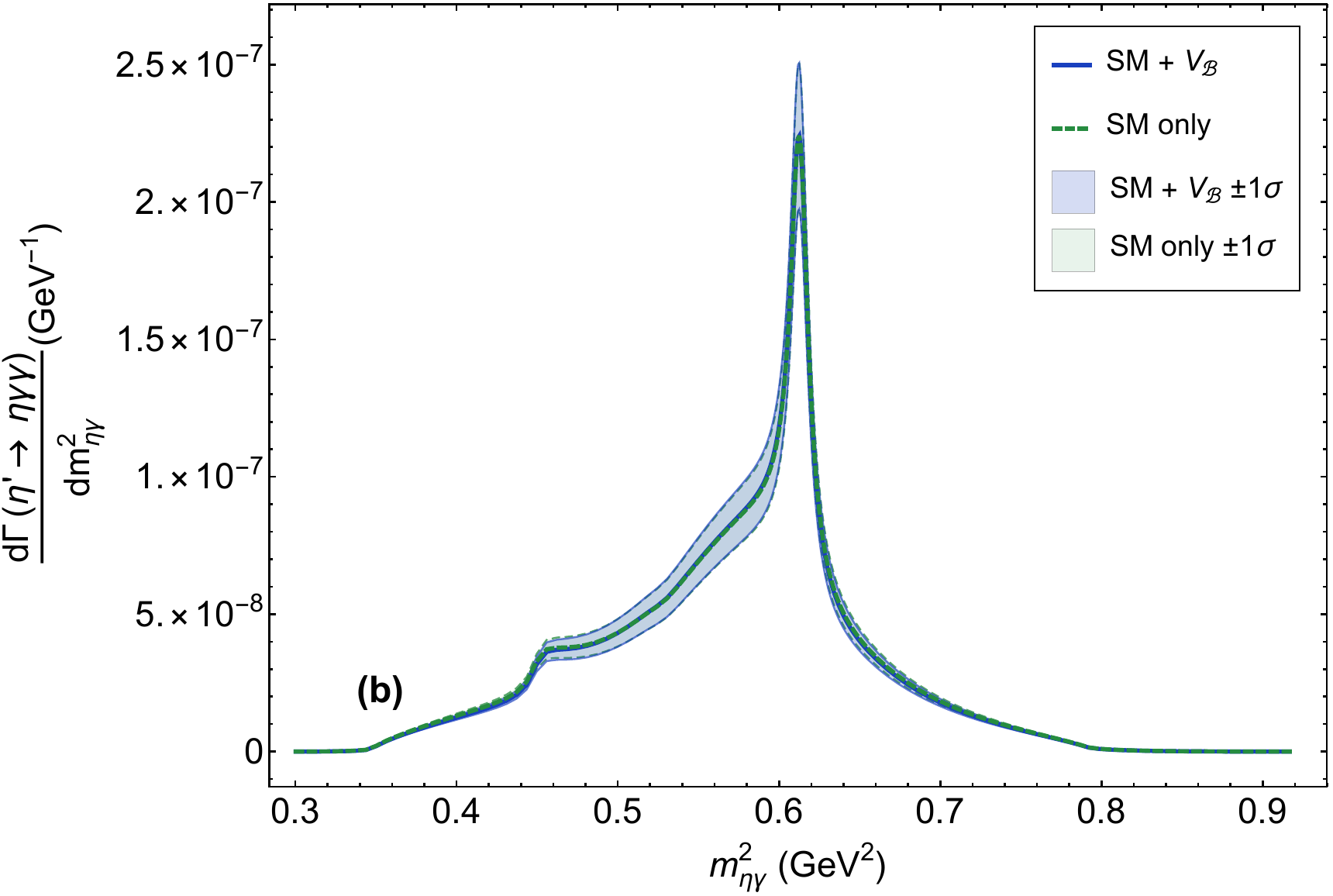}

\caption{Invariant-mass distributions for
$\eta^\prime\rightarrow\eta\gamma\gamma$ in the on-shell approximation $M_X = M_{\eta\gamma\gamma}=m_{\eta^\prime}$. Left \textbf{(a)}: the $\gamma\gamma$ invariant-mass spectrum. Right \textbf{(b)}: the $\eta\gamma$ invariant-mass spectrum.}

    \label{fig:eta_prime_eta_spectra}
\end{figure*}

The single parameter in our model \(C_{\text{eff}}\), which determines the contribution of $V_\mathcal{B}$ for all three decays, is obtained by minimizing $\chi^2$:

\begin{equation}
\chi^2 = \sum_k \frac{\left(\frac{\mathrm{d}\Gamma_{\text{theory}}^{\eta\rightarrow\pi^0\gamma\gamma}\left[{{\cal M}^{\mathrm{VMD}} + \mathcal{M}^{L\sigma M}
+ \mathcal{M}_{V_{\mathcal{B}}}\left(C_{\text{eff}}\right)}\right]}{\mathrm{d}m^2_{\gamma\gamma}} - \frac{\mathrm{d}\Gamma_{\text{MAMI}}^{\eta\rightarrow\pi^0\gamma\gamma}}{\mathrm{d}m^2_{\gamma\gamma}}  \right)_k^2}{\sigma_k^2},
\end{equation}
which yields\footnote{This value differs by $\approx 10\%$ from $C_{\mathrm{eff}}=\left(9.47\pm0.53\right)\times10^{-5}\,\mathrm{GeV}^{-8}$ reported in~\cite{Balytskyi:2026}. The difference originates from an incorrect assignment of one of the $B_{ij}$ terms in the numerical implementation of the squared amplitude, identified after publication, while these coefficients themselves are correct. In addition, the present analysis employs an improved numerical treatment in which the input uncertainties are propagated through 100 reproducibly seeded Monte Carlo realizations. The corrected implementation and updated numerical procedure do not affect the qualitative conclusions of~\cite{Balytskyi:2026}. The corresponding corrected results are summarized in Tables~\ref{tab:eta_pi0_summary} and~\ref{tab:eta_prime_pi0_summary}. The numerical code is publicly available at: 
\href{https://github.com/BalytskyiJaroslaw/Hypothesis}{https://github.com/BalytskyiJaroslaw/Hypothesis}.
}:
\begin{equation}
C_{\mathrm{eff}}
=
\left(
8.54\pm1.08
\right)
\times10^{-5}\,\mathrm{GeV}^{-8}
\label{FittedCoupling}
\end{equation}

The resulting spectra in the on-shell approximation,
$M_X=m_{\eta^{\left(\prime\right)}}$, are shown in
Fig.~\ref{fig:eta_pi0_spectra} for
$\eta\rightarrow\pi^0\gamma\gamma$, in
Fig.~\ref{fig:eta_prime_pi0_spectra} for
$\eta^\prime\rightarrow\pi^0\gamma\gamma$, and in
Fig.~\ref{fig:eta_prime_eta_spectra} for
$\eta^\prime\rightarrow\eta\gamma\gamma$. As discussed in~\cite{Balytskyi:2026}, $\eta\rightarrow\pi^0\gamma\gamma$ is the decay most sensitive to the $V_{\mathcal B}$ effects. For
$\eta^\prime\rightarrow\pi^0\gamma\gamma$, the $V_{\mathcal B}$ contribution increases the decay width by approximately $10\%$, while for $\eta^\prime\rightarrow\eta\gamma\gamma$ it is negligibly small. Accordingly, in the following discussion of the off-shell contribution, we focus on the first two decays.

\begin{figure*}[t]
    \centering
    \includegraphics[width=0.49\textwidth]
        {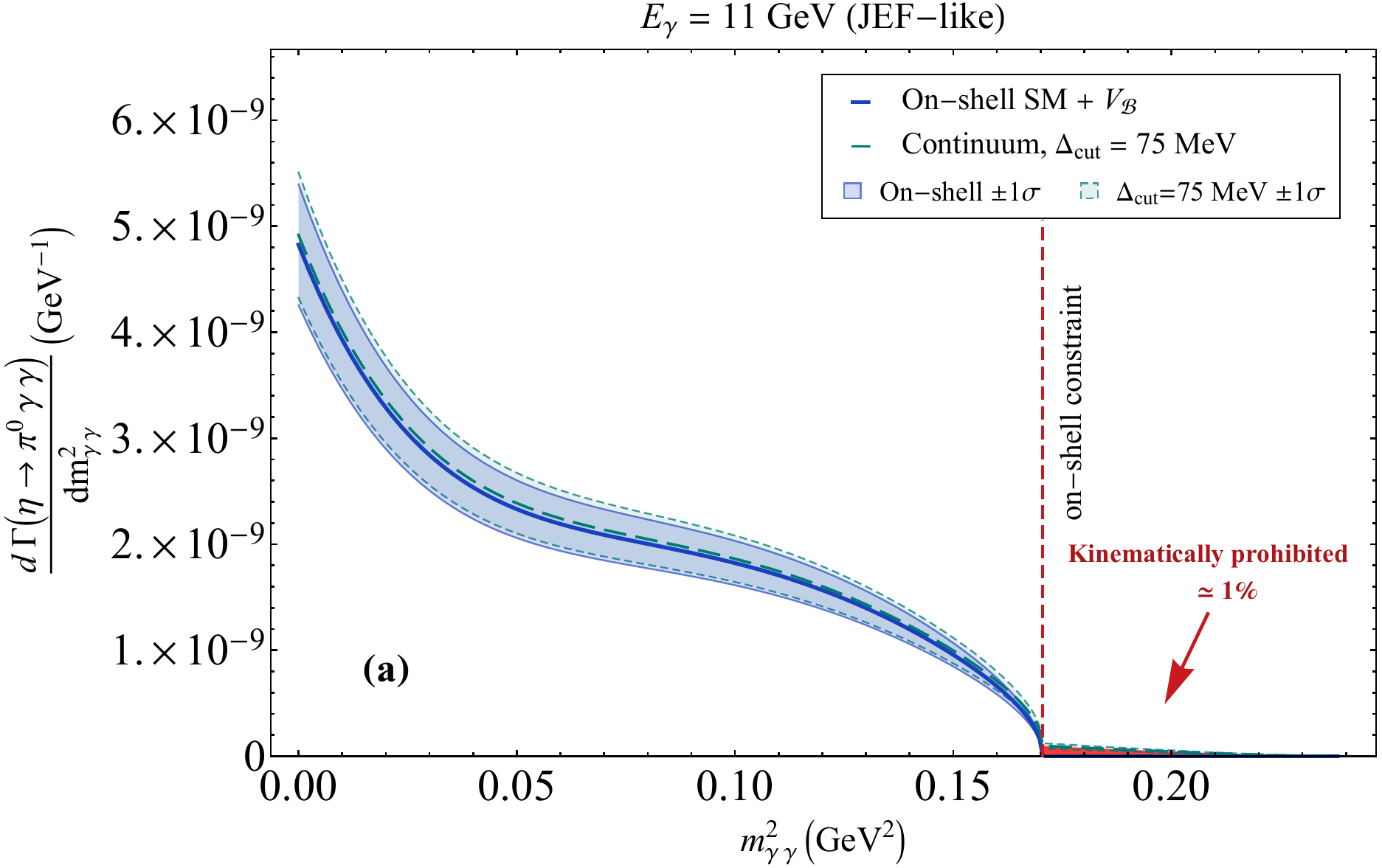}
    \hfill
    \includegraphics[width=0.49\textwidth]
        {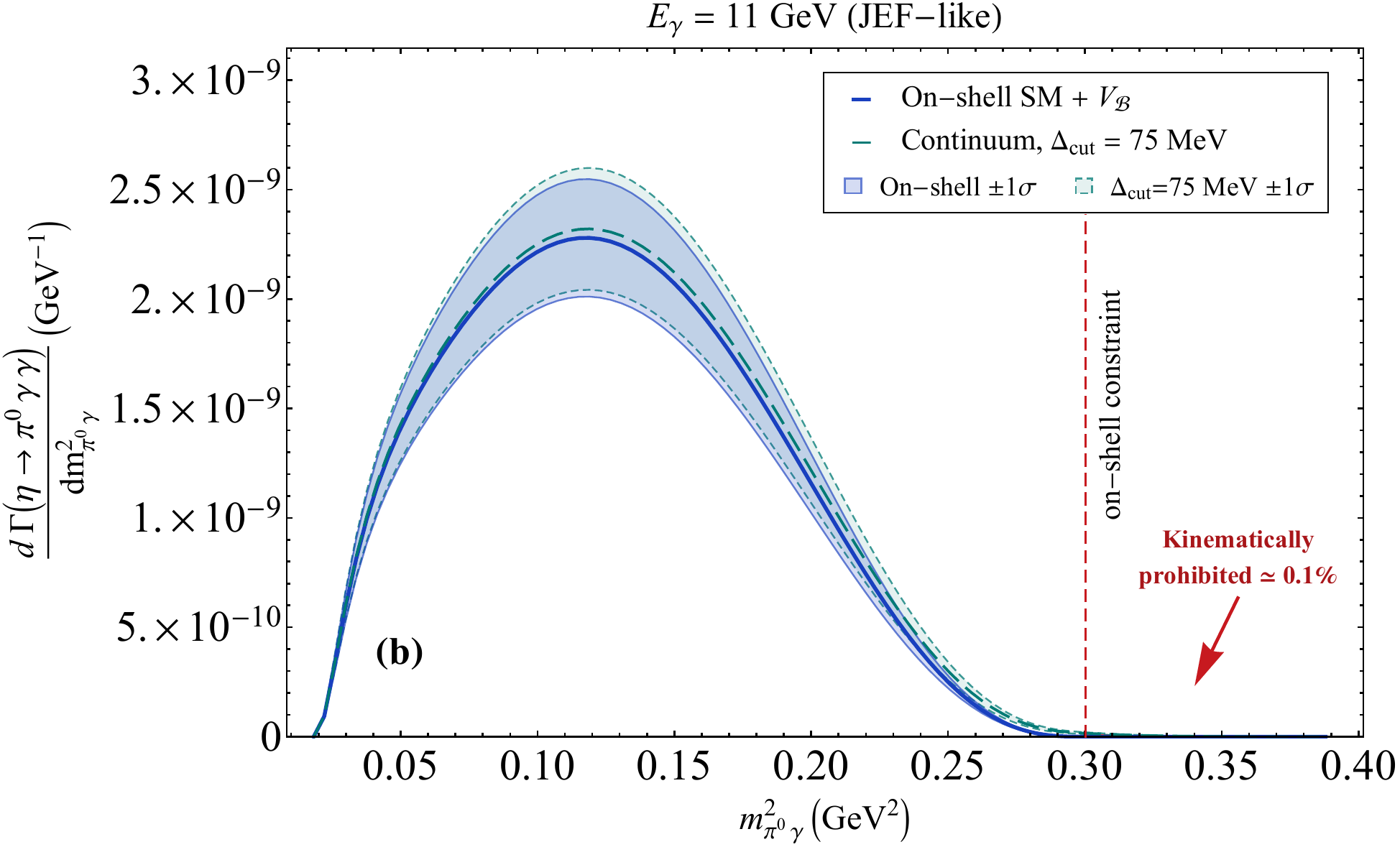}

\caption{Kinematic distributions for $\eta\to\pi^{0}\gamma\gamma$, evaluated both under the on-shell assumption, and with off-shell kinematics included, for the mass-window selection $\Delta_{\mathrm{cut}}=75~\mathrm{MeV}$ at the JEF-like incident-photon energy, $E_\gamma=11~\mathrm{GeV}$. Left \textbf{(a)}: the $\gamma\gamma$ invariant-mass spectrum, for which $\approx 1\%$ of the events lie beyond the corresponding kinematic Rubicon, when off-shell kinematics is included. Right \textbf{(b)}: the $\pi^{0}\gamma$ invariant-mass spectrum, for which the corresponding Rubicon fraction is much smaller, $\approx 0.1\%$.}
    \label{fig:eta_pi0_75MeV_spectra_JEF}
\end{figure*}

\begin{figure*}[t]
    \centering
    \includegraphics[width=0.49\textwidth]
        {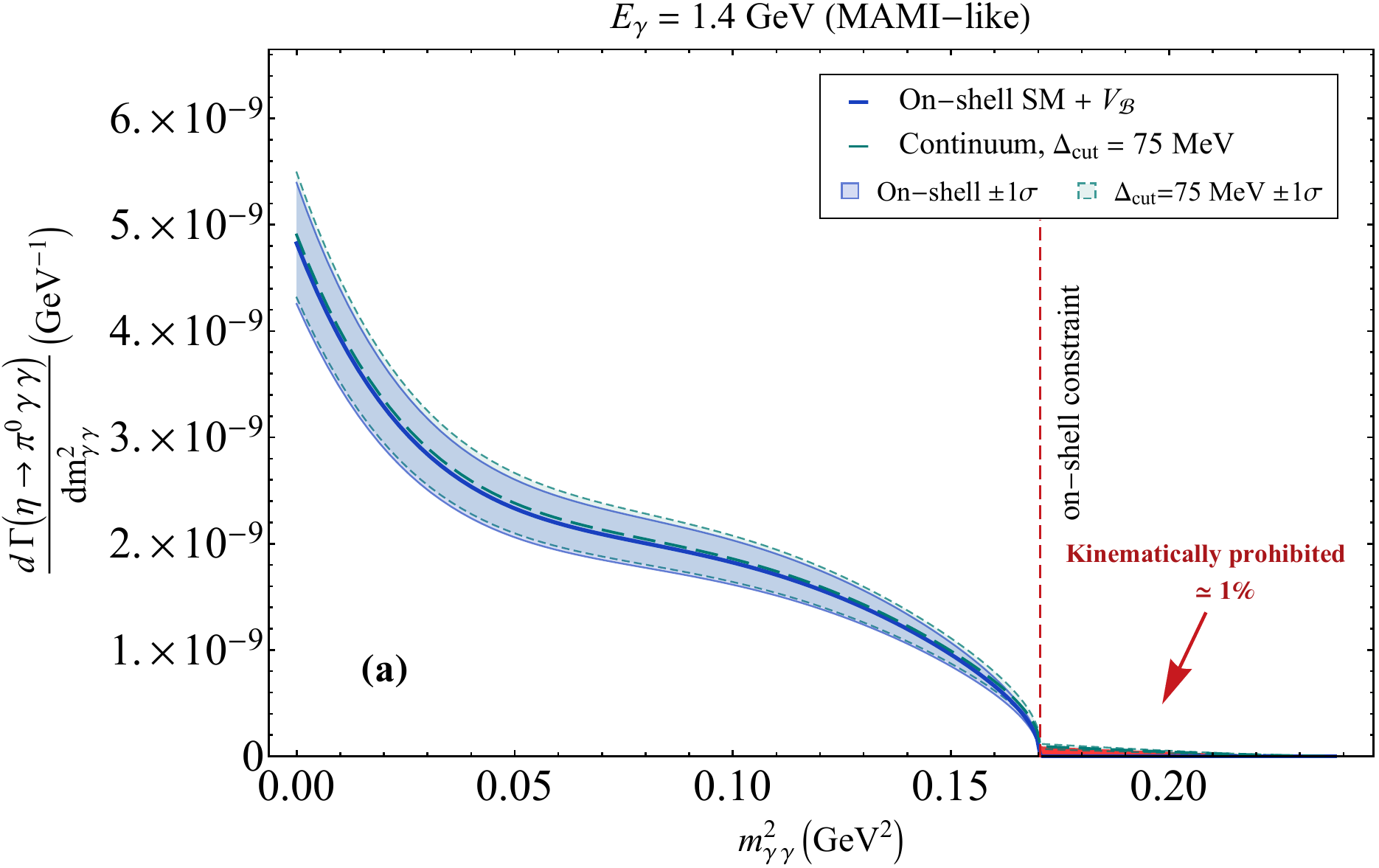}
    \hfill
    \includegraphics[width=0.49\textwidth]
        {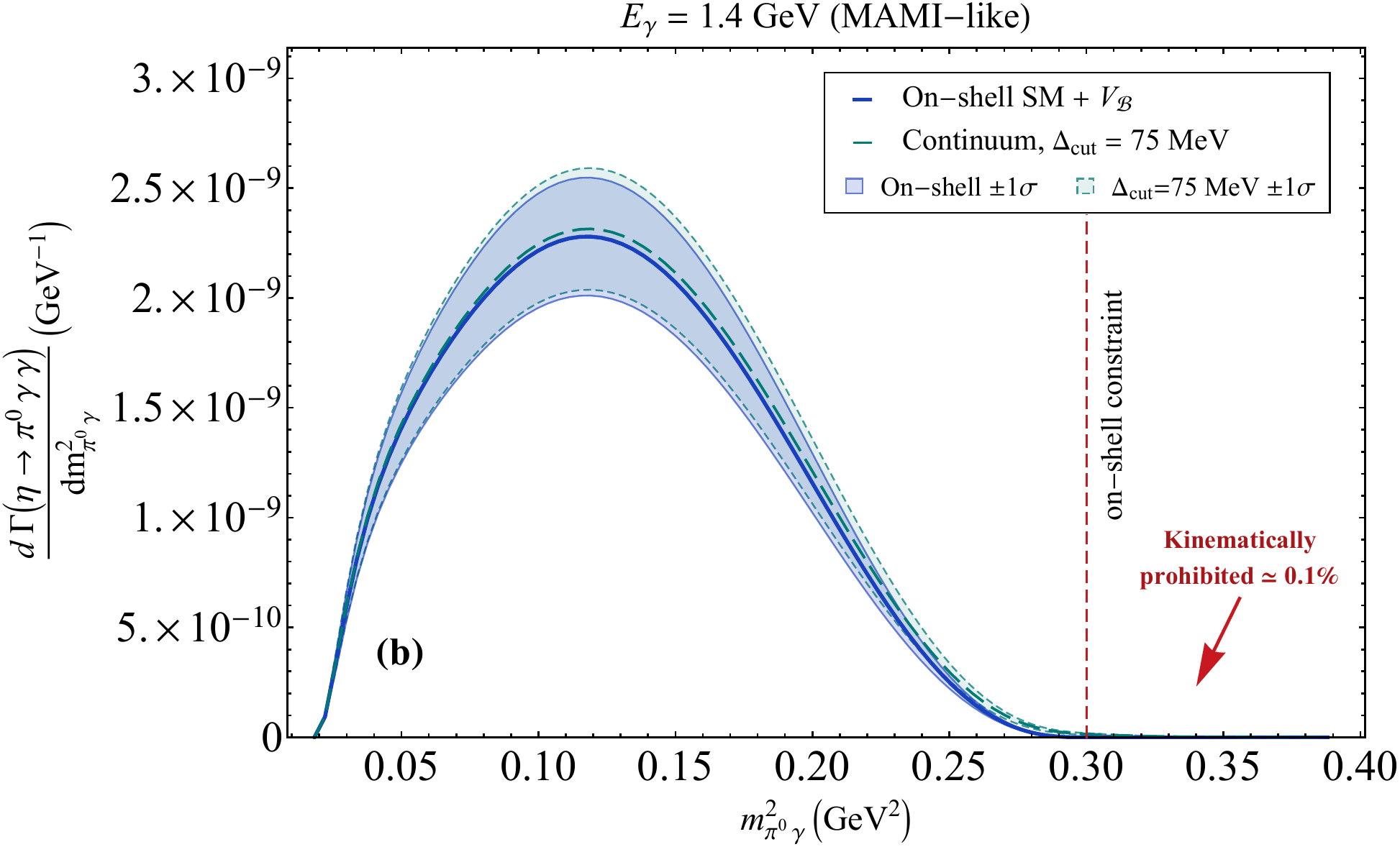}

\caption{Kinematic distributions for $\eta\to\pi^{0}\gamma\gamma$, evaluated both under the on-shell assumption, and with off-shell kinematics included, for the mass-window selection $\Delta_{\mathrm{cut}}=75~\mathrm{MeV}$ at the MAMI-like incident-photon energy, $E_\gamma=1.4~\mathrm{GeV}$. Left \textbf{(a)}: the $\gamma\gamma$ invariant-mass spectrum, for which $\approx 1\%$ of the events lie beyond the corresponding kinematic Rubicon, when off-shell kinematics is included. Right \textbf{(b)}: the $\pi^{0}\gamma$ invariant-mass spectrum, for which the corresponding Rubicon fraction is much smaller, $\approx 0.1\%$.}
    \label{fig:eta_pi0_75MeV_spectra_MAMI}
\end{figure*}

We also note that, in the on-shell approximation,
$M_X=m_{\eta^{\left(\prime\right)}}$, the Dalitz-plot integral of the
pole-associated terms in Eq.~\eqref{eq:pole-continuum-decomposition},
comprising the pure Standard Model contribution and its interference
with the $V_{\mathcal B}$ amplitude, accounts for
$\left(91.0\pm1.6\right)\%$ of the total Dalitz-integrated rate for
$\eta\rightarrow\pi^0\gamma\gamma$,
$\left(99.0\pm0.2\right)\%$ for
$\eta^\prime\rightarrow\pi^0\gamma\gamma$, and
$\left(99.92\pm0.02\right)\%$ for
$\eta^\prime\rightarrow\eta\gamma\gamma$.

Therefore, the modification of these decay widths induced by $V_{\mathcal B}$ is mainly due to its interference with the Standard Model amplitude, while the contribution of the pure continuum, $\lvert\mathcal{M}_{V_{\mathcal B}}\rvert^2$, remains subdominant. This hierarchy is expected since $1.5~\mathrm{GeV} \lesssim m_{V_{\mathcal B}}\lesssim 5.0~\mathrm{GeV}$, which places the $V_{\mathcal B}$ above the mass scale of the light vector mesons entering the Standard Model amplitude. The hypothesis of the $V_\mathcal{B}$ model is that its comparatively strong coupling compensates for this mass suppression,
making the interference term dominant~\cite{Balytskyi:2026}.

The remaining pure-continuum fractions are $\left(9.0\pm1.6\right)\%$ for
$\eta\rightarrow\pi^0\gamma\gamma$, and $\left(1.0\pm0.2\right)\%$ for
$\eta^\prime\rightarrow\pi^0\gamma\gamma$. Although subdominant in the total contribution to the decay rates, this continuum term extends beyond the on-shell kinematic boundaries and is therefore responsible for the kinematic fingerprints discussed below.

Using the same coupling in Eq.~\eqref{FittedCoupling}, we now apply it for the two distinct incident photon energies, 
near-threshold relevant for MAMI, $E_{\gamma}^{\text{MAMI}} = 1.4\ \rm{GeV}$, and higher energy relevant for JEF, $E_{\gamma}^{\text{JEF}} = 11\ \rm{GeV}$.

For $\eta\rightarrow\pi^0\gamma\gamma$, the numerical results are summarized in Table~\ref{tab:eta_pi0_summary}. The corresponding invariant-mass distributions for the representative selection cut $\Delta_{\mathrm{cut}}=75~\mathrm{MeV}$ are shown in Figs.~\ref{fig:eta_pi0_75MeV_spectra_JEF} and~\ref{fig:eta_pi0_75MeV_spectra_MAMI}. The key features can be summarized as follows: 
\begin{itemize}
    \item The effective decay widths and branching fractions predicted for JEF- and MAMI-like kinematics differ by no more than approximately $1\%$, for identical selection windows.
    
    \item Increasing the mass-window cut $\Delta_{\mathrm{cut}}$ over the range considered increases the window-integrated effective decay rate by approximately $3$--$4\%$. 
    
    \item The mean reconstructed-mass shift, $\Delta M_X\equiv\langle M_X\rangle-m_\eta$, is \underline{\emph{positive}} and ranges from approximately $+2$ to $+7~\mathrm{MeV}$.
    
    \item For a reference sample of $1200$ reconstructed signal events, the predicted Rubicon-crossing yield ranges from $4$ to $23$ events, depending on $\Delta_{\mathrm{cut}}$, and is approximately $10$ events for the benchmark selection $\Delta_{\mathrm{cut}}=75~\mathrm{MeV}$.
    
\end{itemize}

\begin{table*}[t]
\centering
\caption{
Summary of predictions for
$\eta\rightarrow\pi^0\gamma\gamma$.
Here, $\Delta M_X\equiv\langle M_X\rangle-m_\eta$ and
$f_{\mathrm R}\equiv
f_{\gamma\gamma}^{\mathrm{out}}
+2f_{\pi^0\gamma}^{\mathrm{out}}$.
The expected number of Rubicon-crossing events in a sample of $1200$
events is
$N_{\mathrm R}^{(1200)}=1200f_{\mathrm R}/100$.
Parenthetical uncertainties refer to the last displayed digits.
}
\label{tab:eta_pi0_summary}
\small
\renewcommand{\arraystretch}{1.28}
\setlength{\tabcolsep}{2.8pt}
\setlength{\arrayrulewidth}{0.4pt}
\resizebox{\linewidth}{!}{%
\begin{tabular}{@{}|c|c|c|c|c|c|c|c|c|@{}}
\hline\hline
Setup
&
$\Delta_{\rm cut}$
&
$\Gamma$
&
$\mathcal{B}$
&
$\Delta M_X$
&
$f_{\gamma\gamma}^{\mathrm{out}}$
&
$f_{\pi^0\gamma}^{\mathrm{out}}$
&
$f_{\mathrm R}$
&
$N_{\mathrm R}^{(1200)}$
\\[-1pt]
&
$\left[\mathrm{MeV}\right]$
&
$\left[10^{-10}\,\mathrm{GeV}\right]$
&
$\left[10^{-4}\right]$
&
$\left[\mathrm{MeV}\right]$
&
$\left[\%\right]$
&
$\left[\%\right]$
&
$\left[\%\right]$
&
$\left[\mathrm{events}\right]$
\\
\hline
SM
&
---
&
$1.67(14)$
&
$1.27(11)$
&
$0$
&
$0$
&
$0$
&
$0$
&
$0$
\\
On-shell $V_{\mathcal B}$
&
---
&
$3.40(41)$
&
$2.59(30)$
&
$0$
&
$0$
&
$0$
&
$0$
&
$0$
\\
\hline
\multirow{3}{*}{\shortstack{JEF-like\\$11~\mathrm{GeV}$}}
&
$50$
&
$3.44(42)$
&
$2.63(31)$
&
$+1.59(26)$
&
$0.320(53)$
&
$0.0158(26)$
&
$0.351(59)$
&
$4.22(70)$
\\
&
$75$
&
$3.51(44)$
&
$2.68(32)$
&
$+3.81(62)$
&
$0.733(120)$
&
$0.0821(134)$
&
$0.897(147)$
&
$10.8(18)$
\\
&
$100$
&
$3.60(46)$
&
$2.75(34)$
&
$+7.35(117)$
&
$1.39(22)$
&
$0.265(42)$
&
$1.92(30)$
&
$23.0(37)$
\\
\hline
\multirow{3}{*}{\shortstack{MAMI-like\\$1.4~\mathrm{GeV}$}}
&
$50$
&
$3.44(42)$
&
$2.63(31)$
&
$+1.50(25)$
&
$0.309(52)$
&
$0.0152(25)$
&
$0.339(57)$
&
$4.07(68)$
\\
&
$75$
&
$3.49(43)$
&
$2.67(32)$
&
$+3.56(58)$
&
$0.695(114)$
&
$0.0773(127)$
&
$0.850(139)$
&
$10.2(17)$
\\
&
$100$
&
$3.58(45)$
&
$2.73(33)$
&
$+6.79(109)$
&
$1.30(21)$
&
$0.245(39)$
&
$1.79(29)$
&
$21.4(34)$
\\
\hline\hline
\end{tabular}%
}
\end{table*}

Here $f_{\mathrm R}\equiv f_{\gamma\gamma}^{\mathrm{out}}+2f_{\pi^0\gamma}^{\mathrm{out}}$ denotes the combined fraction that goes beyond the kinematic Rubicon. We adopt $N_{\mathrm{sig}}=1200$ as a reference since both KLOE and MAMI reconstructed signal samples of approximately this size~\cite{Babusci:2026,Nefkens:2014}. JEF, however, is projected to accumulate a substantially larger sample~\cite{JEFproposal,Gan:JEFII2026}. It would therefore be valuable for JEF to repeat the extraction for several values of $\Delta_{\mathrm{cut}}$, since the resulting cut dependence of the effective yield, mean reconstructed-mass shift, and Rubicon fractions would provide an additional discriminator between the on-shell pole and off-shell continuum contributions in Eq.~\eqref{eq:pole-continuum-decomposition}. Such a scan should be performed without imposing the constraint $M_X=m_\eta$, which would suppress this very dependence.

\begin{figure*}[t]
    \centering
    \includegraphics[width=0.49\textwidth]
{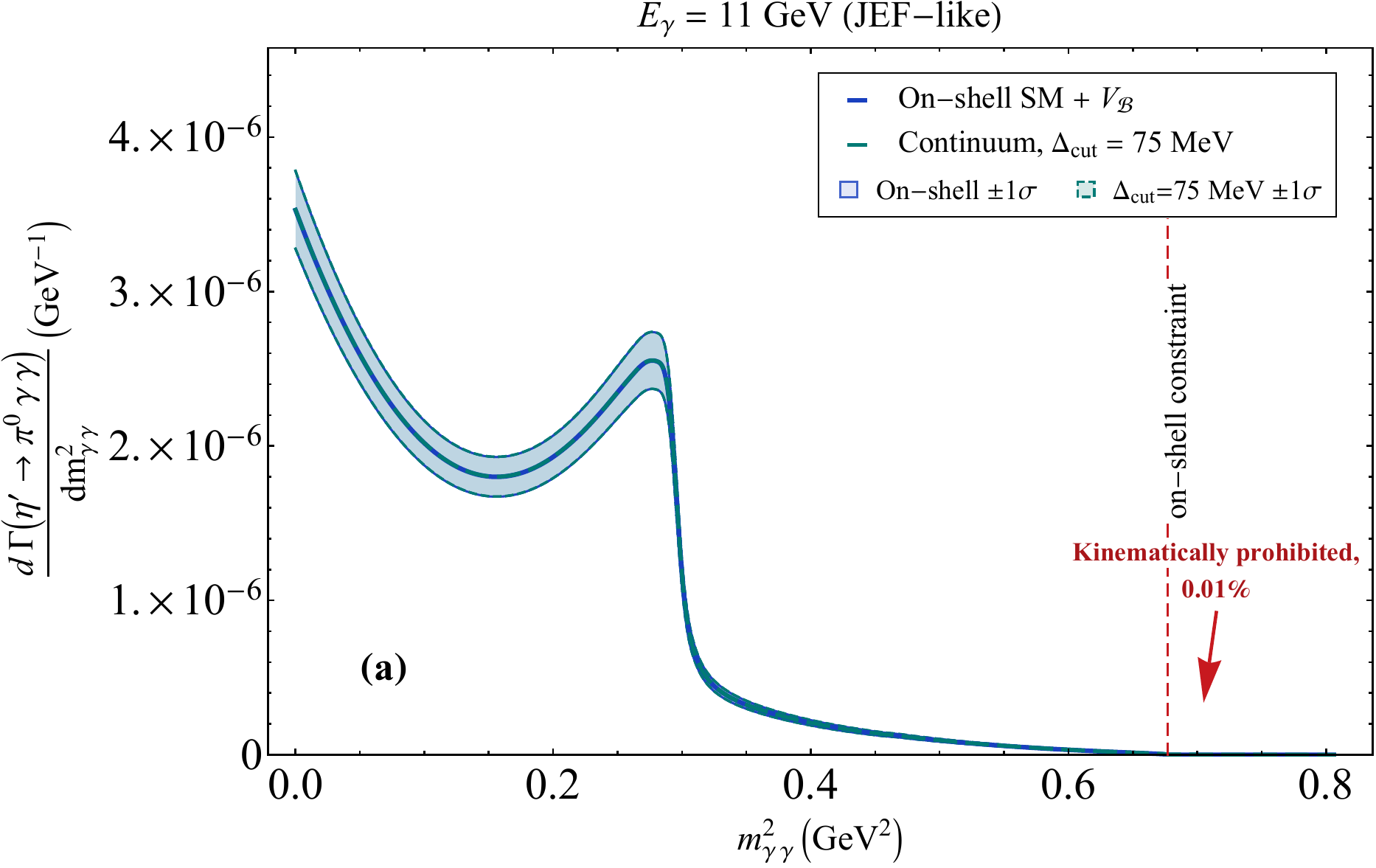}
    \hfill
    \includegraphics[width=0.49\textwidth]    {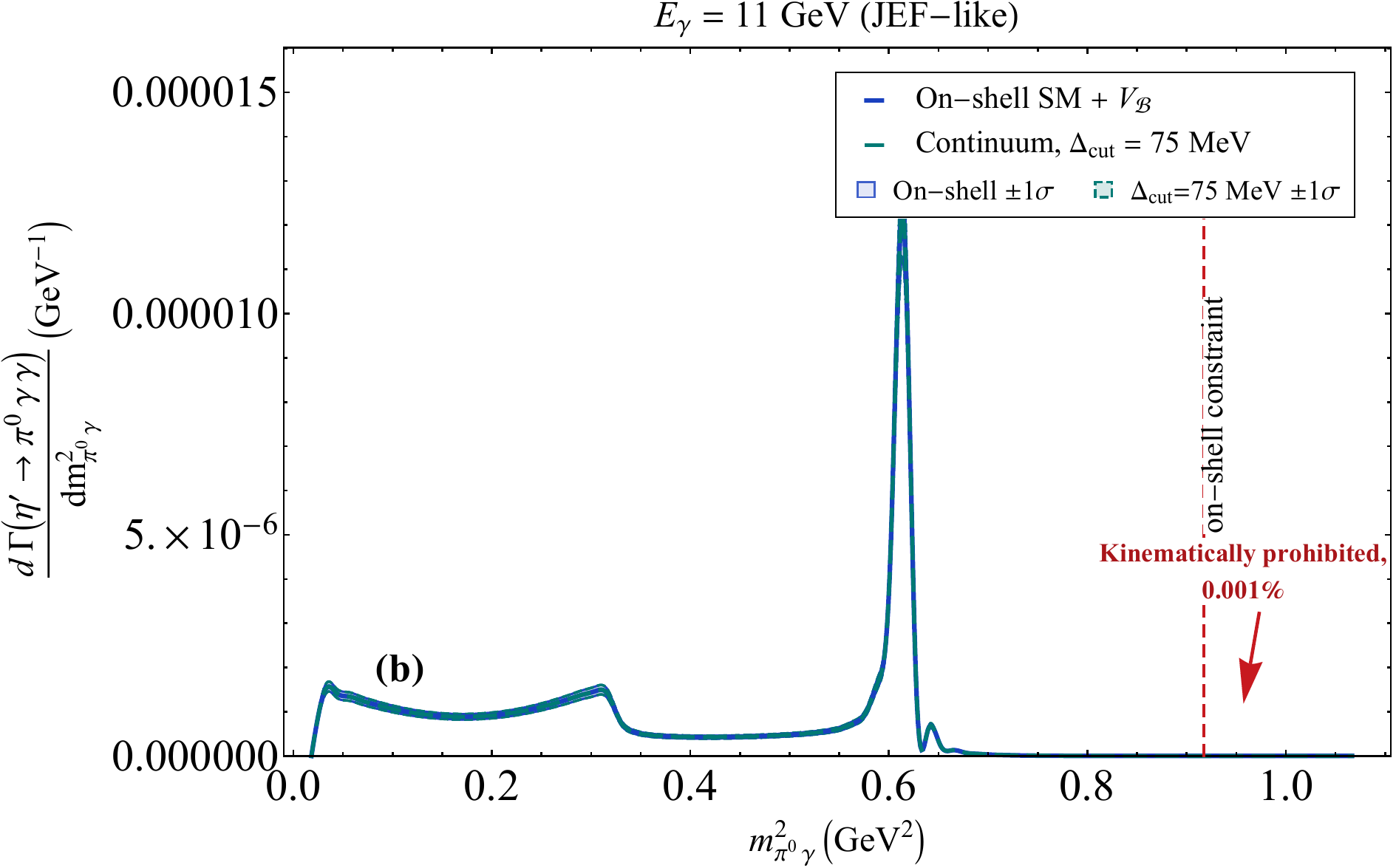}
    \caption{Kinematic distributions for $\eta^\prime\to\pi^{0}\gamma\gamma$, evaluated both under the on-shell assumption and with off-shell kinematics included, for the mass-window selection $\Delta_{\mathrm{cut}}=75~\mathrm{MeV}$ at the JEF-like incident-photon energy, $E_\gamma=11~\mathrm{GeV}$. Left \textbf{(a)}: the $\gamma\gamma$ invariant-mass spectrum, for which $\approx 0.01\%$ of the events lie beyond the corresponding kinematic Rubicon, when off-shell kinematics is included, which is approximately 100 times less than in the case of $\eta\rightarrow\pi^0\gamma\gamma$. Right \textbf{(b)}: the $\pi^{0}\gamma$ invariant-mass spectrum, for which the corresponding Rubicon fraction is even smaller, $\approx 0.001\%$.}
    \label{fig:eta_prime_pi0_spectra_JEF_75MeV}
\end{figure*}

\begin{table*}[t]
\centering
\caption{
Summary of predictions for $\eta^\prime\rightarrow\pi^0\gamma\gamma$
in JEF-like kinematics.
Here, $\Delta M_X\equiv\langle M_X\rangle-m_{\eta^\prime}$ and
$f_{\mathrm R}\equiv
f_{\gamma\gamma}^{\mathrm{out}}
+2f_{\pi^0\gamma}^{\mathrm{out}}$.
The expected number of Rubicon-crossing events in a reference sample
of $3500$ reconstructed events is
$N_{\mathrm R}^{(3500)}=3500f_{\mathrm R}/100$.
Parenthetical uncertainties refer to the last displayed digits.
}
\label{tab:eta_prime_pi0_summary}
\small
\renewcommand{\arraystretch}{1.28}
\setlength{\tabcolsep}{2.8pt}
\setlength{\arrayrulewidth}{0.4pt}
\resizebox{\linewidth}{!}{%
\begin{tabular}{@{}|c|c|c|c|c|c|c|c|c|@{}}
\hline\hline
Setup
&
$\Delta_{\rm cut}$
&
$\Gamma$
&
$\mathcal{B}$
&
$\Delta M_X$
&
$f_{\gamma\gamma}^{\mathrm{out}}$
&
$f_{\pi^0\gamma}^{\mathrm{out}}$
&
$f_{\mathrm R}$
&
$N_{\mathrm R}^{(3500)}$
\\[-1pt]
&
$\left[\mathrm{MeV}\right]$
&
$\left[10^{-7}\,\mathrm{GeV}\right]$
&
$\left[10^{-3}\right]$
&
$\left[\mathrm{MeV}\right]$
&
$\left[\%\right]$
&
$\left[\%\right]$
&
$\left[\%\right]$
&
$\left[\mathrm{events}\right]$
\\
\hline
SM
&
---
&
$6.75(49)$
&
$3.59(28)$
&
$0$
&
$0$
&
$0$
&
$0$
&
$0$
\\
On-shell $V_{\mathcal B}$
&
---
&
$7.30(53)$
&
$3.88(30)$
&
$0$
&
$0$
&
$0$
&
$0$
&
$0$
\\
\hline
\multirow{3}{*}{\shortstack{JEF-like\\$11~\mathrm{GeV}$}}
&
$50$
&
$7.31(53)$
&
$3.89(30)$
&
$+0.090(21)$
&
$0.0052(12)$
&
$0.000145(34)$
&
$0.0055(13)$
&
$0.191(45)$
\\
&
$75$
&
$7.31(53)$
&
$3.89(30)$
&
$+0.207(49)$
&
$0.0115(27)$
&
$0.00075(18)$
&
$0.0130(31)$
&
$0.46(11)$
\\
&
$100$
&
$7.32(53)$
&
$3.89(30)$
&
$+0.381(90)$
&
$0.0213(50)$
&
$0.00242(57)$
&
$0.0261(62)$
&
$0.91(22)$
\\
\hline\hline
\end{tabular}%
}
\end{table*}

For $\eta^\prime\rightarrow\pi^0\gamma\gamma$, the invariant-mass distributions obtained with the selection cut $\Delta_{\mathrm{cut}}=75~\mathrm{MeV}$ at a JEF-like incident photon energy, $E_\gamma=11~\mathrm{GeV}$, are shown in Fig.~\ref{fig:eta_prime_pi0_spectra_JEF_75MeV}, while the numerical predictions are summarized in Table~\ref{tab:eta_prime_pi0_summary}. 

A reference sample of $3500$ reconstructed events was chosen, similar to the number of signal events collected by BESIII~\cite{ablikim2017observation}, because $\eta^\prime\rightarrow\pi^0\gamma\gamma$ has more than an order of magnitude higher branching ratio than $\eta\rightarrow\pi^0\gamma\gamma$. Nevertheless, even for the relatively large choice of $\Delta_{\mathrm{cut}}=100~\mathrm{MeV}$, the expected yield is only about one Rubicon-crossing event. This suppression occurs because the $V_{\mathcal B}$-induced decay width increase is on the order of $\sim\rm{eV}$ relative to the Standard Model being $\sim\rm{keV}$, which is dominated by resonant on-shell $\omega$ exchange~\cite{Balytskyi:2026}.

Therefore, among the channels considered, $\eta\rightarrow\pi^0\gamma\gamma$ provides the most decisive test of the $V_\mathcal{B}$ scenario. It simultaneously offers an $\mathcal{O}(1)$ rate enhancement, a measurable upward mass shift of several \rm{MeV}, and roughly ten events in a realistic sample that are beyond the kinematic Rubicon. Moreover, the kinematically forbidden events remain correlated with the recoiling-nucleon kinematics, providing an additional discriminator of the model, as discussed further in Section~\ref{VB_search_strategy}.

\section{Strategy for experimental signature search}
\label{VB_search_strategy}

\begin{figure}[!t]
    \centering
    \includegraphics[width=\textwidth]{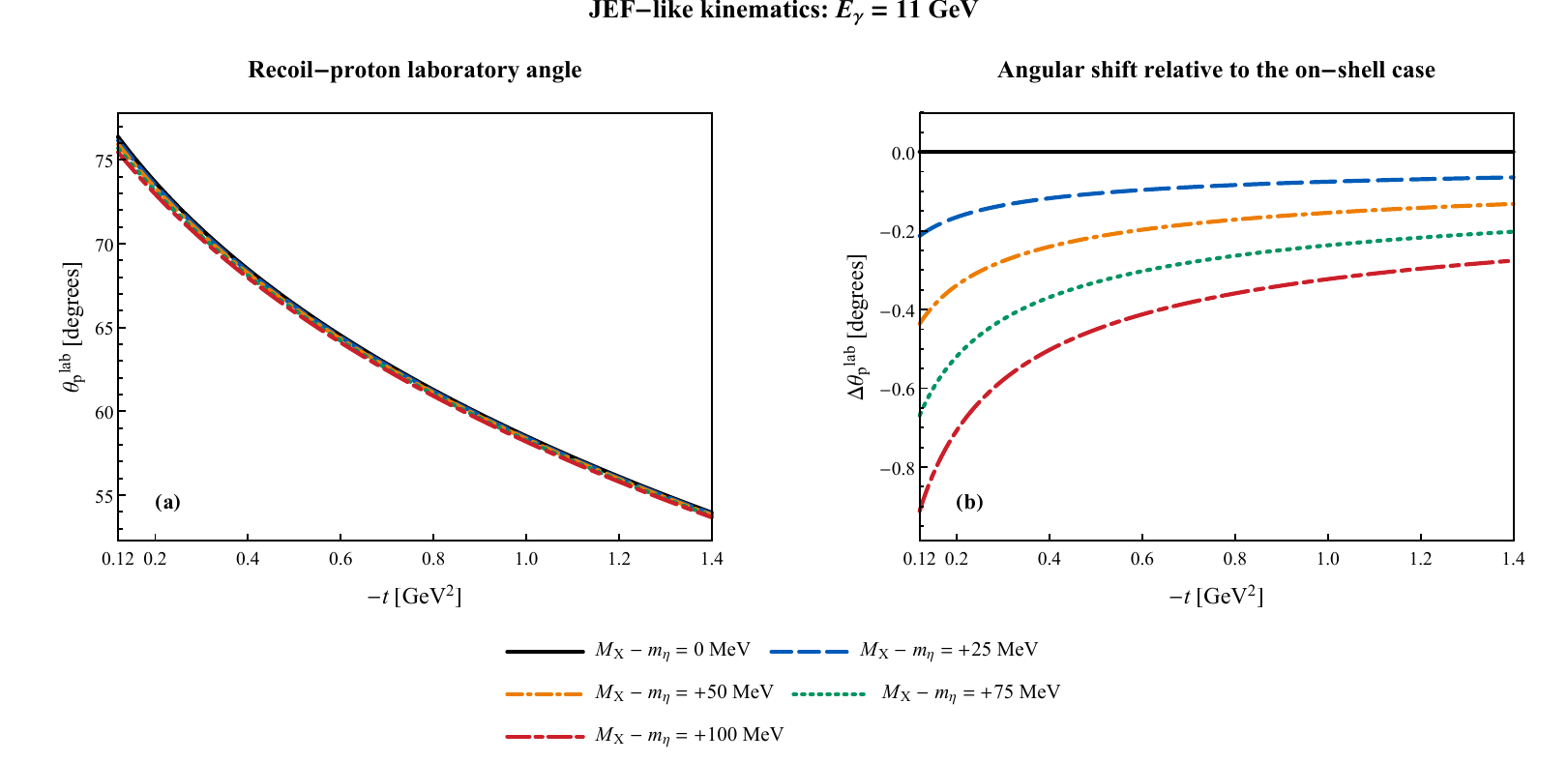}
    \caption{Kinematic correlations between $\theta_p^{\mathrm{lab}}$ and $\lvert t\rvert$ for  JEF-like incident photon energy, $E_\gamma=11~\rm{GeV}$, corresponding to Eqns.~\eqref{eq:theta-proton-t-MX} and~\eqref{eq:shift-wrt-on-shell}.}
    \label{fig:theta-proton-t-MX-JEF}
\end{figure}

\begin{figure}[!t]
    \centering
    \includegraphics[width=\textwidth]{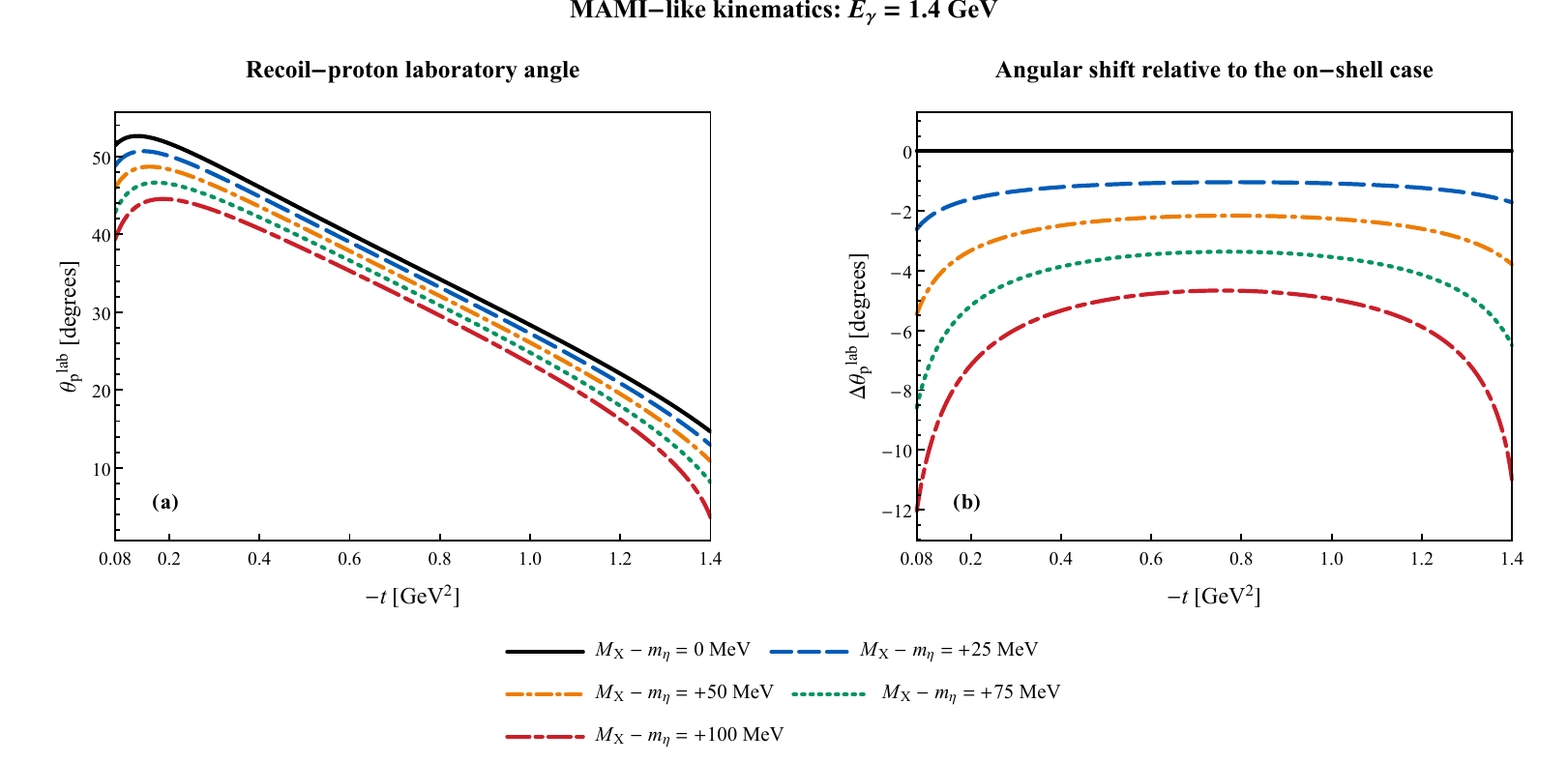}
        \caption{Kinematic correlations between $\theta_p^{\mathrm{lab}}$ and $\lvert t\rvert$ for  MAMI-like incident photon energy, $E_\gamma=1.4~\rm{GeV}$, corresponding to Eqns.~\eqref{eq:theta-proton-t-MX} and~\eqref{eq:shift-wrt-on-shell}.}
        \label{fig:theta-proton-t-MX-MAMI}
\end{figure}

Consider the photoproduction process:
\begin{equation}
    \gamma\left(q_\gamma\right)+p\left(p\right)
    \rightarrow
    X\left(p_X\right)+p\left(p^\prime\right),
    \qquad
    p_X^2=M_X^2,
\end{equation}
where $X$ denotes the reconstructed $\pi^0\gamma\gamma$ system.

In the laboratory frame, in which the target proton is at rest, we choose the incident-photon direction as the positive $\hat{\boldsymbol z}$ axis:
\begin{align}
    q_\gamma
    &=
    \left(E_\gamma,E_\gamma\hat{\boldsymbol z}\right),
    &
    p
    &=
    \left(m_p,\vec{\boldsymbol{0}}\right),
    &
    s
    &=
    \left(q_\gamma+p\right)^2
    =
    m_p^2+2m_pE_\gamma
\end{align}
We parametrize the four-momentum of the recoil proton as:
\begin{equation}
    p^\prime
    =
    \left(
    E_p^{\mathrm{lab}},
    \boldsymbol{p}_p^{\mathrm{lab}}
    \right),
    \qquad
    \theta_p^{\mathrm{lab}}
    =
    \angle\left(
    \boldsymbol{p}_p^{\mathrm{lab}},
    \hat{\boldsymbol z}
    \right)
\end{equation}
Conservation of momentum, $q_\gamma+p=p^\prime+p_X$, gives:
\begin{align}
    M_X^2
    &=
    \left(q_\gamma+p-p^\prime\right)^2 = 
    s+m_p^2
    -2\left(E_\gamma+m_p\right)E_p^{\mathrm{lab}}
    +2E_\gamma
    \left\lvert
    \boldsymbol{p}_p^{\mathrm{lab}}
    \right\rvert
    \cos\theta_p^{\mathrm{lab}}
    \label{eq:MX-recoil-proton}
\end{align}
The four-momentum transfer of the proton is:
\begin{align}
    t
    &\equiv
    \left(p-p^\prime\right)^2
    =
    \left(q_\gamma-p_X\right)^2 =
    2m_p^2-2m_pE_p^{\mathrm{lab}},
\end{align}
and with $\lvert t\rvert=-t$, the proton energy is: 
\begin{equation}
    E_p^{\mathrm{lab}}
    =
    \frac{2m_p^2-t}{2m_p}
    =
    m_p+\frac{\lvert t\rvert}{2m_p},
\end{equation}
and the magnitude of the recoil-proton momentum is: 
\begin{align}
    \left\lvert
    \boldsymbol{p}_p^{\mathrm{lab}}
    \right\rvert
    &=
    \sqrt{
    \left(E_p^{\mathrm{lab}}\right)^2-m_p^2
    }  =
    \sqrt{
    \lvert t\rvert
    +\frac{\lvert t\rvert^2}{4m_p^2}
    }
    =
    \frac{1}{2}
    \sqrt{
    \lvert t\rvert
    \left(
    4+\frac{\lvert t\rvert}{m_p^2}
    \right)
    }.
\end{align}
Substituting these back into Eq.~\eqref{eq:MX-recoil-proton} gives: 
\begin{equation}
    M_X^2
    =
    -\frac{\lvert t\rvert}{m_p}
    \left(E_\gamma+m_p\right)
    +
    2E_\gamma
    \left\lvert
    \boldsymbol{p}_p^{\mathrm{lab}}
    \right\rvert
    \cos\theta_p^{\mathrm{lab}}.
\end{equation}
Therefore, for fixed $M_X$ and incident photon energy, $\theta_p^{\mathrm{lab}}$ and $\lvert t\rvert$ are kinematically correlated: 
\begin{equation}
    \cos\theta_p^{\mathrm{lab}}
    =
    \frac{
    E_\gamma\lvert t\rvert
    +m_p\left(M_X^2+\lvert t\rvert\right)
    }{
    E_\gamma m_p
    \sqrt{
    \lvert t\rvert
    \left(
    4+\dfrac{\lvert t\rvert}{m_p^2}
    \right)
    }
    },
    \label{eq:theta-proton-t-MX}
\end{equation}
and this correlation is stronger for MAMI-like energy due to lower $E_\gamma$: 
\begin{equation}
    \frac{\partial \cos\theta_p^{\mathrm{lab}}}{\partial M_X^2} = \frac{1}{2E_\gamma\left\lvert
    \boldsymbol{p}_p^{\mathrm{lab}}
    \right\rvert}
    \label{eq:MX_derivative}
\end{equation}
The corresponding kinematic relations at representative JEF-like and
MAMI-like photon energies are shown in
Figs.~\ref{fig:theta-proton-t-MX-JEF} and~\ref{fig:theta-proton-t-MX-MAMI}, respectively, for a representative value of $\lvert t\lvert$~\cite{McNicoll:2010,Kashevarov:2017,AlGhoul:2017,Adhikari:2019}. In each figure, we also
show the recoil-angle shift relative to the on-shell $\eta$ production:
\begin{equation}
\Delta\theta_p^{\rm lab}(E_\gamma,t;M_X)
\equiv
\theta_p^{\rm lab}(E_\gamma,t;M_X)
-
\theta_p^{\rm lab}(E_\gamma,t;m_\eta)
\label{eq:shift-wrt-on-shell}
\end{equation}
At MAMI-like energies, the kinematic response is much larger than that at JEF-like energies. For JEF-like energy, the difference between the on-shell and continuous case approximately ranges from $\approx -0.2^\circ$ to $\approx -0.5^\circ$, while for MAMI, this value is much larger, and can reach $\approx -5^\circ$, or even larger magnitudes. A particularly large MAMI response is observed near the upper limit of the displayed $-t$ interval, which is enhanced by the proximity to the kinematic limit.

This large kinematic leverage motivates a reanalysis of existing MAMI data in the $(M_X-m_\eta,\Delta\theta_{p}^{\rm lab})$ variables, using the event-specific tagged-photon energy and momentum transfer. As summarized in Table~\ref{tab:eta_pi0_summary}, approximately $\mathcal{O}(10)$ candidates out of $1.2\times10^3$-event sample could populate the multi-degree, negative-angular-residual region, in case those events survive the experimental acceptance and selection criteria. JEF should perform the same correlated analysis. However, MAMI provides a significantly larger kinematic angular response per displacement in $M_X$, Eq.~\eqref{eq:MX_derivative}, while JEF can be complementary, providing higher statistics.

\begin{figure*}[t]
    \centering
    \includegraphics[width=0.49\textwidth]
{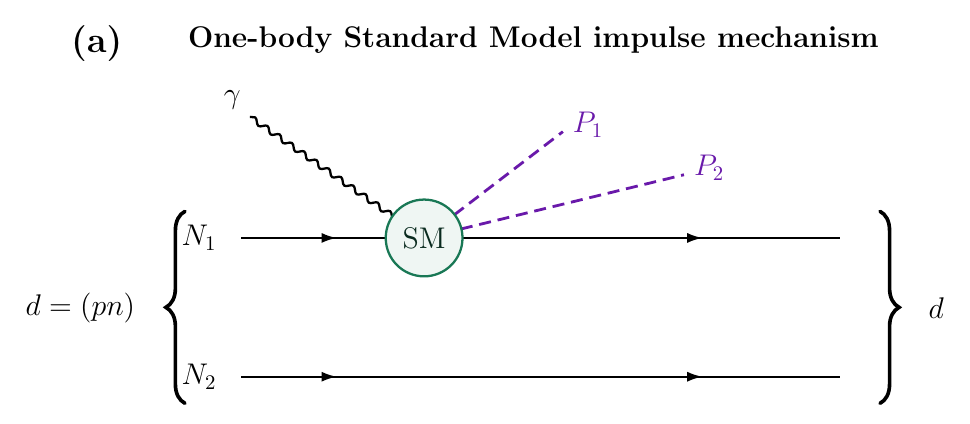}
    \hfill
    \includegraphics[width=0.49\textwidth]    {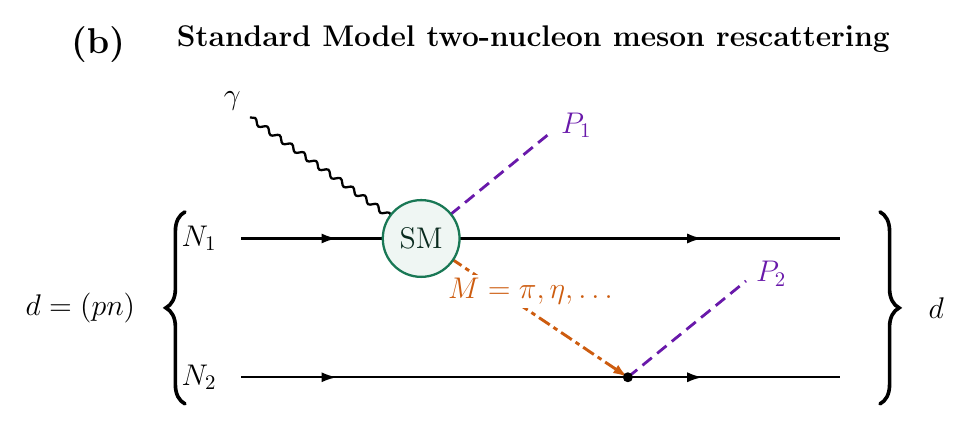}
    \caption{Representative one- and two-nucleon topologies for coherent two-meson photoproduction on the deuteron, $\gamma d\to P_1P_2d$, $(P_1,P_2)=(\pi^0,\eta)$ for $\gamma d\to\pi^0\eta d$ and $(P_1,P_2)=(\pi^0,\pi^0)$ for $\gamma d\to\pi^0\pi^0d$. \textbf{(a)} Conventional one-body impulse mechanism: the incident real photon interacts with one nucleon and produces the two-meson final state, whereas the second nucleon acts as a spectator. The two outgoing nucleons subsequently remain bound in the final deuteron. \textbf{(b)} Representative conventional two-nucleon mechanism involving meson rescattering: the photon-induced interaction on $N_1$ produces $P_1$ together with an exchanged meson $M=\pi,\eta,\ldots$, shown by the orange dash-dotted line, which subsequently interacts with $N_2$ and produces $P_2$.}
    \label{fig:SM_Deuteron_Angular}
\end{figure*}

\begin{figure*}[t]
    \centering
    \includegraphics[width=0.49\textwidth]
{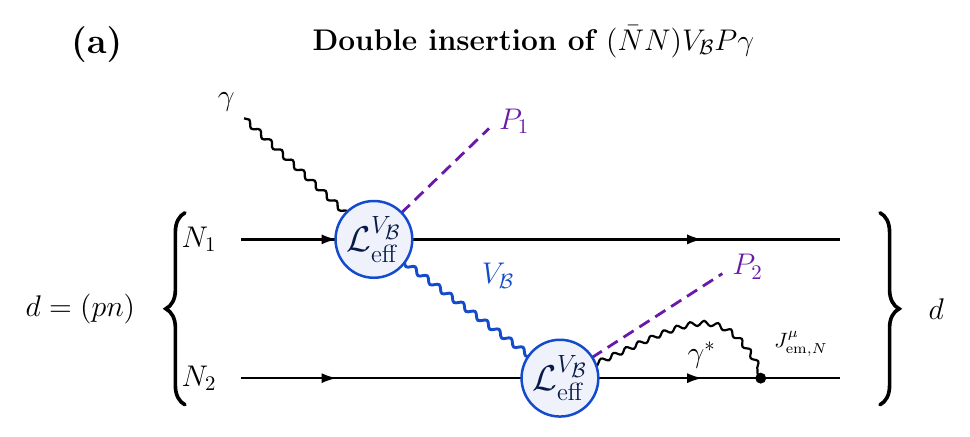}
    \hfill
    \includegraphics[width=0.49\textwidth]    {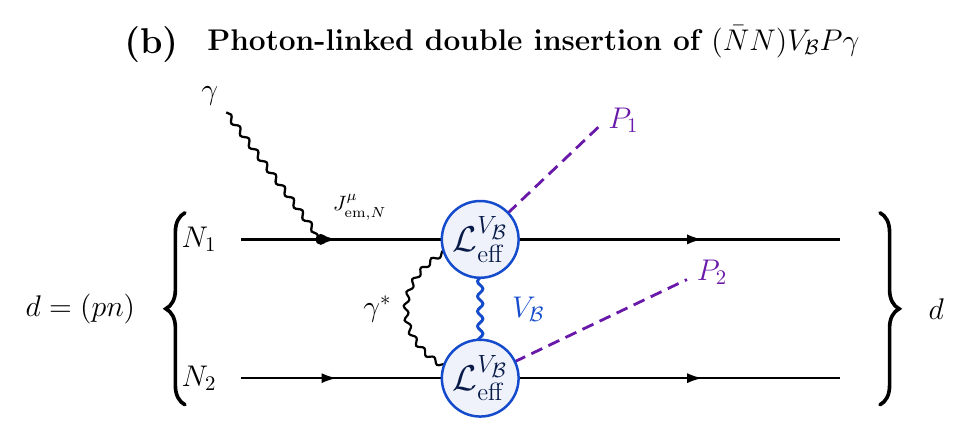}
    \caption{Representative $V_{\mathcal B}$-mediated two-nucleon topologies generated by two insertions of the effective $(\bar NN)V_{\mathcal B}P\gamma$ interaction in Eq.~\eqref{eq:nucleon-operator}. \textbf{(a)} At the first insertion, the incident photon produces $P_1$ and an exchanged $V_{\mathcal B}$, shown by the blue wavy line, on $N_1$. The exchanged vector boson couples at the second insertion on $N_2$, producing $P_2$ and a virtual photon $\gamma^*$ that couples to the ordinary nucleon electromagnetic current $J^\mu_{\mathrm{em},N}$. \textbf{(b)} The incident photon couples to the ordinary nucleon electromagnetic current $J^\mu_{\mathrm{em},N}$, while the two
    $\left(\bar{N}N\right)V_\mathcal{B}P\gamma$ insertions are connected by both the exchanged $V_{\mathcal B}$ and a virtual photon $\gamma^*$. Only representative diagrams are shown. Diagrams obtained by interchanging $N_1\leftrightarrow N_2$ and $P_1\leftrightarrow P_2$, together with the allowed alternative photon attachments and vertex orderings, are implicit.}
    \label{fig:VB_Deuteron_Angular}
\end{figure*}

\section{Possible $V_\mathcal{B}$ contribution to  $\gamma  d \rightarrow \pi^0  \eta  d $ and $\gamma  d \rightarrow \pi^0  \pi^0  d$ processes}
\label{VB_Angular_Distributions}

The conventional Standard Model diagrams relevant to these processes are illustrated in Fig.~\ref{fig:SM_Deuteron_Angular}, panels \textbf{(a)} and \textbf{(b)}. For $\gamma d\to\pi^0\eta d$, in the impulse approximation, the contribution in panel \textbf{(a)} is dominated by the sequential process $\gamma N\to\Delta^*(1700)\to\eta\Delta(1232)\to\eta\pi^0N$, where the second nucleon acts as a spectator. The Standard Model two-nucleon contributions, schematically represented in panel \textbf{(b)}, include $s$- and $p$-wave pion rescattering, $\eta$ rescattering, and a charged-pion conversion mechanism in which a charged pion produced on one nucleon subsequently interacts with the second nucleon to produce an $\eta$~\cite{MartinezTorres:2023}. For $\gamma d\to\pi^0\pi^0d$, the Standard Model calculations~\cite{Fix:2005DoublePion,Egorov:2015Pi0Pi0} employ a one-body impulse approximation, corresponding to panel \textbf{(a)}.

At the same time, Fig.~\ref{fig:VB_Deuteron_Angular}, panels \textbf{(a)} and \textbf{(b)}, illustrate how two insertions of the effective $V_{\mathcal B}$ operator in Eq.~\eqref{eq:nucleon-operator} can connect the two nucleon lines and generate an additional two-nucleon interaction. Such contributions are potentially relevant in view of the unexpectedly flat angular dependence observed in the coherent photoproduction of two mesons, which motivated the suggestion in ~\cite{Ishikawa:2026review}: ``The rather uniform distributions may suggest existence of two-baryon correlated states (dibaryons) in the intermediate states.''

Unlike the rare-decay process in Fig.~\ref{fig:SM_VB_cross}, these topologies involve both nucleons of the deuteron, rather than the single external nucleon relevant for the photoproduction and charge-exchange processes considered above. Quantifying their contribution to the observed angular distributions, including their interference with the Standard Model amplitudes, requires a dedicated calculation, which we postpone to future work. Thus, these reactions provide additional channels in which the same $V_{\mathcal B}$ interaction can be tested and constrained.

\section{Discussion, conclusions, and future work}
\label{Conclusions}

The purpose of this work is to investigate the kinematic fingerprints of the hypothetical $V_\mathcal{B}$ mechanism in the $\eta^{(\prime)}\to\pi^0(\eta)\gamma\gamma$ decays when $\eta^{\left(\prime\right)}$ are produced on nuclear targets. Among the channels considered, the effect is most pronounced in $\eta\to\pi^0\gamma\gamma$, making the JEF measurement of this channel potentially a decisive experiment capable of ruling out or confirming this hypothesis. Under the assumptions made here, the model predicts an effective branching fraction at JEF close to, but not completely identical to, that measured by MAMI. If an upcoming JEF measurement is consistent with the KLOE result and incompatible with the predicted enhancement, the proposed mechanism will be ruled out as an explanation of the MAMI--KLOE discrepancy.

An additional test would be provided by a future BESIII measurement of the $\eta\to\pi^0\gamma\gamma$ branching, using $\eta$ mesons produced through $J/\psi\to\gamma\eta$~\cite{ablikim2023improved}, without an external nucleon current. A result compatible with MAMI and incompatible with KLOE would strongly disfavor the proposed environment-dependent explanation of the MAMI--KLOE discrepancy. On the other hand, a result consistent with KLOE and significantly lower than the MAMI value would support the qualitative pattern predicted by the $V_{\mathcal B}$ model, although it would not in itself confirm the model.

For the hypothesized \rm{GeV}-scale $V_{\mathcal B}$ mass and the parameter choices adopted here, we find that the dominant new contribution to the decay rate arises through interference with the Standard Model amplitude, which is dominated by the exchange of vector mesons $V\in\{\omega,\rho^0,\phi\}$. Approximately $91\%$ of the total predicted rate, the Standard Model and $V_{\mathcal B}$-induced contributions combined, remains effectively on shell.

\textbf{\underline{If}} JEF observes an effective branching ratio close to that measured by MAMI,  $\approx 2.6$ times the KLOE value, the mechanism can be tested further through its predicted off-shell component, which accounts for the remaining approximately $9\%$ of the total rate and generates tails extending beyond the kinematic boundaries of an on-shell $\eta$ decay (the kinematic Rubicon).

Our results can be summarized as follows: 

\begin{itemize}

  \item For identical selection windows, the $V_{\mathcal B}$ model predicts a slightly larger effective branching fraction in JEF-like than in MAMI-like kinematics, although the difference remains below $1\%$. 
  
  \item For a representative sample of $1200$ reconstructed
  $\eta\to\pi^0\gamma\gamma$ events, the $V_\mathcal{B}$ model predicts approximately $10$ events beyond the kinematic Rubicon in either the $\gamma\gamma$ or $\pi^0\gamma$ invariant-mass distribution, which can serve as a clean signature of the $V_{\mathcal B}$ mechanism, since a genuinely on-shell $\eta$ pole has exactly zero kinematic support in this region before detector effects.

  \item The associated recoil-proton angular displacement is substantially larger in MAMI-like than in JEF-like kinematics. At MAMI energies, the shift relative to the on-shell $\eta$ hypothesis can reach approximately $-5^\circ$, or even larger magnitudes, especially towards the upper end of the accessible $-t$ interval.

  \item In both MAMI-like and JEF-like kinematics,  the $V_\mathcal{B}$ model predicts that the  mean reconstructed-mass shift is \underline{\emph{positive}} of similar magnitude, $\langle M_{\pi^0\gamma\gamma}\rangle-m_\eta\simeq + 2$--$7~\mathrm{MeV}$.
  
\end{itemize}

Qualitatively similar but much smaller effects are predicted for $\eta^\prime\to\pi^0\gamma\gamma$, while for $\eta^\prime\to\eta\gamma\gamma$, they are negligible at the current level of experimental accuracy.

In addition, GlueX has recently analyzed exclusive $\pi^0$ and $\eta$ photoproduction on $^{208}\mathrm{Pb}$, reconstructing the mesons through their two-photon decay channels~\cite{GrahamHoward2026}. The model predicts that the effective $\eta\to\pi^0\gamma\gamma$ branching increases strongly with the nuclear mass number $\rm{A}$, potentially exceeding the value measured by MAMI~\cite{Nefkens:2014}. Therefore, a systematic comparison using light and heavy nuclear targets could provide a decisive test of the model.

Finally, the same $V_{\mathcal B}$ interaction proposed to address the discrepancies in $\eta^{(\prime)}\to\pi^0(\eta)\gamma\gamma$ decays may also contribute to coherent $\gamma d\to\eta\pi^0 d$ and $\gamma d\to\pi^0\pi^0 d$ photoproduction, where discrepancies between measured angular distributions and existing hadronic calculations have been reported. These channels could therefore provide an additional probe of the proposed mechanism. We do \underline{not} claim here that the $V_{\mathcal B}$ interaction improves agreement with the experimental data. We note instead that its possible contribution warrants a dedicated quantitative investigation in future work.

\section*{Code and reproducibility of our results}

All numerical code used to generate the predictions and figures presented in this work is publicly available in the accompanying GitHub repository~\cite{NumericalCode}, to facilitate the reproduction and further use of our results.

\appendix

\section{Standard Model benchmark}
\label{Appendix:SM_benchmark}

For completeness, we summarize the Standard Model benchmark used
throughout this work. For $P_i(P)\to P_f(k)\gamma(q_1,\epsilon_1)\gamma(q_2,\epsilon_2)$, with $P_i=\{\eta,\eta^\prime\}$ and $P_f=\{\pi^0,\eta\}$, the amplitude is written as the coherent sum: 
\begin{equation}
\mathcal M_{\rm SM}
=
\mathcal M_{\rm VMD}
+
\mathcal M_{\rm L\sigma M},
\end{equation}
and therefore
\begin{equation}
\begin{aligned}
|\mathcal M_{\rm SM}|^2
={}&|\mathcal M_{\rm VMD}|^2
+2\,\mathrm{Re}
 \left(
 \mathcal M_{\rm VMD}\mathcal M_{\rm L\sigma M}^{*}
 \right)
+|\mathcal M_{\rm L\sigma M}|^2 
\end{aligned}
\end{equation}
The two gauge-invariant Lorentz structures are defined as:
\begin{equation}
\begin{aligned}
\{a\} \equiv{}&
(\epsilon_1\!\cdot\!\epsilon_2)(q_1\!\cdot\!q_2)
-(\epsilon_1\!\cdot\!q_2)(\epsilon_2\!\cdot\!q_1),
\\[2pt]
\{b\} \equiv{}&
(\epsilon_1\!\cdot\!q_2)(\epsilon_2\!\cdot\!P)(P\!\cdot\!q_1)
+(\epsilon_2\!\cdot\!q_1)(\epsilon_1\!\cdot\!P)(P\!\cdot\!q_2)
-(\epsilon_1\!\cdot\!\epsilon_2)
  (P\!\cdot\!q_1)(P\!\cdot\!q_2)
  \nonumber\\
&
-(\epsilon_1\!\cdot\!P)(\epsilon_2\!\cdot\!P)
  (q_1\!\cdot\!q_2)
\end{aligned}
\end{equation}
The dominant VMD contribution is:
\begin{equation}
\label{VMD}
\begin{aligned}
\mathcal M_{\rm VMD}^{P_i\to P_f\gamma\gamma}
={}&
\sum_{V=\rho^0,\omega,\phi}
g_{VP_i\gamma}g_{VP_f\gamma}
\Bigg[
\frac{(P\!\cdot\!q_2-m_{P_i}^2)\{a\}-\{b\}}
     {D_V(t)} +
\frac{(P\!\cdot\!q_1-m_{P_i}^2)\{a\}-\{b\}}
     {D_V(u)}
\Bigg],
\end{aligned}
\end{equation}
where
\begin{equation}
t=(P-q_2)^2=m_{P_i}^2-2P\!\cdot q_2,
\qquad
u=(P-q_1)^2=m_{P_i}^2-2P\!\cdot q_1,
\end{equation}
and the propagator of the vector meson is:

\begin{equation}\label{BW}
D_V(t)=m_V^2-t-i\,m_V\Gamma_V,
\end{equation}
where $V\in\{\omega,\rho,\phi\}$. The $\omega$- and $\phi$-meson widths are kept constant, while the broad $\rho$ resonance is described using an energy-dependent width. Instead of the Roos--Pi\v{s}\'{u}t prescription~\cite{roos1969} adopted in~\cite{escribano2020theoretical}, we follow the parametrization of Lichard and Vojik~\cite{lichard2006}:
\begin{equation}
\Gamma_{\rho}(s)
=
\Gamma_{\rho}\frac{m_{\rho}}{\sqrt{s}}
\left(
\frac{s-4m_{\pi^+}^{2}}
     {m_{\rho}^{2}-4m_{\pi^+}^{2}}
\right)^{3/2}
\theta\!\left(s-4m_{\pi^+}^{2}\right),
\end{equation}
where $\Gamma_{\rho}$ denotes the on-shell width, while the step function enforces the charged-pion threshold. This parametrization provides a description of the CMD-2, SND, and KLOE data that is comparable to or better than that obtained with conventional parametrization~\cite{lichard2006}.

We adopt the phenomenological $\mathrm{SU}(3)$-breaking parametrization of the $VP\gamma$ couplings from~\cite{Bramon:2000fr,Escribano:2020jdy}. For compactness, we define $c_{P,V}\equiv\cos\varphi_{P,V}$,
$s_{P,V}\equiv\sin\varphi_{P,V}$, and
$r_s\equiv z_{\rm S}\overline m/m_s$. The couplings are: 
\begin{equation}
\label{CouplingConstants}
\begin{gathered}
g_{\rho^0\pi^0\gamma}=\frac{g}{3},\quad
g_{\rho^0\eta\gamma}=g z_{\rm NS}c_P,\quad
g_{\rho^0\eta^\prime\gamma}=g z_{\rm NS}s_P,\quad
g_{\omega\pi^0\gamma}=g c_V,\quad
g_{\phi\pi^0\gamma}=g s_V,
\\[2pt]
g_{\omega\eta\gamma}
=\frac{g}{3}\left(z_{\rm NS}c_Pc_V-2r_s s_Ps_V\right),\qquad
g_{\phi\eta\gamma}
=\frac{g}{3}\left(z_{\rm NS}c_Ps_V+2r_s s_Pc_V\right),
\\[2pt]
g_{\omega\eta^\prime\gamma}
=\frac{g}{3}\left(z_{\rm NS}s_Pc_V+2r_s c_Ps_V\right),\qquad
g_{\phi\eta^\prime\gamma}
=\frac{g}{3}\left(z_{\rm NS}s_Ps_V-2r_s c_Pc_V\right)
\end{gathered}
\end{equation}
Numerically,  we employ: 
\begin{equation}
\label{fit4}
\begin{aligned}
g&=0.70(1)~{\rm GeV}^{-1},&
\varphi_P&=41.4(5)^\circ,&
z_{\rm NS}&=0.83(2),\\
r_s&=0.65(1),&
\varphi_V&=3.3(1)^\circ ,
\end{aligned}
\end{equation}
which were fitted to reproduce the measured $V\to P\gamma$ and $P\to V\gamma$~\cite{Bramon:2000fr,Escribano:2020jdy}.

In our calculations, the numerically subleading L$\sigma$M contribution to the Standard Model amplitude is kept exactly as implemented in~\cite{escribano2020theoretical}. For completeness, the corresponding expressions are summarized below:
\begin{equation}
\label{AKpKmpi0etaChPTLsM}
\begin{aligned}
{\cal A}_{K^+K^-\to\pi^0\eta}^{\mbox{\scriptsize L$\sigma$M}} =
\frac{1}{2f_\pi f_K}\bigg\{&
(s-m_\eta^2)\frac{m_K^2-m_{a_0}^2}{D_{a_0}(s)}\cos\varphi_P
+\frac{1}{6}\Big[(5m_\eta^2+m_\pi^2-3s)\cos\varphi_P\\
&-\sqrt{2}(m_\eta^2+4m_K^2+m_\pi^2-3s)\sin\varphi_P\Big]\bigg\},
\end{aligned}
\end{equation}
\begin{equation}
\label{AKpKmpi0etapChPTLsM}
\begin{aligned}
{\cal A}_{K^+K^-\to\pi^0\eta'}^{\mbox{\scriptsize L$\sigma$M}} =
\frac{1}{2f_\pi f_K}\bigg\{&
(s-m_{\eta'}^2)\frac{m_K^2-m_{a_0}^2}{D_{a_0}(s)}\sin\varphi_P
+\frac{1}{6}\Big[(5m_{\eta'}^2+m_\pi^2-3s)\sin\varphi_P\\
&+\sqrt{2}(m_{\eta'}^2+4m_K^2+m_\pi^2-3s)\cos\varphi_P\Big]\bigg\},
\end{aligned}
\end{equation}
\begin{equation}
\label{AKpKmetaetapChPTLsM}
\begin{aligned}
{\cal A}_{K^+K^-\to\eta\eta'}^{\mbox{\scriptsize L$\sigma$M}} =
&\frac{s-m_K^2}{2f_K}\bigg[
\frac{g_{\sigma\eta\eta'}}{D_\sigma(s)}
(\cos\varphi_S-\sqrt{2}\sin\varphi_S)
+\frac{g_{f_0\eta\eta'}}{D_{f_0}(s)}
(\sin\varphi_S+\sqrt{2}\cos\varphi_S)\bigg]\\
&-\frac{s-m_K^2}{4f_\pi f_K}
\left[1-2\left(\frac{2f_K}{f_\pi}-1\right)\right]\sin(2\varphi_P)\\
&-\frac{1}{4f_\pi^2}\bigg[
\left(s-\frac{m_\eta^2+m_{\eta'}^2}{3}-\frac{8m_K^2}{9}-\frac{2m_\pi^2}{9}\right)
\left(\sqrt{2}\cos(2\varphi_P)+\frac{1}{2}\sin(2\varphi_P)\right)\\
&\hspace{30mm}
+\frac{4}{9}(2m_K^2-m_\pi^2)
\left(2\sin(2\varphi_P)-\frac{1}{\sqrt{2}}\cos(2\varphi_P)\right)
\bigg],
\end{aligned}
\end{equation}
\begin{equation}
\label{ApippimetaetapChPTLsM}
{\cal A}_{\pi^+\pi^-\to\eta\eta'}^{\mbox{\scriptsize L$\sigma$M}} =
\frac{s-m_\pi^2}{f_\pi}\bigg[
\frac{g_{\sigma\eta\eta'}}{D_\sigma(s)}\cos\varphi_S
+\frac{g_{f_0\eta\eta'}}{D_{f_0}(s)}\sin\varphi_S
\bigg]
+\frac{2m_\pi^2-s}{2f_\pi^2}\sin(2\varphi_P)
\end{equation}
The involved scalar couplings are: 
\begin{equation}
\label{gsigmaetaetap}
\begin{aligned}
g_{\sigma\eta\eta'} =
\frac{\sin(2\varphi_P)}{2f_\pi}\bigg\{&
(m_\eta^2\cos^2\varphi_P+m_{\eta'}^2\sin^2\varphi_P-m_{a_0}^2)
\left[\cos\varphi_S+\sqrt{2}\sin\varphi_S
\left(2\frac{f_K}{f_\pi}-1\right)\right]\\
&-(m_{\eta'}^2-m_\eta^2)
\left[\cos\varphi_S\cos(2\varphi_P)
-\frac{1}{2}\sin\varphi_S\sin(2\varphi_P)\right]\bigg\},
\end{aligned}
\end{equation}
\begin{equation}
\label{gf0etaetap}
\begin{aligned}
g_{f_0\eta\eta'} =
\frac{\sin(2\varphi_P)}{2f_\pi}\bigg\{&
(m_\eta^2\cos^2\varphi_P+m_{\eta'}^2\sin^2\varphi_P-m_{a_0}^2)
\left[\sin\varphi_S-\sqrt{2}\cos\varphi_S
\left(2\frac{f_K}{f_\pi}-1\right)\right]\\
&-(m_{\eta'}^2-m_\eta^2)
\left[\sin\varphi_S\cos(2\varphi_P)
+\frac{1}{2}\cos\varphi_S\sin(2\varphi_P)\right]\bigg\}
\end{aligned}
\end{equation}
Following~\cite{escribano2006,escribano2020theoretical}, the scalar mixing angle is numerically fixed to $\varphi_S=-8^\circ$,  and is defined as: 
\begin{equation}
\begin{cases}
\ket{\sigma}=\cos\varphi_S\,\ket{\sigma_{\rm NS}}-\sin\varphi_S\,\ket{\sigma_{\rm S}},\\[3pt]
\ket{f_0}=\sin\varphi_S\,\ket{\sigma_{\rm NS}}+\cos\varphi_S\,\ket{\sigma_{\rm S}},
\end{cases}
\qquad
\ket{\sigma_{\rm NS}}=\frac{\ket{u\bar u}+\ket{d\bar d}}{\sqrt{2}},
\qquad
\ket{\sigma_{\rm S}}=\ket{s\bar s}
\end{equation}
Only the charged-kaon loop contributes to the L$\sigma$M amplitude for $\eta^{(\prime)}\to\pi^0\gamma\gamma$:
\begin{equation}
{\cal M}^{\mathrm{L}\sigma\mathrm M}_{\eta^{(\prime)}\to\pi^0\gamma\gamma} =
\frac{2\alpha}{\pi}\frac{L(s_K)}{m_{K^+}^2}\{a\}
\times{\cal A}^{\mathrm{L}\sigma\mathrm M}_{K^+K^-\to\pi^0\eta^{(\prime)}}
\end{equation}
Both charged-pion and charged-kaon loops contribute to $\eta'\to\eta\gamma\gamma$:
\begin{equation}
\begin{aligned}
{\cal M}^{\mathrm{L}\sigma\mathrm M}_{\eta'\to\eta\gamma\gamma} =
&\frac{2\alpha}{\pi}\frac{L(s_\pi)}{m_{\pi^+}^2}\{a\}
\times{\cal A}^{\mathrm{L}\sigma\mathrm M}_{\pi^+\pi^-\to\eta\eta'}+\frac{2\alpha}{\pi}\frac{L(s_K)}{m_{K^+}^2}\{a\}
\times{\cal A}^{\mathrm{L}\sigma\mathrm M}_{K^+K^-\to\eta\eta'}
\end{aligned}
\end{equation}
The loop function is:
\begin{equation}
L(z)=-\frac{1}{2z}-\frac{2}{z^2}f\left(\frac{1}{z}\right),\qquad
f(z)=
\begin{cases}
\dfrac{1}{4}\left(\log\dfrac{1+\sqrt{1-4z}}{1-\sqrt{1-4z}}-i\pi\right)^2,
& z<\dfrac{1}{4}\\[2ex]
-\left[\arcsin\left(\dfrac{1}{2\sqrt z}\right)\right]^2,
& z>\dfrac{1}{4}
\end{cases},
\end{equation}
where $s_K=s/m_{K^+}^2$, $s_\pi=s/m_{\pi^+}^2$, and
$s=(q_1+q_2)^2=2q_1\cdot q_2$ is the diphoton invariant mass squared.
The $a_0(980)$ resonance contributes to both $\eta^{(\prime)}\to\pi^0\gamma\gamma$ decays, and its renormalized mass is fixed to $m_{a_0}=980~\mathrm{MeV}$. For $\eta'\to\eta\gamma\gamma$, the relevant scalar states are $\sigma(500)$ and $f_0(980)$, with $m_\sigma=498~\mathrm{MeV}$ and $m_{f_0}=990~\mathrm{MeV}$. Numerically, we use $f_\pi=92.07~\mathrm{MeV}$ and $f_K=110.10~\mathrm{MeV}$.
Each scalar resonance is described by the dressed one-loop propagator:
\begin{equation}
\label{ScalarResonances}
D_R(s)=s-m_R^2+\operatorname{Re}\Pi_R(s)-\operatorname{Re}\Pi_R(m_R^2)
+i\operatorname{Im}\Pi_R(s)
\end{equation}
For the $a_0$ resonance, the couplings in the isospin limit are:
\begin{equation}
\begin{cases}
g_{a_0K\bar K}^2=2g_{a_0K^+K^-}^2
=\dfrac{1}{2}\left(\dfrac{m_K^2-m_{a_0}^2}{f_K}\right)^2\\[3ex]
g_{a_0\pi\eta}^2
=\left(\dfrac{m_\eta^2-m_{a_0}^2}{f_\pi}\cos\varphi_P\right)^2
\end{cases}
\end{equation}
For $i=\{\pi,K\}$, the kinematic quantities are
$\beta_i=\sqrt{1-4m_i^2/s}$,
$\bar\beta_i=\sqrt{4m_i^2/s-1}$,
$\theta_i=\theta(s-4m_i^2)$, and
$\bar\theta_i=\theta(4m_i^2-s)$. For the $\pi\eta$ channel,
$\beta^\pm_{\pi\eta}=\sqrt{1-(m_\pi\pm m_\eta)^2/s}$,
$\bar\beta^\pm_{\pi\eta}=\sqrt{(m_\pi\pm m_\eta)^2/s-1}$,
$\theta_{\pi\eta}=\theta[s-(m_\pi+m_\eta)^2]$,
$\bar\theta_{\pi\eta}=\theta[s-(m_\pi-m_\eta)^2]\theta[(m_\pi+m_\eta)^2-s]$, and
$\bar{\bar\theta}_{\pi\eta}=\theta[(m_\pi-m_\eta)^2-s]$.
The real and imaginary parts of the $a_0$ self-energy are:
\begin{equation}
\label{pi0Rea0}
\begin{aligned}
\operatorname{Re}\Pi_{a_0}(s)=&
\frac{g_{a_0K\bar K}^2}{16\pi^2}
\left[2-\beta_K\log\left(\frac{1+\beta_K}{1-\beta_K}\right)\theta_K
-2\bar\beta_K\arctan\left(\frac{1}{\bar\beta_K}\right)\bar\theta_K\right]\\
&+\frac{g_{a_0\pi\eta}^2}{16\pi^2}
\bigg[2-\frac{m_\eta^2-m_\pi^2}{s}\log\left(\frac{m_\eta}{m_\pi}\right)
-\beta^+_{\pi\eta}\beta^-_{\pi\eta}
\log\left(\frac{\beta^-_{\pi\eta}+\beta^+_{\pi\eta}}
{\beta^-_{\pi\eta}-\beta^+_{\pi\eta}}\right)\theta_{\pi\eta}\\
&\qquad
-2\bar\beta^+_{\pi\eta}\beta^-_{\pi\eta}
\arctan\left(\frac{\beta^-_{\pi\eta}}{\bar\beta^+_{\pi\eta}}\right)
\bar\theta_{\pi\eta}
+\bar\beta^+_{\pi\eta}\bar\beta^-_{\pi\eta}
\log\left(\frac{\bar\beta^+_{\pi\eta}+\bar\beta^-_{\pi\eta}}
{\bar\beta^+_{\pi\eta}-\bar\beta^-_{\pi\eta}}\right)
\bar{\bar\theta}_{\pi\eta}\bigg],
\end{aligned}
\end{equation}
\begin{equation}
\label{pi0Ima0}
\operatorname{Im}\Pi_{a_0}(s)=
-\frac{g_{a_0K\bar K}^2}{16\pi}\beta_K\theta_K
-\frac{g_{a_0\pi\eta}^2}{16\pi}
\beta^+_{\pi\eta}\beta^-_{\pi\eta}\theta_{\pi\eta}
\end{equation}
For the $\sigma(500)$ resonance:
\begin{equation}
\begin{cases}
g_{\sigma\pi\pi}^2=\dfrac{3}{2}g_{\sigma\pi^+\pi^-}^2
=\dfrac{3}{2}\left(\dfrac{m_\pi^2-m_\sigma^2}{f_\pi}\cos\varphi_S\right)^2\\[3ex]
g_{\sigma K\bar K}^2=2g_{\sigma K^+K^-}^2
=\dfrac{1}{2}\left[\dfrac{m_K^2-m_\sigma^2}{f_K}
(\cos\varphi_S-\sqrt{2}\sin\varphi_S)\right]^2
\end{cases}
\end{equation}
For the $f_0(980)$ resonance:
\begin{equation}
\begin{cases}
g_{f_0\pi\pi}^2=\dfrac{3}{2}g_{f_0\pi^+\pi^-}^2
=\dfrac{3}{2}\left(\dfrac{m_\pi^2-m_{f_0}^2}{f_\pi}\sin\varphi_S\right)^2\\[3ex]
g_{f_0K\bar K}^2=2g_{f_0K^+K^-}^2
=\dfrac{1}{2}\left[\dfrac{m_K^2-m_{f_0}^2}{f_K}
(\sin\varphi_S+\sqrt{2}\cos\varphi_S)\right]^2
\end{cases}
\end{equation}
The corresponding $\sigma$ and $f_0$ self-energies are:
\begin{equation}
\label{EtaEtaPrimeSigmaR}
\begin{aligned}
\operatorname{Re}\Pi_\sigma(s)=&
\frac{g_{\sigma\pi\pi}^2}{16\pi^2}
\left[2-\beta_\pi\log\left(\frac{1+\beta_\pi}{1-\beta_\pi}\right)\theta_\pi
-2\bar\beta_\pi\arctan\left(\frac{1}{\bar\beta_\pi}\right)\bar\theta_\pi\right]\\
&+\frac{g_{\sigma K\bar K}^2}{16\pi^2}
\left[2-\beta_K\log\left(\frac{1+\beta_K}{1-\beta_K}\right)\theta_K
-2\bar\beta_K\arctan\left(\frac{1}{\bar\beta_K}\right)\bar\theta_K\right],
\end{aligned}
\end{equation}
\begin{equation}
\label{EtaEtaPrimeSigmaI}
\operatorname{Im}\Pi_\sigma(s)=
-\frac{g_{\sigma\pi\pi}^2}{16\pi}\beta_\pi\theta_\pi
-\frac{g_{\sigma K\bar K}^2}{16\pi}\beta_K\theta_K,
\end{equation}
\begin{equation}
\label{EtaEtaPrimeRf0}
\begin{aligned}
\operatorname{Re}\Pi_{f_0}(s)=&
\frac{g_{f_0\pi\pi}^2}{16\pi^2}
\left[2-\beta_\pi\log\left(\frac{1+\beta_\pi}{1-\beta_\pi}\right)\theta_\pi
-2\bar\beta_\pi\arctan\left(\frac{1}{\bar\beta_\pi}\right)\bar\theta_\pi\right]\\
&+\frac{g_{f_0K\bar K}^2}{16\pi^2}
\left[2-\beta_K\log\left(\frac{1+\beta_K}{1-\beta_K}\right)\theta_K
-2\bar\beta_K\arctan\left(\frac{1}{\bar\beta_K}\right)\bar\theta_K\right],
\end{aligned}
\end{equation}
\begin{equation}
\label{EtaEtaPrimeIf0}
\operatorname{Im}\Pi_{f_0}(s)=
-\frac{g_{f_0\pi\pi}^2}{16\pi}\beta_\pi\theta_\pi
-\frac{g_{f_0K\bar K}^2}{16\pi}\beta_K\theta_K
\end{equation}

Descriptions of doubly radiative decays $\eta^{(\prime)}\to\pi^0(\eta)\gamma\gamma$ in the Standard Model cover a wide range of theoretical frameworks. For $\eta\to\pi^0\gamma\gamma$, these include the Vector Meson Dominance (VMD) model~\cite{oppo1967models,baracca1970general},
chiral perturbation theory ($\chi$PT)~\cite{ametller1992chiral}, and
extensions incorporating $C$-odd axial-vector
resonances~\cite{ko1993contributions,ko1995eta}. Other treatments employ unitarized chiral amplitudes~\cite{oset2003eta,oset2008eta}, dispersive methods~\cite{danilkin2017theoretical}, including a coupled-channel Muskhelishvili--Omnès analysis~\cite{lu2020interaction}, quark-loop
(box-diagram) calculations~\cite{ng1993,nemoto1996}, and both the original
and extended formulations of the Nambu--Jona-Lasinio
model~\cite{belkov1995,bellucci1995,bijnens1995,volkov2026decays}.

For the related decays, $\eta^\prime\to\pi^0\gamma\gamma$ and $\eta^\prime\to\eta\gamma\gamma$, early theoretical predictions were presented in~\cite{jora2010,escribano2012,Balytskyi:2018pzb,Balytskyi:2018uxb}, while a comprehensive treatment of all three channels, $\eta^{(\prime)}\to\pi^0(\eta)\gamma\gamma$, within the VMD--L$\sigma$M framework was subsequently performed in~\cite{escribano2020theoretical}. This framework was recently extended for the $\eta^{(\prime)}\to\pi^0\gamma\gamma$ channels to include the contribution of the $a_2(1320)$ tensor meson~\cite{escribano2025assessment}. Independent analyses have also been performed using VMD~\cite{Schaefer:2023stm} and, more recently, the Nambu--Jona-Lasinio model~\cite{volkov2026decays}.

It is useful to note the historical evolution of the VMD--L$\sigma$M description of these decays. An early implementation of this framework yielded $\mathrm{BR}(\eta\to\pi^0\gamma\gamma)=2.1\times10^{-4}$ (see Table~I in~\cite{escribano2012}), close to the experimental value available at that time~\cite{Prakhov:2005,Prakhov:2008}, and later measurement by MAMI~\cite{Nefkens:2014}. However, the same analysis predicted $\mathrm{BR}(\eta^\prime\to\pi^0\gamma\gamma)=6.5\times10^{-3}$, and $\mathrm{BR}(\eta^\prime\to\eta\gamma\gamma)=2.6\times10^{-4}$ (see Table~I in~\cite{escribano2012}), both of which were subsequently found to be incompatible with the corresponding BESIII results, \(\mathrm{BR}_{\textrm{BESIII}}^{\eta^\prime \to \pi^0 \gamma\gamma} = \left(3.20 \pm 0.07 \pm 0.23\right) \times 10^{-3}\)~\cite{ablikim2017observation} and \(\mathrm{BR}_{\textrm{BESIII}}^{\eta^\prime \to \eta \gamma\gamma} < 1.33 \times 10^{-4}\), respectively~\cite{ablikim2019search}.

All three channels were subsequently revisited in~\cite{escribano2020theoretical}, and the main outcome was the impossibility to describe all three decays within the same set of parameters:``...the corresponding branching ratios cannot be reproduced simultaneously.'' In other words, parameters compatible with the MAMI branching ratio for $\eta\to\pi^0\gamma\gamma$~\cite{Nefkens:2014} were incompatible with the BESIII results for $\eta^\prime\to\pi^0\gamma\gamma$ and $\eta^\prime\to\eta\gamma\gamma$~\cite{ablikim2017observation,ablikim2019search}, and vice versa. According to the updated fit in~\cite{escribano2020theoretical}: ``Interestingly, our predictions for the $\eta\rightarrow\pi^0\gamma\gamma$ are found to be approximately a factor of 2 smaller than the experimental measurements, whereas our theoretical predictions for the $\eta^\prime\rightarrow\pi^0\gamma\gamma$ and $\eta^\prime\rightarrow\eta\gamma\gamma$ are in good agreement with recent measurements performed by BESIII.''

This tension motivated~\cite{Balytskyi:2023} to introduce a production-independent leptophobic-vector contribution to the intrinsic $\eta^{(\prime)}\to\pi^0(\eta)\gamma\gamma$ amplitudes through the exchange of an intermediate vector boson $\mathcal{B}$ via $\eta\to\mathcal{B}\gamma$. However, KLOE-2~\cite{Babusci:2026} subsequently measured a lower branching ratio $\eta\to\pi^0\gamma\gamma$ consistent with the revised VMD--L$\sigma$M~\cite{escribano2020theoretical} benchmark, while remaining in significant tension with the MAMI~\cite{Nefkens:2014} result. Therefore, any production-independent modification would affect both KLOE and MAMI measurements identically and could not accommodate both measurements simultaneously. This observation motivated the production-dependent $V_{\mathcal B}$ mechanism proposed in~\cite{Balytskyi:2026}, whose contribution is activated specifically in the presence of an external nucleon current and thereby accounts for the production-dependent pattern shown in Fig.~\ref{EnvironmentPatternFig}.

\section{Phase space factorization}
\label{Appendix:PhaseSpaceFactorization}

The following derivation of the recursive phase-space factorization follows~\cite{Byckling:1969sx,BycklingKajantie:1973}. The Lorentz-invariant phase-space measure for \(n\) final-state particles with total four-momentum \(P\) is given by: 
\begin{equation}
\mathrm d\Phi_n\left(P;p_1,\ldots,p_n\right)
=
(2\pi)^4
\delta^{(4)}\left(
P-\sum_{i=1}^{n}p_i
\right)
\prod_{i=1}^{n}
\frac{\mathrm d^3\boldsymbol p_i}
{(2\pi)^3\,2E_i},
\label{eq:n-body-phase-space}
\end{equation}
where \(p_i^0=E_i\geq0\) and \(p_i^2=m_i^2\).
Consider grouping the first \(k\) particles into a composite system
\(X\), with four-momentum and invariant mass: 
\begin{equation}
q\equiv\sum_{i=1}^{k}p_i,
\qquad
q^2\equiv M_X^2,
\qquad
q^0\equiv E_q>0
\label{eq:cluster-momentum}
\end{equation}
We treat \(q^0\) and \(\boldsymbol q\) as independent components of the four-momentum \(q\), and replace the integration variable \(q^0\) by \(M_X^2=q^2\). At fixed \(\boldsymbol q\): 
\begin{equation}
E_q=\sqrt{\boldsymbol q^{\,2}+M_X^2},\qquad
M_X^2=(q^0)^2-\boldsymbol q^{\,2},
\qquad
\left.
\frac{\partial M_X^2}{\partial q^0}
\right|_{\boldsymbol q}
=2q^0=2E_q,
\end{equation}
and therefore: 
\begin{equation}
\mathrm d^4q
=
\mathrm d^3\boldsymbol q\,\mathrm dq^0
=
\frac{\mathrm dM_X^2}{2E_q}\,
\mathrm d^3\boldsymbol q
\label{eq:q-measure-factorization}
\end{equation}
We now insert the identity:
\begin{equation}
1=
\int\mathrm d^4q\,
\delta^{(4)}\left(
q-\sum_{i=1}^{k}p_i
\right)
\label{eq:cluster-identity}
\end{equation}
into Eq.~\eqref{eq:n-body-phase-space}. Using the second delta
function to replace \(\sum_{i=1}^{k}p_i\) by \(q\) in the overall
momentum-conservation constraint gives:
\begin{align}
\mathrm d\Phi_n\left(P;p_1,\ldots,p_n\right)
&=
(2\pi)^4
\int\mathrm d^4q\,
\delta^{(4)}\left(
P-q-\sum_{j=k+1}^{n}p_j
\right)
\delta^{(4)}\left(
q-\sum_{i=1}^{k}p_i
\right)
\nonumber\\
&\quad\times
\prod_{i=1}^{k}
\frac{\mathrm d^3\boldsymbol p_i}
{(2\pi)^3\,2E_i}
\prod_{j=k+1}^{n}
\frac{\mathrm d^3\boldsymbol p_j}
{(2\pi)^3\,2E_j}.
\label{eq:phase-space-before-factorization}
\end{align}
Substituting Eq.~\eqref{eq:q-measure-factorization} and
redistributing the factors of \(2\pi\), this becomes
\begin{align}
\mathrm d\Phi_n\left(P;p_1,\ldots,p_n\right)
&=
\int\frac{\mathrm dM_X^2}{2\pi}
\Bigg[
(2\pi)^4
\delta^{(4)}\left(
P-q-\sum_{j=k+1}^{n}p_j
\right)
\frac{\mathrm d^3\boldsymbol q}
{(2\pi)^3\,2E_q}
\prod_{j=k+1}^{n}
\frac{\mathrm d^3\boldsymbol p_j}
{(2\pi)^3\,2E_j}
\Bigg]
\nonumber\\
&\quad\times
\Bigg[
(2\pi)^4
\delta^{(4)}\left(
q-\sum_{i=1}^{k}p_i
\right)
\prod_{i=1}^{k}
\frac{\mathrm d^3\boldsymbol p_i}
{(2\pi)^3\,2E_i}
\Bigg].
\label{eq:phase-space-bracket-factorization}
\end{align}
The first and second expressions in square brackets are,
respectively:
\begin{align}
\mathrm d\Phi_{n-k+1}
\left(P;q,p_{k+1},\ldots,p_n\right)
&=
(2\pi)^4
\delta^{(4)}\left(
P-q-\sum_{j=k+1}^{n}p_j
\right)
\frac{\mathrm d^3\boldsymbol q}
{(2\pi)^3\,2E_q}
\prod_{j=k+1}^{n}
\frac{\mathrm d^3\boldsymbol p_j}
{(2\pi)^3\,2E_j},
\\[0.3em]
\mathrm d\Phi_k\left(q;p_1,\ldots,p_k\right)
&=
(2\pi)^4
\delta^{(4)}\left(
q-\sum_{i=1}^{k}p_i
\right)
\prod_{i=1}^{k}
\frac{\mathrm d^3\boldsymbol p_i}
{(2\pi)^3\,2E_i}
\end{align}
Therefore, the \(n\)-body phase space obeys the recursive
factorization formula:
\begin{equation}
\mathrm d\Phi_n\left(P;p_1,\ldots,p_n\right)
=
\frac{\mathrm dM_X^2}{2\pi}\,
\mathrm d\Phi_{n-k+1}
\left(P;q,p_{k+1},\ldots,p_n\right)
\mathrm d\Phi_k\left(q;p_1,\ldots,p_k\right),
\label{eq:recursive-phase-space-factorization}
\end{equation}
where integration over \(M_X^2\) is assumed when evaluating the full phase space. In the absence of additional cuts, its physical range is:
\begin{equation}
\left(\sum_{i=1}^{k}m_i\right)^2
\leq M_X^2\leq
\left(
\sqrt{P^2}-\sum_{j=k+1}^{n}m_j
\right)^2
\label{eq:cluster-mass-range}
\end{equation}
Applied to nucleon-assisted production of the
$\pi^0(\eta)\gamma\gamma$ subsystem, the general identity yields the
four-body factorization in Eq.~\eqref{eq:four-body-phase-space-factorization}.
A further decomposition with respect to the $\pi^0\gamma$ subsystem
gives Eq.~\eqref{eq:pi0-gamma-body-phase-space-factorization}.


\begin{thebibliography}{99}


\bibitem{Gan:2020aco}
L.~Gan, B.~Kubis, E.~Passemar, and S.~Tulin,
\href{https://doi.org/10.1016/j.physrep.2021.11.001}
{Phys.\ Rep.\ \textbf{945}, 1 (2022)}.

\bibitem{Babusci:2026}
D.~Babusci \textit{et al.} (KLOE-2 Collaboration),
\href{https://doi.org/10.1007/JHEP01(2026)091}
{J.\ High Energy Phys.\ \textbf{01} (2026) 091}.

\bibitem{DiMicco:2006}
B.~Di Micco \textit{et al.} (KLOE Collaboration),
\href{https://hdl.handle.net/11590/121954}
{Acta Phys.\ Slovaca \textbf{56}, 403 (2006)}.

\bibitem{PDG:2026}
F.~Takahashi \textit{et al.} (Particle Data Group),
\href{https://pdg.lbl.gov/}
{to be published in Int.\ J.\ Mod.\ Phys.\ A \textbf{41},
2630011 (2026)}.

\bibitem{Nefkens:2014}
B.~M.~K.~Nefkens \textit{et al.} (A2 at MAMI Collaboration),
\href{https://doi.org/10.1103/PhysRevC.90.025206}
{Phys.\ Rev.\ C \textbf{90}, 025206 (2014)}.

\bibitem{Prakhov:2005}
S.~Prakhov \textit{et al.},
\href{https://doi.org/10.1103/PhysRevC.72.025201}
{Phys.\ Rev.\ C \textbf{72}, 025201 (2005)}.

\bibitem{Prakhov:2008}
S.~Prakhov \textit{et al.},
\href{https://doi.org/10.1103/PhysRevC.78.015206}
{Phys.\ Rev.\ C \textbf{78}, 015206 (2008)}.

\bibitem{Knecht:2004}
N.~Knecht \textit{et al.},
\href{https://doi.org/10.1016/j.physletb.2004.03.038}
{Phys.\ Lett.\ B \textbf{589}, 14 (2004)}.

\bibitem{Balytskyi:2026}
Y.~Balytskyi,
\href{https://doi.org/10.1007/JHEP02(2026)053}
{J.\ High Energy Phys.\ \textbf{02} (2026) 053}.



\bibitem{Alde:1984}
D.~Alde \textit{et al.}
(Serpukhov-Brussels-Annecy (LAPP) and Soviet-CERN Collaborations),
\href{https://doi.org/10.1007/BF01547921}
{Z.\ Phys.\ C \textbf{25}, 225 (1984)}.

\bibitem{Landsberg:1985}
L.~G.~Landsberg,
\href{https://doi.org/10.1016/0370-1573(85)90129-2}
{Phys.\ Rep.\ \textbf{128}, 301 (1985)}.

\bibitem{ablikim2017observation}
M.~Ablikim \textit{et al.} (BESIII Collaboration),
\href{https://doi.org/10.1103/PhysRevD.96.012005}
{Phys.\ Rev.\ D \textbf{96}, 012005 (2017)}.

\bibitem{ablikim2019search}
M.~Ablikim \textit{et al.} (BESIII Collaboration),
\href{https://doi.org/10.1103/PhysRevD.100.052015}
{Phys.\ Rev.\ D \textbf{100}, 052015 (2019)}.

\bibitem{alde1987neutral}
D.~Alde \textit{et al.}
(Serpukhov-Brussels-Los Alamos-Annecy (LAPP) Collaboration),
\href{https://doi.org/10.1007/BF01630597}
{Z.\ Phys.\ C \textbf{36}, 603 (1987)}.

\bibitem{escribano2020theoretical}
R.~Escribano, S.~Gonzalez-Solis, R.~Jora, and E.~Royo,
\href{https://doi.org/10.1103/PhysRevD.102.034026}
{Phys.\ Rev.\ D \textbf{102}, 034026 (2020)}.

\bibitem{ablikim2023improved}
M.~Ablikim \textit{et al.} (BESIII Collaboration),
\href{https://doi.org/10.1103/PhysRevD.108.092002}
{Phys.\ Rev.\ D \textbf{108}, 092002 (2023)}.

\bibitem{JEFproposal}
H.~Al Ghoul \textit{et al.}
(GlueX Collaboration and other participants),
``Update to the JEF Proposal (PR12-14-004),''
Jefferson Lab Proposal C12-14-004 (2017),
\url{https://www.jlab.org/exp_prog/proposals/17/C12-14-004.pdf}.

\bibitem{Gan:JEFII2026}
L.~Gan \textit{et al.},
``JEF-II: An Extension of the JLab Eta Factory Experiment,''
Jefferson Lab Proposal E12-24-006A (2026),
\url{https://www.jlab.org/exp_prog/proposals/26/E12-24-006A.pdf}.

\bibitem{Ishikawa:2022}
T.~Ishikawa \textit{et al.},
\href{https://doi.org/10.1103/PhysRevC.105.045201}
{Phys.\ Rev.\ C \textbf{105}, 045201 (2022)}.

\bibitem{Figueiredo:2026}
A.~J.~C.~Figueiredo \textit{et al.},
\href{https://doi.org/10.1016/j.physletb.2026.140189}
{Phys.\ Lett.\ B \textbf{873}, 140189 (2026)}.

\bibitem{MartinezTorres:2023}
A.~Mart\'inez Torres, K.~P.~Khemchandani, and E.~Oset,
\href{https://doi.org/10.1103/PhysRevC.107.025202}
{Phys.\ Rev.\ C \textbf{107}, 025202 (2023)}.

\bibitem{Ishikawa:2024review}
T.~Ishikawa,
\href{https://doi.org/10.12693/APhysPolA.146.687}
{Acta Phys.\ Pol.\ A \textbf{146}, 687 (2024)}.

\bibitem{Ishikawa:2026review}
T.~Ishikawa,
\href{https://doi.org/10.22323/1.500.0090}
{Proc.\ Sci.\ HADRON2025, 090 (2026)}.


\bibitem{Ishikawa:2017Pi0Pi0}
T.~Ishikawa \textit{et al.},
\href{https://doi.org/10.1016/j.physletb.2017.04.010}
{Phys.\ Lett.\ B \textbf{772}, 398 (2017)}.

\bibitem{Ishikawa:2019Pi0Pi0}
T.~Ishikawa \textit{et al.},
\href{https://doi.org/10.1016/j.physletb.2018.12.050}
{Phys.\ Lett.\ B \textbf{789}, 413 (2019)}.

\bibitem{Jude:2022Pi0Pi0}
T.~C.~Jude \textit{et al.},
\href{https://doi.org/10.1016/j.physletb.2022.137277}
{Phys.\ Lett.\ B \textbf{832}, 137277 (2022)}.

\bibitem{Fix:2005DoublePion}
A.~Fix and H.~Arenh\"ovel,
\href{https://doi.org/10.1140/epja/i2005-10067-5}
{Eur.\ Phys.\ J.\ A \textbf{25}, 115 (2005)}.

\bibitem{Egorov:2015Pi0Pi0}
M.~Egorov and A.~Fix,
\href{https://doi.org/10.1016/j.nuclphysa.2014.10.002}
{Nucl.\ Phys.\ A \textbf{933}, 104 (2015)}.




\bibitem{joseph2026}
M.~Joseph, S.~Liebersbach, A.~A.~Madathil, and G.~Marques-Tavares,
\href{https://doi.org/10.48550/arXiv.2603.04513}
{arXiv:2603.04513 [hep-ph] (2026)}.

\bibitem{epelbaum2002}
E.~Epelbaum, U.-G.~Mei{\ss}ner, W.~Gl\"ockle, and C.~Elster,
\href{https://doi.org/10.1103/PhysRevC.65.044001}
{Phys.\ Rev.\ C \textbf{65}, 044001 (2002)}.

\bibitem{kamiya2015}
Y.~Kamiya, K.~Itagaki, M.~Tani, G.~N.~Kim, and S.~Komamiya,
\href{https://doi.org/10.1103/PhysRevLett.114.161101}
{Phys.\ Rev.\ Lett.\ \textbf{114}, 161101 (2015)}.

\bibitem{xu2013}
J.~Xu, B.-A.~Li, L.-W.~Chen, and H.~Zheng,
\href{https://doi.org/10.1088/0954-3899/40/3/035107}
{J.\ Phys.\ G \textbf{40}, 035107 (2013)}.

\bibitem{dev2020}
P.~S.~B.~Dev, R.~N.~Mohapatra, and Y.~Zhang,
\href{https://doi.org/10.1088/1475-7516/2020/08/003}
{J.\ Cosmol.\ Astropart.\ Phys.\ \textbf{08}, 003 (2020)}.

\bibitem{hardy2025}
E.~Hardy, A.~Sokolov, and H.~Stubbs,
\href{https://doi.org/10.1007/JHEP04(2025)013}
{J.\ High\ Energy\ Phys.\ \textbf{04}, 013 (2025)}.

\bibitem{epelbaum2009}
E.~Epelbaum, H.-W.~Hammer, and U.-G.~Mei{\ss}ner,
\href{https://doi.org/10.1103/RevModPhys.81.1773}
{Rev.\ Mod.\ Phys.\ \textbf{81}, 1773 (2009)}.

\bibitem{winkler2019}
M.~W.~Winkler,
\href{https://doi.org/10.1103/PhysRevD.99.015018}
{Phys.\ Rev.\ D \textbf{99}, 015018 (2019)}.

\bibitem{boiarska2019}
I.~Boiarska, K.~Bondarenko, A.~Boyarsky, V.~Gorkavenko,
M.~Ovchynnikov, and A.~Sokolenko,
\href{https://doi.org/10.1007/JHEP11(2019)162}
{J.\ High\ Energy\ Phys.\ \textbf{11}, 162 (2019)}.

\bibitem{batell2019}
B.~Batell, A.~Freitas, A.~Ismail, and D.~McKeen,
\href{https://doi.org/10.1103/PhysRevD.100.095020}
{Phys.\ Rev.\ D \textbf{100}, 095020 (2019)}.

\bibitem{kling2023}
F.~Kling, S.~Li, H.~Song, S.~Su, and W.~Su,
\href{https://doi.org/10.1007/JHEP08(2023)001}
{J.\ High\ Energy\ Phys.\ \textbf{08}, 001 (2023)}.

\bibitem{strassler2007}
M.~J.~Strassler and K.~M.~Zurek,
\href{https://doi.org/10.1016/j.physletb.2007.06.055}
{Phys.\ Lett.\ B \textbf{651}, 374 (2007)}.

\bibitem{kuwahara2023}
T.~Kuwahara and S.-R.~Yuan,
\href{https://doi.org/10.1007/JHEP06(2023)208}
{J.\ High Energy Phys.\ \textbf{06} (2023) 208}.

\bibitem{asadi2026}
P.~Asadi, A.~Batz, and G.~D.~Kribs,
\href{https://doi.org/10.48550/arXiv.2606.30760}
{arXiv:2606.30760 [hep-ph] (2026)}.



\bibitem{McNicoll:2010}
E.~F.~McNicoll et al. [Crystal Ball Collaboration at MAMI],
\href{https://doi.org/10.1103/PhysRevC.82.035208}
{Phys.\ Rev.\ C \textbf{82} (2010) 035208},
\href{https://doi.org/10.1103/PhysRevC.84.029901}
{Erratum: Phys.\ Rev.\ C \textbf{84} (2011) 029901}.

\bibitem{Kashevarov:2017}
V.~L.~Kashevarov et al. [A2 Collaboration at MAMI],
\href{https://doi.org/10.1103/PhysRevLett.118.212001}
{Phys.\ Rev.\ Lett.\ \textbf{118} (2017) 212001}.

\bibitem{AlGhoul:2017}
H.~Al Ghoul et al. [GlueX Collaboration],
\href{https://doi.org/10.1103/PhysRevC.95.042201}
{Phys.\ Rev.\ C \textbf{95} (2017) 042201}.

\bibitem{Adhikari:2019}
S.~Adhikari et al. [GlueX Collaboration],
\href{https://doi.org/10.1103/PhysRevC.100.052201}
{Phys.\ Rev.\ C \textbf{100} (2019) 052201}.




\bibitem{GrahamHoward2026}
S.~A.~Graham-Howard, \href{https://meetings-archive.aps.org/smt/2026/apr-r90/1/}
{APS Global Physics Summit 2026, contribution APR-R90.1 (2026)}.

\bibitem{NumericalCode} \href{https://github.com/BalytskyiJaroslaw/Hypothesis}{https://github.com/BalytskyiJaroslaw/Hypothesis}

\bibitem{lichard2006}
P.~Lichard and M.~Vojik,
\href{https://doi.org/10.48550/arXiv.hep-ph/0611163}
{arXiv:hep-ph/0611163}.

\bibitem{roos1969}
M.~Roos and J.~Pi\v{s}\'{u}t,
\href{https://doi.org/10.1016/0550-3213(69)90042-X}
{Nucl.\ Phys.\ B \textbf{10} (1969) 563--577}.

\bibitem{Bramon:2000fr}
A.~Bramon, R.~Escribano and M.~D.~Scadron,
\href{https://doi.org/10.1016/S0370-2693(01)00161-7}
{Phys.\ Lett.\ B \textbf{503} (2001) 271--276}.

\bibitem{Escribano:2020jdy}
R.~Escribano and E.~Royo,
\href{https://doi.org/10.1016/j.physletb.2020.135534}
{Phys.\ Lett.\ B \textbf{807} (2020) 135534}.

\bibitem{escribano2006}
R.~Escribano, \href{https://doi.org/10.1103/PhysRevD.74.114020}
{Phys.\ Rev.\ D \textbf{74} (2006) 114020}.

\bibitem{oppo1967models}
G.~Oppo and S.~Oneda,
\href{https://doi.org/10.1103/PhysRev.160.1397}
{Phys.\ Rev.\ \textbf{160} (1967) 1397}.

\bibitem{baracca1970general}
A.~Baracca and A.~Bramon,
\href{https://doi.org/10.1007/BF02819090}
{Nuovo Cim.\ A \textbf{69} (1970) 613}.

\bibitem{ametller1992chiral}
L.~Ametller, J.~Bijnens, A.~Bramon and F.~Cornet,
\href{https://doi.org/10.1016/0370-2693(92)90561-H}
{Phys.\ Lett.\ B \textbf{276} (1992) 185}.

\bibitem{ko1993contributions}
P.~Ko,
\href{https://doi.org/10.1103/PhysRevD.47.3933}
{Phys.\ Rev.\ D \textbf{47} (1993) 3933}.

\bibitem{ko1995eta}
P.~Ko,
\href{https://doi.org/10.1016/0370-2693(95)00284-R}
{Phys.\ Lett.\ B \textbf{349} (1995) 555}.

\bibitem{oset2003eta}
E.~Oset, J.~R.~Pel\'aez and L.~Roca,
\href{https://doi.org/10.1103/PhysRevD.67.073013}
{Phys.\ Rev.\ D \textbf{67} (2003) 073013}.

\bibitem{oset2008eta}
E.~Oset, J.~R.~Pel\'aez and L.~Roca,
\href{https://doi.org/10.1103/PhysRevD.77.073001}
{Phys.\ Rev.\ D \textbf{77} (2008) 073001}.

\bibitem{danilkin2017theoretical}
I.~Danilkin, O.~Deineka and M.~Vanderhaeghen,
\href{https://doi.org/10.1103/PhysRevD.96.114018}
{Phys.\ Rev.\ D \textbf{96} (2017) 114018}.

\bibitem{lu2020interaction}
J.~Lu and B.~Moussallam,
\href{https://doi.org/10.1140/epjc/s10052-020-7969-8}
{Eur.\ Phys.\ J.\ C \textbf{80} (2020) 436}.

\bibitem{ng1993}
J.~N.~Ng and D.~J.~Peters,
\href{https://doi.org/10.1103/PhysRevD.47.4939}
{Phys.\ Rev.\ D \textbf{47} (1993) 4939}.

\bibitem{nemoto1996}
Y.~Nemoto, M.~Oka and M.~Takizawa,
\href{https://doi.org/10.1103/PhysRevD.54.6777}
{Phys.\ Rev.\ D \textbf{54} (1996) 6777}.


\bibitem{belkov1995}
A.~A.~Bel'kov, A.~V.~Lanyov and S.~Scherer,
\href{https://doi.org/10.1088/0954-3899/22/10/004}
{J.\ Phys.\ G \textbf{22} (1996) 1383}.

\bibitem{bellucci1995}
S.~Bellucci and C.~Bruno,
\href{https://doi.org/10.1016/0550-3213(95)00380-B}
{Nucl.\ Phys.\ B \textbf{452} (1995) 626}.

\bibitem{bijnens1995}
J.~Bijnens, A.~Fayyazuddin and J.~Prades,
\href{https://doi.org/10.1016/0370-2693(96)00455-8}
{Phys.\ Lett.\ B \textbf{379} (1996) 209}.

\bibitem{volkov2026decays}
M.~K.~Volkov, A.~A.~Pivovarov and K.~Nurlan,
\href{https://doi.org/10.1134/S1547477126700019}
{Phys.\ Part.\ Nucl.\ Lett.\ \textbf{23} (2026) 259--265}.

\bibitem{escribano2012}
R.~Escribano,
\href{https://doi.org/10.22323/1.157.0079}
{PoS \textbf{QNP2012} (2012) 079}.

\bibitem{jora2010}
R.~Jora,
\href{https://doi.org/10.1016/j.nuclphysbps.2010.10.058}
{Nucl.\ Phys.\ B Proc.\ Suppl.\ \textbf{207--208} (2010) 224}.

\bibitem{Balytskyi:2018pzb}
Y.~Balytskyi,
\href{https://doi.org/10.48550/arXiv.1804.02607}
{arXiv:1804.02607 [hep-ph]}.

\bibitem{Balytskyi:2018uxb}
Y.~Balytskyi,
\href{https://lettersinhighenergyphysics.com/index.php/LHEP/article/view/156}
{LHEP \textbf{2020} (2020) 156}.

\bibitem{Schaefer:2023stm}
H.~Sch\"afer, M.~Zanke, Y.~Korte and B.~Kubis, \href{https://doi.org/10.1103/PhysRevD.108.074025}
{Phys.\ Rev.\ D \textbf{108} (2023) 074025}.

\bibitem{escribano2025assessment}
R.~Escribano, S.~Gonzàlez-Solís and E.~Royo,
\href{https://doi.org/10.1103/Y661-ZT9K}
{Phys.\ Rev.\ D \textbf{112} (2025) 114009}.








\bibitem{Balytskyi:2023}
Y.~Balytskyi, \href{https://doi.org/10.1016/j.physletb.2023.137668}
{Phys.\ Lett.\ B \textbf{838} (2023) 137668}.


\bibitem{Byckling:1969sx}
E.~Byckling and K.~Kajantie,
\href{https://doi.org/10.1016/0550-3213(69)90271-5}
{Nucl. Phys. B \textbf{9}, 568 (1969)}.

\bibitem{BycklingKajantie:1973}
E.~Byckling and K.~Kajantie,
\textit{Particle Kinematics}
(Wiley, London, 1973).


\end{thebibliography}
\end{document}